\documentclass[
aps,%
12pt,%
final,%
notitlepage,%
twoside,%
twocolumn,
nobibnotes,%
nofootinbib,
superscriptaddress,%
noshowpacs,%
centertags,
floatfix, 
preprintnumbers, %
tightenlines, %
a4paper %
]
{revtex4}

\usepackage{graphics}
\usepackage{epsfig}

\usepackage{amsmath} 

\usepackage{subfigure}

\begin{document}

\title{{\normalsize ``RADIOASTRON'' --- A TELESCOPE WITH A SIZE OF 300\,000~KM:
 \protect\\
MAIN PARAMETERS AND FIRST OBSERVATIONAL RESULTS}}

\author{\firstname{N.~S.}~\surname{Kardashev}}
\affiliation{Astro Space Center, Lebedev Physical Institute, Moscow,
Russia}

\author{\firstname{V.~V.}~\surname{Khartov}}
\affiliation{Lavochkin Scientific and Production Association, 24
Leningradskaya, Khimki, Moscow region, 141400, Russia}

\author{\firstname{V.~V.}~\surname{Abramov}}
\affiliation{Institute of Radio Technology and Electronics, Russian Academy
of Sciences, Moscow, Russia}

\author{\firstname{V.~Yu.}~\surname{Avdeev}}
\affiliation{Astro Space Center, Lebedev Physical Institute, Moscow,
Russia}

\author{\firstname{A.~V.}~\surname{Alakoz}}
\affiliation{Astro Space Center, Lebedev Physical Institute, Moscow,
Russia}

\author{\firstname{Yu.~A.}~\surname{Aleksandrov}}
\affiliation{Astro Space Center, Lebedev Physical Institute, Moscow,
Russia}

\author{\firstname{S.}~\surname{Ananthakrishnan}}
\affiliation{Giant Metrewave Radio Telescope, Tata Institute of Fundamental
Research, P.B. 6, Narayangoan, Tal-Junnar, Pune, Maharashtra, India}

\author{\firstname{V.~V.}~\surname{Andreyanov}}
\affiliation{Astro Space Center, Lebedev Physical Institute, Moscow,
Russia}

\author{\firstname{A.~S.}~\surname{Andrianov}}
\affiliation{Astro Space Center, Lebedev Physical Institute, Moscow,
Russia}

\author{\firstname{N.~M.}~\surname{Antonov}}
\affiliation{Astro Space Center, Lebedev Physical Institute, Moscow,
Russia}

% ...

\received{5 July 2012}
\accepted{12 July 2012}

\begin{abstract}
The Russian Academy of Sciences and Federal Space Agency, together with the
participation of many international organizations, worked toward the launch
of the {\em RadioAstron} orbiting space observatory with its onboard
10-m reflector radio telescope from the Baikonur cosmodrome on July 18, 2011.
Together with some of the largest ground-based radio telescopes and a set
of stations for tracking, collecting, and reducing the data obtained,
this space radio telescope forms a multi-antenna ground--space radio
interferometer with extremely long baselines, making it possible for the
first time to study various objects in the Universe with angular resolutions
a million times better than is possible with the human eye. The project is
targeted at systematic studies of compact radio-emitting sources and their
dynamics. Objects to be studied include supermassive black holes, accretion
disks, and relativistic jets in active galactic nuclei, stellar-mass black holes,
neutron stars and hypothetical quark stars, regions of formation of stars
and planetary systems in our and other galaxies, interplanetary and
interstellar plasma, and the gravitational field of the Earth. The results
of ground-based and inflight tests of the space radio telescope carried out
in both autonomous and ground--space interferometric regimes are reported.
The derived characteristics are in agreement with the main requirements of
the project. The astrophysical science program has begun.
\end{abstract}

\preprint{Astronomy Reports, 2013, Vol. 57, No.~3, pp.~153--194.}

\maketitle

\section{INTRODUCTION}

A method for obtaining very high angular resolution in radio astronomy and
a specific scheme for the realization of this method are presented in~[1--3].
It was noted that radio interferometers on Earth and in space could operate
with very long baselines between antennas, with independent registration of
the signals at each antenna. Such radio interferometers were first operated
in 1967 in Canada [4] and the USA [5]. The first trans-continental
interferometers were realized in 1968--1969, between telescopes in the USA
and Sweden~[6], and also between the Deep Space Network antennas in the USA
and Australia [7, 8]. Some of the first observations with trans-continental
radio interferometers were carried out jointly by radio astronomers in the
USSR and USA in 1969, using the 43-m Green Bank radio telescope (USA) and
the 22-m Simeiz telescope (USSR) [9, 10]. Such observations were subsequently
carried out between all continents. Modern trans-continental radio
interferometers can achieve angular resolutions of fractions of a
milliarcsecond (mas). These observations show that most active galactic nuclei
(AGNs) possess unresolved components, even on the longest projected ground
baselines (approximately 10\,000~km); see, e.g., [11, 12] and references
therein.

The possibility of creating space interferometers was discussed at a
scientific session of the Division of General Physics and Astronomy of the
USSR Academy of Sciences on December 23, 1970 [13]. The first Earth--Space
interferometer projects emerged at that time. In the 1970s, the Space
Research Institute of the USSR Academy of Sciences (IKI) working jointly with
industrial partners created the first space radio telescope (SRT), which had
a 10-m diameter reflector. This telescope had a trussed, opening construction
with a reticulated reflecting surface and receivers tuned to 12 and 72~cm.
This radio telescope was delivered to the {\em Salyut-6} manned orbital
station by the cargo ship {\em Progress} in Summer 1979, where it was tested
using astronomical objects with the participation of the cosmonauts
V.A.~Lyakhov and V.V.~Ryumin [14, 15]. One of the outcomes of these
experiments was the decision to use a rigid reflecting surface for the
{\em RadioAstron} project.

A decree of the Council of Ministers of the USSR announcing the development
of six spacecraft for astrophysical investigations at the Lavochkin Scientific
and Production Association was made in 1980. These included the decimeter- and
centimeter-wavelength interferometer {\em RadioAstron} (the {\em Spektr-R}
project), as well as the millimeter and submillimeter radio telescope
{\em Millimetron} (the {\em Spektr-M} project) [16]. The technical
specifications for the {\em RadioAstron} project had already been prepared
in 1979. The first international conference on this project took place in
Moscow on December 17-18, 1985. Agreements were signed, and an international
group concerned with the development of onboard radio-astronomy receivers
based on sets of individual technical specifications was formed. These
technical specifications were developed and issued in 1984--1985 by the
Astrophysics Division of IKI, headed by I.S.~Shklovskii. The group included
specialists from the USSR, the Netherlands, the Federal Republic of Germany,
Australia, Finland, and India. In the early 1990s, the flight models of
the first receivers at 1.35, 6.2, and 18~cm and onboard blocks of input
low-noise amplifiers (LNAs) for the 92-cm receiver were delivered to the
Astro Space Center of the Lebedev Physical Institute (ASC; formed in 1990
from the IKI Astrophysics Division and the Radio Astronomy Station of the
Lebedev Physical Insitute in Pushchino). The 18-cm receiver and 92-cm
amplifier blocks form part of the complex of scientific equipment used
with the {\em RadioAstron} SRT in flight today.

The first successful space interferometer was realized in 1986--1988 using
the 5-m diameter antenna of the NASA TDRSS geostationary satellite (USA), 
which operated at 2 and 13~cm, together with several
ground-based radio telescopes [17, 18]. The first SRT specially designed
for interferometry was the {\em HALCA} satellite of the {\em VSOP} project,
launched by Japan in 1997 [19, 20]. This 8-m diameter antenna was mounted
on a satellite that orbited the Earth in an elliptical orbit with a period
of 6.3~hr and a maximum distance from the center of the Earth of 28\,000~km.
This SRT successfully functioned at wavelengths of 6 and 18~cm until 2003.
Both of these space interferometers confirmed not only the possibility, but
also the scientific necessity of further developing ground--space radio
Very Long Baseline Interferometry (VLBI), in particular of enhancing the
angular resolution obtained by increasing the size of the orbit and of
expanding the range of wavelengths observed. All of this experience was
taken into account when preparing the {\em RadioAstron} project.

The {\em RadioAstron} SRT is a 10-m diameter reflecting antenna equipped
with a complex of 1.35, 6.2, 18, and 92~cm receivers. A {\em Navigator}
module space platform was used to install the {\em RadioAstron} antenna
and equipment complex into the {\em Spektr-R} spacecraft [21--25]. The
arrangement of the SRT and equipment complex in the {\em Navigator} module
is shown in Fig.~\ref{fig:srtnavigator}\footnote{All figures referred to in the Introduction
(Figs.~\ref{fig:srtnavigator}, 2, 4, 5, 7--10) are discussed in more detail in later
sections of this paper. Figs.~4a--4l and Figs.~7a--7g are presented as color
inserts.}. A general block schematic of the antenna and equipment complex
of the SRT is shown in Fig.~2. The precision carbon-fiber panels of the main
antenna of the SRT were manufactured and tested in Russia, and then at the
European Space Research and Technology Center (ESTEC) of the European Space
Agency in 1994 (Nordwijk, the Netherlands; Fig.~4a). Tests of the model
SRT and the equipment complex of the interferometer (Fig.~4b) were carried
out from Autumn 2003 through Summer 2004 at the Pushchino Radio Astronomy
Observatory (PRAO) of the ASC. The main parameters of the model SRT were
measured during these tests using observations of astronomical radio sources,
and test observations in an interferometric regime were carried out using
the SRT together with the PRAO 22-m radio telescope. This 22-m radio telescope
was subsequently outfitted with additional equipment enabling its use as a
ground station for tracking the {\em Spektr-R} spacecraft in flight. The
last ground tests of the SRT with the {\em Navigator} module occured at the
Lavochkin Association (Figs.~4c,d). At the suggestion of the International
Grote Reber Foundation, a memorial plate with a portrait of the pioneer
radio astronomer Grote Reber (1911--2002) was installed on the SRT (Fig.~4e).
A poster with an image of symbols of the organizations and countries
participating in the {\em RadioAstron} project was placed on the fairing
of the {\em Zenit-3F} rocket used to launch the {\em Spektr-R} spacecraft
(Fig.~4f). Figs.~4g--4i show the transport of the rocket carrier with the
{\em Spektr-R} spacecraft and the {\em Fregat} booster to the launch
position.

The launch of the {\em Zenit-3F} rocket with the {\em Spektr-R} spacecraft
took place on July 18, 2011 at 5~hr 31~min 17.91~s Moscow daylight saving
time, from the 45th launch pad of the Baikonur cosmodrome (Figs.~4j--4k). On
that same day at 14:25, the booster and spacecraft, which had separated from
it, were photographed using a 45.5-cm optical telescope in New Mexico, at the
request of the Keldysh Institute of Applied Mathematics (IAM) of the Russian
Academy of Sciences (Fig.~5). The SRT was successfully deployed on July 23,
2011 (a general view of the {\em Spektr-R} spacecraft in space is shown in
Fig.~4l). After this, it was possible to begin the inflight tests planned
for the first six months of flight: verifying the functioning of the
service systems and the scientific equipment of the spacecraft, measuring
and updating the characteristics of the orbit, measuring the main parameters
of the SRT, searching for fringes in the ground--space interferometer signal,
and beginning the Early Science Program (ESP) of astrophysical investigations.

Let us now present a brief history of key astronomical observations in the 
first half year of the inflight tests of the SRT. The radio-astronomy receivers
were successfully turned on for the first time in mid-September 2011, and
regular tests of the onboard scientific equipment were begun. Radiometric
measurements of the parameters of the SRT using radio-astronomical methods
and observations of various astronomical objects during operation of the
SRT in a single-dish regime began on September 27, 2011 (Figs.~7a,
7b, 8a--8c). The adjustment and testing of the high-data-rate radio
channel for transmitting data between the SRT and the ground tracking
station in Pushchino in an interferometric regime were conducted in parallel.
Measurements at 92, 18, 6.2, and 1.35~cm began with observations of the
Cassiopeia~A supernova remnant (Figs.~7a, 8a, 8b), then went on to
observations of Jupiter, the Moon, the Crab Nebula (Fig.~7b), the Seyfert
galaxy 3C~84, and the quasars 3C~273 and 3C~279, as well as cosmic masers
(Figs.~9a--9c) and pulsars (Figs.~7f, 7g, 10). Tests of the ground--space
radio interferometer at 18, 6.2, 92, and then 1.35~cm began with observations
of the quasar 0212${+}$735 at 18~cm on November 15, 2011 (Fig.~7c). These
tests were conducted at various distances of the SRT from the Earth, from
the minimum distance to the maximum distance of about 330\,000~km, and
using observations of various extragalactic and Galactic objects: quasars
and galaxies, pulsars, and molecular maser sources radiating in narrow radio
lines (Figs.~7c--7f).

Further, we describe the construction of the SRT and the configuration of the
onboard science complex (Section~2); the launch and inflight tests of the
{\em Spektr-R} spacecraft and ground control complex (Section~3); the
parameters of the orbit and the means used to measure and refine them
(Section~4); measurement of the main parameters of the SRT based on
astronomical sources (Section~5); and verification of the functioning of the
ground--space interferometer (first fringes) and the first observational
results (Section~6). In conclusion, we list directions for further studies.
The Appendix presents a possible interpretation of the antenna measurements
at 1.35~cm.

\section{CONSTRUCTION OF THE SRT AND CONFIGURATION OF THE ONBOARD SCIENCE
COMPLEX}

The automated {\em Spektr-R} spacecraft was designed to carry an SRT to be
used as an orbiting element in ground--space VLBI experiments. It includes the
{\em Navigator} service module (with the Lavochkin Association as
the lead organization) [21] and a science complex containing the scientific
equipment to be used for the international {\em RadioAstron} project
(with the ASC as the lead organization) together with the 10-m parabolic
antenna itself (jointly developed by the Lavochkin Association and the
ASC) [22, 23]. In addition, the spacecraft carried the scientific
equipment associated with the {\em Plazma-F} project (with IKI as the lead
organization), designed for studies of cosmic plasma along the orbit of
{\em Spektr-R} (this equipment and experiment are described in [26, 27]).

\subsection{{Construction of the Antenna}}

The design of the SRT antenna was based on the need to fit the deployable
10-m reflector in its folded state into the payload compartment of the rocket
underneath the fairing, which has a specified internal diameter
of 3.8~m, and also to ensure the required precision of the reflecting
surface after its deployment. According to the technical specifications, the
maximum allowed deviation (tolerance) of the dish surface of the radio
telescope from the profile for an ideal paraboloid of rotation under all
conditions is ${\pm} 2$~mm~[28]. The reflecting surface is formed of the
central part of the dish, with a diameter of 3~m, and 27 radial petal
segments, which open synchronously in orbit.

\begin{figure*}[t!]
\setcounter{figure}{0}
\centering
\includegraphics[width=1.0\textwidth,angle=0,clip=true,trim=0 2.5cm 0 0]{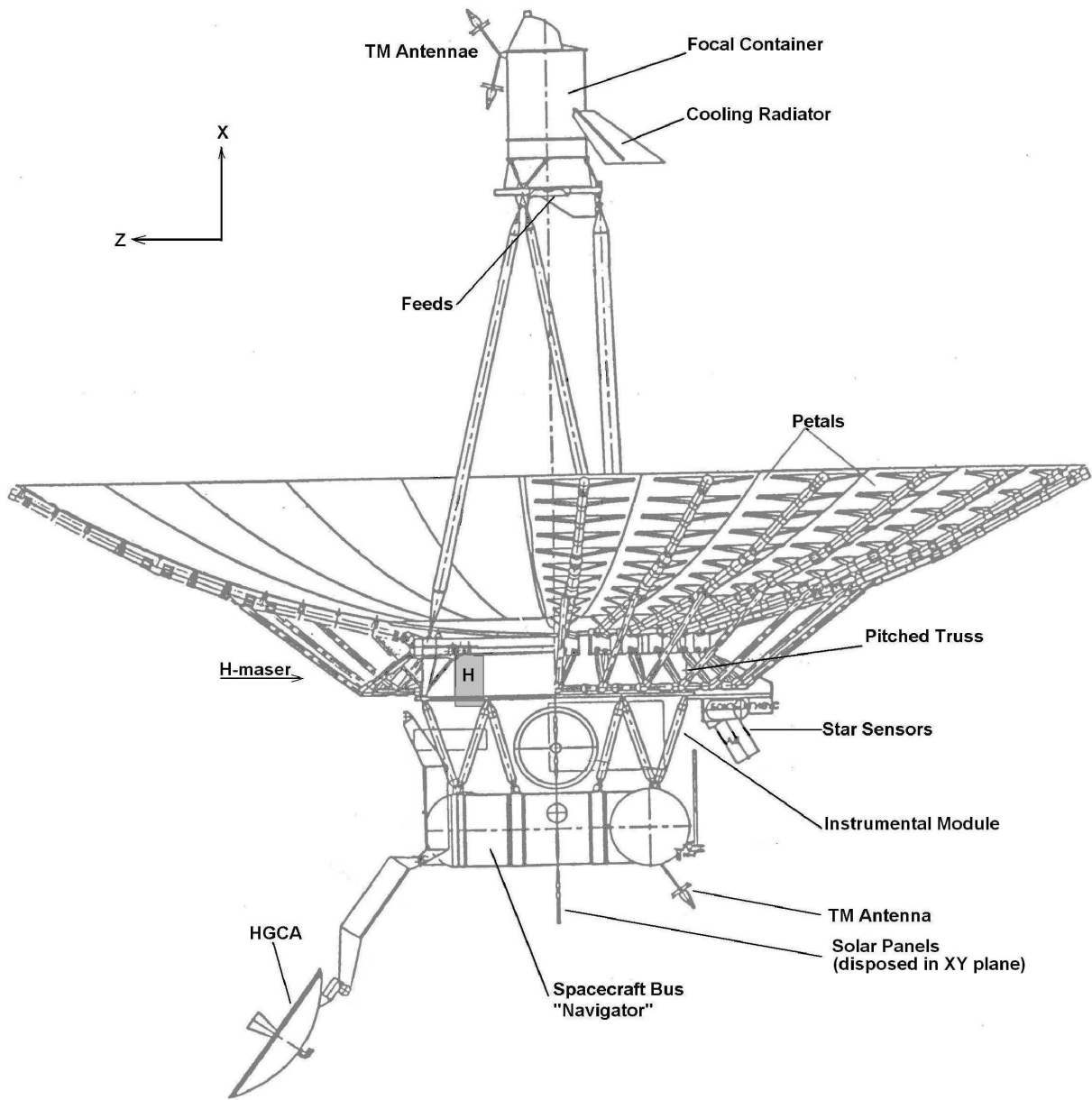}
\caption{Arrangement of the SRT in the {\em Navigator} basic module.}
\label{fig:srtnavigator}
\end{figure*}

A general schematic of the components of the SRT in the {\em Spektr-R}
spacecraft is presented in Fig.~\ref{fig:srtnavigator}. The main structural elements of the dish
are the following:

-- focal module truss (serves to regulate the position of the feedhorns);

-- reflector truss (fastens the focal module to the focal container);

-- cylindrical compartment (designed to fix the central dish and the reflector-petal
opening mechanism, and also to house the two onboard hydrogen masers);

-- a transitional truss between the SRT and the {\em Navigator} service
module (used for the installation of the scientific-equipment container).

The petal positions were aligned on the ground before launch, to allow the
creation of the precision reflecting surface upon deployment. This was carried
out in two stages. In the first, each petal was adjusted individually using
adjustment screws at 45 points on a specialized weight-unloading support,
taking into account the mass and the position of the axis of rotation of the
petal. In the second, the positions of the petals were aligned after assembling
the reflector, by varying the lengths of struts fixed to the positions of the
petals in the open state. The central dish was fixed on a cylindrical compartment using
nine regulating support units. Measurements showed that, after the alignment
on the ground, in the presence of backlash and taking into account
uncertainties in the manufacture and weight distribution, the maximum
deviation of the reflector surface from the theoretical shape of a paraboloid
did not exceed ${\pm} 1$~mm.

A thermal regulation system (TRS) for the petals, cylindrical compartment, focal 
and scientific
containers, focal unit, and onboard hydrogen masers was designed, to ensure
reliable functioning of the instrumentation complex and minimization of
thermal structural deformations~[29]. A cold plate with the blocks of LNAs
for the 1.35, 6.2, and 18-cm receivers mounted on it was connected to the
antenna-feed assembly (AFA) and installed in the focal unit of the SRT; the TRS
radiator of the cold plate was installed in the shaded side of the focal
container (Fig.~\ref{fig:srtnavigator}). The TRS of the cold plate was designed to provide the
required thermal regime for the LNAs and the central part of the AFA:
maintaining the temperatures of the LNA sites between 90 and 150~K, and the
sites of the antenna feeds (for 1.35, 6.2, and 18~cm) between 130 and 200~K,
throughout the normal operation of the SRT. The geometrical area of shadowing
of the SRT dish by the TRS cold-plate radiator does not exceed 1~m$^2$. The
maximum thermal energy deposited onto the cold plate from the LNAs is no
greater than 0.3~W. Heat flow due to thermal connections with
other structural elements of the SRT is from 5 to 15~W (this varies primarily
with the position of the SRT relative to the Sun). There is a thermal
connection between the LNAs and AFA along waveguides and cables. According
to housekeeping data, the temperature regimes for the cylindrical compartment, 
containers, focal unit, and onboard hydrogen masers of the SRT in flight 
correspond to the projected requirements.

\subsection{{Onboard Science Complex}}

The onboard science complex was constructed starting in 1985, as a
collaboration between Soviet and foreign organizations. This was carried
out in accordance with the general technical requirements for the design
of scientific equipment for the {\em Spektr-R} spacecraft and the technical
specifications for the specific scientific instruments developed by the lead
organization for the {\em RadioAstron} project --- the Astrophysics
Division of IKI, which became the Astro Space Center of the Lebedev Physical
Institute in 1990. The spacecraft was designed at the Lavochkin Association.

\begin{figure*}[t!]
\setcounter{figure}{1}
\includegraphics[width=1.0\textwidth,angle=0,clip=true,trim=0 1.9cm 0 4.6cm]{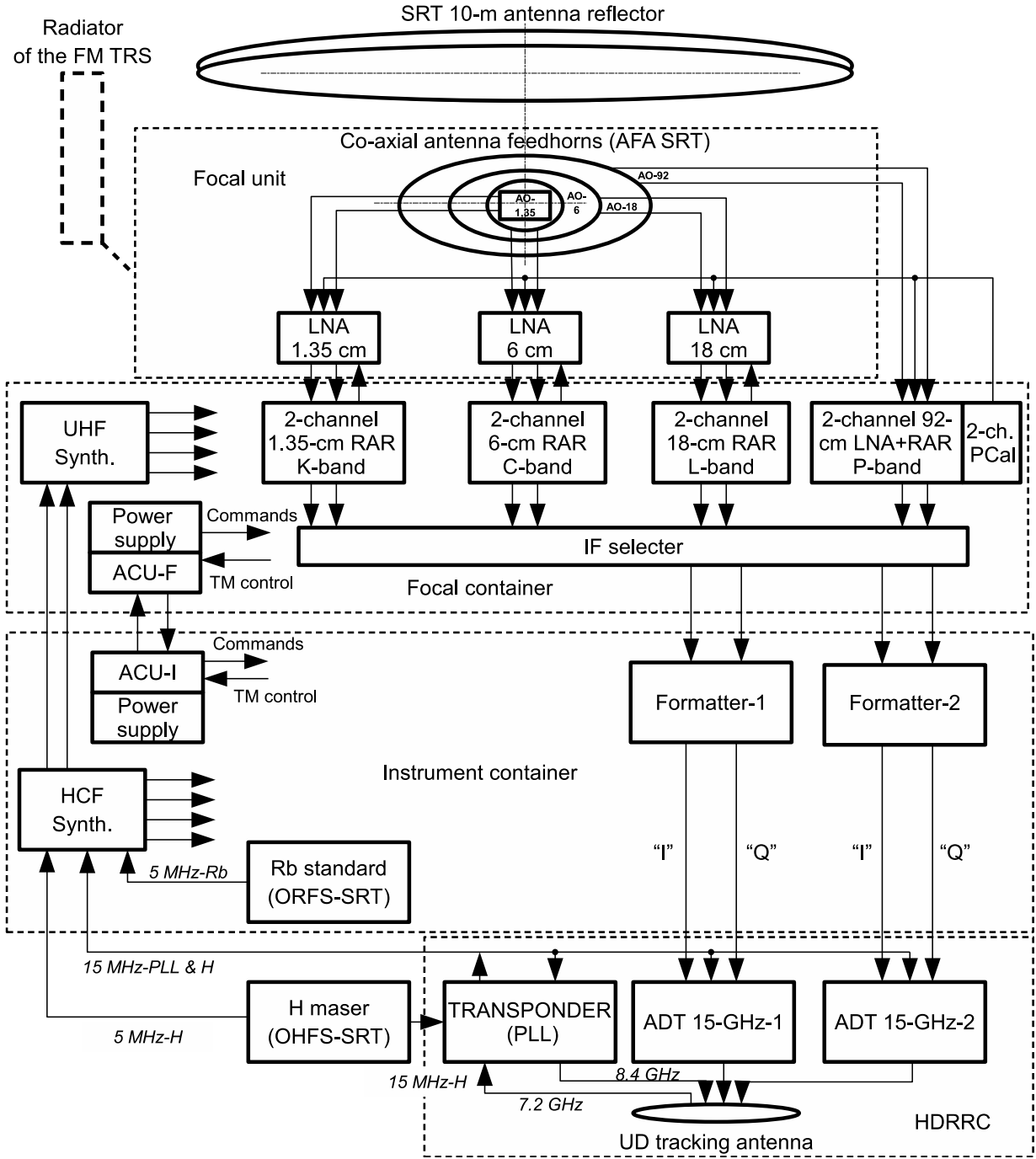}
\caption{General block schematic of the SRT. LNA represents a low-noise
amplifier; FM TRS, the focal-module thermal-regulation system; RAR, a
radio-astronomy receiver; PCal, the pulsed phase-calibration block; IF, an
intermediate frequency; TM, telemetry; Rb standard, the rubidium frequency
standard; H maser, the onboard hydrogen frequency standard (two copies); HCF
Synth., the heterodyne and clock frequency synthesizer; UHF Synth., the heterodyne
ultra-high frequency synthesizer; ACU-F and ACU-I, the analysis and control
units for the focal and instrument containers, respectively; HDRRC, the
high-data-rate radio complex; PLL, the phase link loop; I and Q, the
interferometric data fluxes in these Stokes parameters; ADT 15-GHz-1 and 2,
the astronomical data transmitters at 15 GHz; UD tracking antenna, the
two-dish, unidirectional, 1.5-m HDRRC antenna.}
\end{figure*}

The first onboard receivers began to be delivered to the ASC in the early
1990s. Ground radio-astronomical tests of the SRT were carried out at the
PRAO in 2003--2004 (see Fig.~4b in the color insert), and acceptance
tests of the entire onboard complex of scientific and service instruments
together with the spacecraft were conducted in 2009--2011. A functional
schematic of the onboard science complex is presented in Fig.~2.

The science complex consists of the following instruments and blocks,
located in the corresponding modules, shown in Figs.~\ref{fig:srtnavigator} and 2.

1. The block of co-axial antenna feeds operating at the radio-astronomy
bands 1.35, 6.2, 18, and 92~cm in right- and left-circular polarizations
is located in the thermally stabilized, cooled focal unit of the focal module,
together with the LNA blocks for the 1.35, 6.2, and 18-cm receivers.

2. The radio-astronomy receivers operating at the four wavelengths indicated
above, for both incident polarizations (denoted RAR in Fig.~2; with individual
sources of secondary electric power) are located in the thermally stablized,
hermetic focal container. The 92-cm LNA is located inside the 92-cm receiver.
Structurally, the 92-cm receiver is joined to the block containing the pulse
phase calibration units for all the receiver wavelengths. The output signals
of the receivers at the intermediate frequency (IF) arrive at the IF selecter,
which patches the output IF signals to the corresponding input frequency
converters of the formatter for further conversion to lower frequencies. The
focal container also houses a frequency synthesizer, consisting of two heterodyne
ultra-high frequency synthesizer blocks (UHF Synth-1 and 2) with their sources
of secondary electrical power and two analysis and control units
(ACU-F) with a power-switching unit (Fig.~2).

3. Two onboard hydrogen frequency standards (OHFSs; H maser in Fig.~2) and
the (scientific) instrument container are installed in the instrument module.
Two onboard rubidium frequency standards (ORFS; Rb standard in Fig.~2), a
frequency synthesizer with a double block forming the heterodyne and clock
frequency synthesizer (HCF Synth.) and two associated power sources, two
analog--digital converters for the signals from the formatter block, and
the two analysis and control units of the scientific container (ACU-I)
with their power supplies are housed in the thermally stable, hermetic
scientific container.

The {\em Spektr-R} spacecraft with the SRT onboard is a unique piece of
space science instrumentation. The entire complex of onboard equipment and
the telescope are designed for a single task: multi-frequency observations of
very weak radio emission at centimeter and decimeter wavelengths located
far below the intrinsic noise levels of the receiver systems, and the
multi-stage conversion of these signals with the very highest
available phase stability into a videoband from 0 to 16~MHz,
providing high-speed recording and transmission of the onboard data on the
Earth. To successfully carry out this task, all instruments along the ``onboard
receiver--frequency converter--onboard transmitter--ground receiver station''
serial line must function faultlessly, since the loss of signal or
phase stability for even one of these elements (not to mention the possible
failure of an element) leads to a loss of all information and a loss of very
expensive observing time for all the radio telescopes forming the multi-antenna
ground--space radio interferometer. For this reason, all the units and
instruments have backup copies, so that there is functional redundancy in the
onboard science complex, making it possible to compose this serial
line of a large number of combinations of instruments and units. All this
appreciably enhances the reliability of the operation of the complex as a
whole. A positive effect of this approach was already exhibited during the
inflight tests of the SRT: only one of the two onboard hydrogen masers
tested in flight proved to have the required characteristics. This unit
has been functioning continuously in orbit for more than a year.

\textbf{2.2.1. {Antenna Feed Assembly.}} The antenna-feed assembly (AFA) has a
special construction, and is installed at the focus of the antenna dish. It
consists of four co-axial feeds (one inside the other), in accordance with
the wavelengths of the receivers. The 1.35, 6.2, and 18-cm feeds are cooled
to about 150~K by a passive cooling system (see Section 2.1). These feeds
are connected to the cooled LNA blocks by co-axial cables (waveguides for
1.35~cm), which also provide thermal contact between the feeds and the cold
plate. The 92-cm feed is not cooled, and is at the temperature of the
ambient space; it is thermally isolated from the cooled feeds to reduce
the heat flow from the 92-cm to the other feeds. The 6.2, 18 and
92-cm feeds are resonance ``traveling wave'' feeds, with the signals divided
into right- and left-circular polarizations along co-axial outputs. The
1.35-cm feed forms the open end of a waveguide with a circular apeture, with
a circular-polarization splitter that makes a transition
into two rectangular waveguides at the output.

\textbf{2.2.2. {Receiver Complex.}} This complex consists of onboard
receivers operating at four wavelengths:

-- P band, with a central frequency of 324~MHz and a ${\pm} 7$~MHz bandwidth,
receiver P-SRT-92,

-- L band, with a central frequency of 1664~MHz and a ${\pm} 30$~MHz bandwidth,
receiver P-SRT-18,

-- C band, with a central frequency of 4832~MHz and a ${\pm} 55$~MHz bandwidth,
receiver P-SRT-6M,

-- K band, with a central frequency of 22\,232~MHz, together with seven
sub-bands for multi-frequency synthesis~[30, 31], covering the frequency
range 18\,372--25\,120~MHz, receiver P-SRT-1.35M.

The frequencies for the eight K-band sub-bands with widths of ${\pm} 75$~MHz
each for multi-frequency synthesis have the following names and central
frequencies, separated by 960~MHz: $F_{-4}$ is 18\,392~MHz, $F_{-3}$ is
19\,352~MHz, $F_{-2}$ is 20\,312 MHz, $F_{-1}$ is 21\,272~MHz, $F_0$ is
22\,232~MHz, $F_1$ is 23\,192~MHz, $F_2$ is 24\,152~MHz, and $F_3$ is
25\,112~MHz. In addition, four sub-bands can be formed for spectral
observations of narrow radio lines, with the central frequencies 22\,232~MHz,
22\,200~MHz, 22\,168~MHz and 22\,136~MHz.

All the receivers are designed to amplify, filter, and convert the noise
signals and the continuous spectrum of the indicated bands into output
signals at intermediate frequencies in the interval approximately from
405 to 555~MHz, and for the narrow-line signals into output signals at
intermediate frequencies near 400~MHz. Each of the receivers consists of two
independent, identical channels labeled 1 and 2,  whose inputs are the
left- and right-circularly polarized signals from the antenna-feed assembly.
These channels are separated into three separate blocks: the LNA block, the
receiver block, and the power-supply block. For backup of the power-supply
block, channels 1 and 2 can be connected to either their own channel (1 or
2, under the command DIRECT), or to the other receiver channel (2 or 1, under
the command CROSS). Both channels for the 1.35-cm receiver are supplied from
the main or backup power supplies, chosen by an external command. The LNAs
for L, C, and K bands are separate from the receivers, and are arranged on
the cold plate in the focal unit of the radio telescope
(Figs.~\ref{fig:srtnavigator} and 2),
where they are radiatively cooled to temperatures $130 \pm 20$~K. All the
receiver and associated power-supply blocks are located in the hermetic,
thermostatically regulated focal container at temperatures from ${+}5^{\circ}$C
to ${+}35^{\circ}$C. The P-band LNA is located in the thermostatically
controlled receiver block, at a temperature of ${+}30^{\circ}\pm3^{\circ}$C.
The construction of the P, L, and C-band receivers is based on the same
design with a single frequency conversion, while the K-band receiver has
two frequency conversions. The central frequency of the intermediate
frequencies at L and C bands is 512~MHz, and at P band 524~MHz. The paths of
the output intermediate frequencies of all receiver channels include a step
attenuator, which introduces an attenuation of 0--31~dB to establish the
required levels of the output signals at the intermediate frequency during
the ground tests and observations in space.

In addition to the output signal at the intermediate frequency, which is
fed to the IF selecter and is used further in the interferometric regime,
there are two radiometric signals with amplitudes from 0--6~dB in each of
the orthogonal polarizations at the receiver output, which are detected by
a square-law detector at the intermediate frequency: an analog signal
(converted into an 8-bit signal by the housekeeping telemetry system for transmission
to the Earth) and a digital signal (12~bit). The transmission bandwidths
of the radiometric paths to the detectors at the minus 3~dB level are equal
to 14, 60, 110, and 150~MHz at~P, L, C, and K bands, respectively; the
signal-averaging time depends on the band, and is about 1~s. The bandwidth
allocated from the IF output signal of the receiver for use in the
interferometric regime is formed in the formatter (see below); this bandwidth
depends on the observing band, and comprises from 4 to 32~MHz (two sub-bands,
upper and lower, of 16~MHz each).

In each polarization channel, there is a two-level calibration noise generator
for amplitude calibration, whose signal is summed with the external
phase-calibration signal from the pulsed phase-calibration block, which is
located in the P-band receiver and is fed to the input of the LNA block,
providing calibration of both receiver channels simultaneously. The high
level of the noise generator is close to the noise temperature of the channel,
and is used in antenna measurements. The low level of the noise generator
(which is a factor of ten lower) is used for calibration during observations
of weak radio sources. The pulsed periodic signal used for the phase
calibration, whose repetition frequency is 1~MHz, is used during observations
in the interferometric regime. A thermostatic control system is used to
enhance the stability of the amplification and the signal level of the noise
generator. The receiver blocks and noise-generator blocks within them are
separately thermostatically regulated, and the LNA blocks exposed to
open space are likewise thermally stabilized. The temperatures are monitored
through the telemetry parameters.

\textbf{2.2.3. {Onboard Frequency Standards.}} Frequency (phase) stability
is of key importance in VLBI, and is determined in first instance by the
frequency standard used, whose signal acts as a primary reference signal
for realizing the necessary subsequent frequency conversions. The SRT is
designed to function with reference signals from three sources: 1) the onboard
hydrogen frequency standard (OHFS; 5~MHz or 15~MHz), 2) the 15-MHz signal of
the phase-synchronization loop of the high-data-rate radio complex (HDRRC),
which is synchronized by the signal from a ground hydrogen maser at the
tracking station, and 3) the onboard rubidium frequency standard (ORFS; 5~MHz).

A hydrogen maser device was launched vertically in a rocket to a height of 10\,000~km
and successfully operated during two hours of flight in 1976. Its purpose
was to measure the gravitational potential and test the predictions of
relativistic gravitational theory as part of the {\em Gravity Probe A}
experiment~[32]. The European Space Agency launched the {\em GIOVE-B}
navigational satellite with three atomic clocks on board into Earth orbit in
2008; a passive hydrogen maser was used as a primary reference, and two
rubidium oscillators as secondary references~[33]. The {\em RadioAstron}
onboard hydrogen maser is
the first {\textit{active}} onboard hydrogen frequency standard in a near-Earth
orbit, and has now successfully been used to realize the orbital program
for more than a year. Therefore, the results of its inflight tests as part
of the SRT have special value, both scientifically and practically. A number
of specialized problems not usual for maser frequency standards were solved
during its construction at the Vremya-Ch Joint Stock Company:

-- degassing of the thermostats by the vacuum of space;

-- the need to enhance the structural stability of the
resonator and storage bulb for the hard conditions to which the instruments
are subject during launch;

-- temperature stabilization of the standard using the system for thermal
regulation of the instrument base, and thermal isolation of the structure
of the OHFS using multi-layer vacuum insulation;

-- carrying out ground tests in the absence of the vacuum of space;

-- a number of engineering problems associated with the use of instruments
in the vacuum of space.

In addition, new problems arose, associated with the higher stability of the
onboard masers and the appearance of new destabilizing factors in space
flight, such as the gravitational and relativistic shifts of the standard
frequency due to the motion of the spacecraft in its orbit. Since this is
the first experience using a hydrogen maser under such unusual conditions,
other unforeseen problems are also likely to appear.

\textbf{2.2.4. {Reference-Frequency Generator.}} The secondary reference
frequencies are generated in the heterodyne and clock frequency generation
blocks, and the heterodyne ultra-high frequencies in the corresponding HUHF
blocks (Fig.~2). The HCF blocks form the 64 and 160~MHz secondary reference
signals, 72-MHz clock-frequency signals, and 40~kHz synch-frequency signals
required for the functioning of the instruments in the science complex,
based on the primary reference signals at 5~MHz or 15~MHz from the OHFS
or the 15~MHz signal from the loop phase link (which will be discussed below).
The HUHF block is a functional continuation of the HCF block, and is
conceptually similar. The heterodyne signals for the 92-cm (at 200 MHz),
18-cm (at 1152 MHz), and 6.2-cm (at 4320 MHz) receivers, and also the 8-MHz
reference signals for the formation of the heterodynes inside the 1.35-cm
receiver and for the pulsed phase-calibration block inside the 92-cm
receiver, are formed from the secondary reference signals from the HCF and
HUHF blocks. This calibration is realized for all the receivers at
intermediate frequencies. The frequency-generation system of the SRT is
described in more detail in~[34].

\textbf{2.2.5. {Intermediate-Frequency Selector.}} The IF selecter patches
any IF outputs from the receivers (four outputs in each of left- and
right-circular polarization) to any two inputs of the main or reserve formatter
block (Fig.~2), apart from combinations of the same polarization in different
ranges (left with left, right with right). The configuration is specified by
the selecter keys, which are established by external commands. In
single-frequency mode, signals from one or two IF outputs from the receivers
of a specified frequency can be patched to the formatter --- in left- and/or
right-circular polarization. The signal from any one IF output can be patched
to two inputs of the formatter in parallel, which is important during test
measurements. In two-frequency mode, two IF signals from the receiver outputs
for any two frequencies can be patched to the formatter, but with the
restriction indicated above concerning combinations of the same polarization
at the different frequencies.

\textbf{2.2.6. {Formatter.}} The formatter

-- converts the signal spectrum from the receiver output from the IF
frequency range to the videofrequency range, and forms the upper and lower
sidebands of the videospectrum from 0 to 16~MHz each (SSB videoconverter);

-- carries out the conversion for transmitting the videodata to the Earth
via the onboard high-data-rate radio transmitter at 15~GHz.

The separation of the upper and lower sidebands of the videospectrum is carried
out according to a SSB-converter scheme with rotations of the signal phases
by $90^{\circ}$, $180^{\circ}$, and $270^{\circ}$, as is required for
reliable formation of the sidebands. The videosignals are converted into
digital form, and are digitally filtered using a seventh-order Butterworth
filter. The use of digital filters ensured high repeatability of the shape
of the amplitude--frequency and phase--frequency characteristics. Filter
bandwidths of 4~MHz or 16~MHz can be chosen.

The conversion chain for the transmission of the signals to the Earth includes:

-- the one-bit (two-level) clipped videosignal and its conversion to digital
form;

-- the parallel, synchronous interrogation of the digital values of the signals
in the upper and lower sidebands;

-- conversion of the parallel sub-streams of data into a denser serial,
high-data-rate stream;

-- the generation of the frame structure of the high-data-rate stream of the
synchronized, serial streams and introduction of the data from the onboard
telemetry system into the frame headers;

-- execution of differential coding of the signals for equalization of the
phase-modulated signal spectrum transmitted through the HDRRC channel.

The result of interrogating the signal values for a single sideband is a
stream with a data rate of $16 \times 2 = 32$~Mbits/s or $4 \times 2 =
8$~Mbits/s for a video bandwidth of 16 or 4~MHz, respectively. To transmit
the entire volume of information (four videobands) with a serial stream,
taking into account the introduction of a ninth parity bit for each transmitted
byte of information, the data rate is $32 \times 4 \times (9/8) = 144$~MHz
and $8 \times 4 \times (9/8) = 36$~MHz for video bandwidths of 16 and 4~MHz,
respectively.

Two IF converters are provided in the formatter system. The digital
information taken from them arrives at the data stream of the corresponding
converter at the high-data-rate stream generator. Note that the videodata
are transferred through the high-data-rate channel using a transmitter with
quadrature phase manipulation of the carrier frequency of 15 GHz. This makes
it possible to simultaneously transmit two characters of information, which
is used in this instrument. Therefore, the clock frequency of the serial
data stream can be lowered by a factor of two, so that it comprises 72 and
18~MHz for video bandwidths of 16 and 4~MHz, respectively. Differential
coding is provided to improve the energetic parameters of the video-data
transmission line in the instrument. As a result, two streams of digital
information are obtained at the output of the instrument after the coding,
but with  ``mixed'' data from the two streams from the converters. These
streams are denoted I and Q, and arrive at the modulator of the 15-GHz
transmitter of the HDRRC.

The I and Q streams are divided into frames with durations of 2.5~ms and
10~ms for the clock frequencies of 72~MHz and 18~MHz, respectively. Within
a frame, the information is transmitted in bytes. A ninth parity bit is
created for each 8 bits, which is transmitted in the data stream. The bits
are rigidly fixed to the source of data, so that they can be identified and
sorted according to the corresponding groups after the arrival of the
stream at the Earth (to reconstruct the sub-streams of the onboard formatter).

For housekeeping purposes, the first 30 bytes in a frame are formed as a
header, which includes a synchronization packet of seven bytes (for precise
determination of the times for the 1st bit and 1st byte of the frame and
the subsequent correct decoding of the binary data), a frame counter (2 bytes)
from the 1st to the 400th frames, and the bytes of certain accompanying
information. The first 10 bytes of the header are used to transmit telemetry
information from the standard onboard telemetry system, which is especially
important in the observing regime with the housekeeping telemetry channel for
command--measurement information turned on (see below). The operational
mode for the converter is chosen using commands transmitted to the instrument
along the address bus using control code words (CCWs).

\textbf{2.2.7. {Analysis and Control Units.}} The onboard science
complex is controlled mainly through the ACU-I and ACU-F instuments,
using pulsed functional (PF) commands and CCWs. In the ACUs, the digital
CCW commands are converted into commands analogous to pulsed
commands. Some instruments (the OHFS, P-SRT-1.35, and P-SRT-Rec) are
controlled directly by CCW commands sent along the address bus.

Monitoring of the functioning of the complex instruments is carried out using
the standard onboard telemetry system. Telemetry signals arrive at this
system directly from the instruments or via the ACU collection system, in
accordance with the requirements of the apparatus. Some of the telemetry
data are generated in the form of digital databases (for example, some of
the data from the 1.35~cm receiver and all the data from the OHFS are
telemetrized in this say).

Currently, the onboard science-equipment complex is providing full
functioning of the SRT in essentially all operational modes, thanks to the
system of functional and instrumental duplication. Most of the duplicated
instruments are located in reserve, as a contingency.

\subsection{{Ground Tests}}

During preparations for the {\em Spektr-R} launch, various tests were
carried out at the ASC in accordance with the requirements for the scientific
equipment to be used. At early stages in the construction of the SRT, the
goal of such tests was to achieve the required technical specifications
for the parameters of individual instruments. Later tests of the onboard
science-equipment complex and the spacecraft were designed to determine the
capabilities for their joint operation in flight.

Starting from the mid-1990s, after the first sets of instruments were
delivered, tests of their electrical coupling and electromagnetic compatibility
were carried out at the ASC. The programs and methods for the tests were
developed as they proceeded, and the functional adequacy of the instruments
and the completeness of the complex of scientific equipment was determined.
Three integrated tests based on a zero-baseline interferometer were carried
out in 1999--2002, during which specific parameters of the interferometer
were obtained and compared with calculated values. A set of receiving
equipment designed for ground radio telescopes was used as the second element
of the interferometer. By the second half of 2002, the entire radio complex
was technologically ready to conduct radio-astronomical tests at a specially
built test facility at the PRAO.

\textbf{2.3.1. {Radio-Astronomy Tests.}} The SRT was assembled at this
test facility on a support structure in 2002--2003. The dish surface was geodetically
adjusted, the electrical assembly of the science-equipment complex and ground
equipment carried out, and the entire complex and test facility functionally
checked. From the end of 2003 through mid-2004, radio-astronomical tests of
the engineering model of the SRT were carried out using actual astronomical
sources (see Fig.~4b in the color insert).

Fluctuations of the sensitivity of the system, the effective area of the
SRT, and the width and shape of the main lobe of the antenna beam were
measured in the radiometric regime. The positions of the first sidelobes
and the level of scattering outside the main lobe of the antenna beam were
determined (including using observations of the Moon); see Table~2 in Section~5.
The focal container of the radio telescope was adjusted to determine the
position of the focus relative to the calculated value, and the difference
between the positions of the geometrical and electrical axes of the SRT was
determined.

Observations of astronomical sources for tests of the SRT in a
radio-interferometric regime were conducted at 6.2 and 1.35~cm using the
PRAO 22-m radio telescope as a second interferometer element. This same
two-element interferometer was used to investigate the interference environment
at all the operational wavelengths of the SRT, and the possibility of
transmitting reference signals from a hydrogen maser. The electromagnetic
compatibility of the 1.35-cm receiver and the 15-GHz HDRRC transmitter was
investigated, and individual elements of the tracking station were tested, in
particular, the S2 and RadioAstron data recorders and the ASC--NRAO decorder.

Although the results of these tests led to difficult decisions about changing
the formatter, antenna-feed assembly, and decoder and the need to further
develop the ACUs and RadioAstron data recorders, the main result of the tests
was that the ASC obtained an operational radio-electronical complex, i.e., a
full set of scientific equipment, for further study.

\textbf{2.3.2. {Zero-Baseline Interferometer Tests.}} After completion of the
radio-astronomical tests, the SRT was disassembled and the entire engineering
model of the science-equipment complex was sent to the ASC for further
zero-baseline interferometer tests. This stage of the testing was continued
during 2005--2008, and the tests were carried out at 6~cm. The first task
of these tests was the practical verification of the compatibility of the
scientific data obtained by the space and ground radio telescopes. This task
was successfully completed in full. The second task was comparison of the
interferometer parameters measured through a data-correlation analysis with
their calculated values. The results of the comparison were satisfactory,
and provided reliable experimental material for determining the interferometer
sensitivity and the required coherence time for the integrated signal.

In mid-2008, a flight model of the onboard science complex was delivered
to the ASC for use in zero-baseline interferometer tests. Tests at 6.2 and
1.35~cm were conducted using this model, but with new onboard P-SRT-6M and
P-SRT-1.35M receivers. The results of these tests showed not only a good
agreement between the calculated and experimental parameter values, but also
stability of these values both in time and for different models. Based on
these test results, and taking into account the dual-channel design of the
onboard receivers, it was decided to simplify and shorten further tests due
to the subsequent unavailability of the ground scientific equipment. The
interchannel correlation function obtained during the correlation
reduction of signals that had passed through corresponding pairs of receiver
channels was adopted as a key parameter estimating the operation of the
onboard science complex. Further, when conducting various grades of factory
tests at the Lavochkin Association, the interchannel correlation function was
used as the main parameter characterizing the state of the onboard science
complex.

This essentially completed the radio-engineering tests of the SRT equipment
at the ASC. The suitability of the apparatus for radio interferometric
observations and its full radio-engineering compatibility was demonstrated.

At the beginning of 2009, the flight model of the onboard science complex
of the SRT was sent to the Lavochkin Association for the final assembly and
integrated and acceptance factory tests of the SRT complex. The assembly of
the entire SRT took place in 2010--2011, together with integrated factory
tests and acceptance tests of the SRT complex. The complex was mounted on
the {\em Navigator} service module in April--June 2011, and integrated tests
of the {\em Spektr-R} spacecraft were successfully completed. Although
these were electrical tests of the SRT complex, in the interests of verifying
the future joint functioning of the scientific equipment and the service
module in flight, the interchannel correlation function was continuously
monitored during these tests. The fully assembled launch vehicle and 
{\em Spektr-R} spacecraft were transported to the
launch position in July 2011, and the {\em Spektr-R} spacecraft was
successfully launched on July 18, 2011. In the following sections, we present
material on inflight tests of the SRT and the transition to the main science
program.

\subsection{{SRT--Ground High-Data-Rate Radio Line}}

The high-data-rate radio line includes the onboard HDRRC and the ground
tracking station, together with the scientific data collected using
the PRAO 22-m radio telescope in Pushchino.

\textbf{2.4.1. {HDC Onboard Complex.}} The onboard HDRRC is designed to
transmit data from the SRT to the ground tracking station at a high rate, and
to synchronize the onboard reference frequency using a signal from a ground
hydrogen maser in one of the operational regimes of the SRT and the HDRRC. The
HDRRC can operate in one of two regimes: ``COHERENT'' or ``H maser''.

In the ``COHERENT'' regime, the HDRRC is used to synchronize the 15-MHz onboard
reference signal for the SRT frequency-generation system, as well as the HDRRC
transmitter signals at 8.4~GHz (a power of 2~W) and 15~GHz (a power of 40~W).
This is done using a hydrogen-maser signal that is transmitted to the
spacecraft from the ground tracking station. In the ``H maser'' regime, the
15-MHz HDRRC transmitter signals are synchronized using a signal from the
onboard hydrogen maser. It is possible for the HDRRC to operate with a lower
transmitter power output (4~W) at 15~GHz.

The HDRRC includes the antenna-feeder system and onboard radio-engineering
complex. The antenna-feeder system includes:

-- a double-reflector, receiving--transmitting, narrow-beam antenna with a diameter
of the primary reflector of 1.5~m;

-- a rotating waveguide junction joined to the drive of this antenna; and

-- a waveguide tract and filters.

The onboard HDRRC radio-engineering complex contains:

-- a transponder phase-synchronization loop at 7.2/8.4 GHz;

-- a radio transmitter at 15~GHz.

\begin{figure*}[t!]
\setcounter{figure}{2}
\includegraphics[width=1.0\textwidth,angle=0,clip=true,trim=2.5cm 2.8cm 1.0cm 2.1cm]{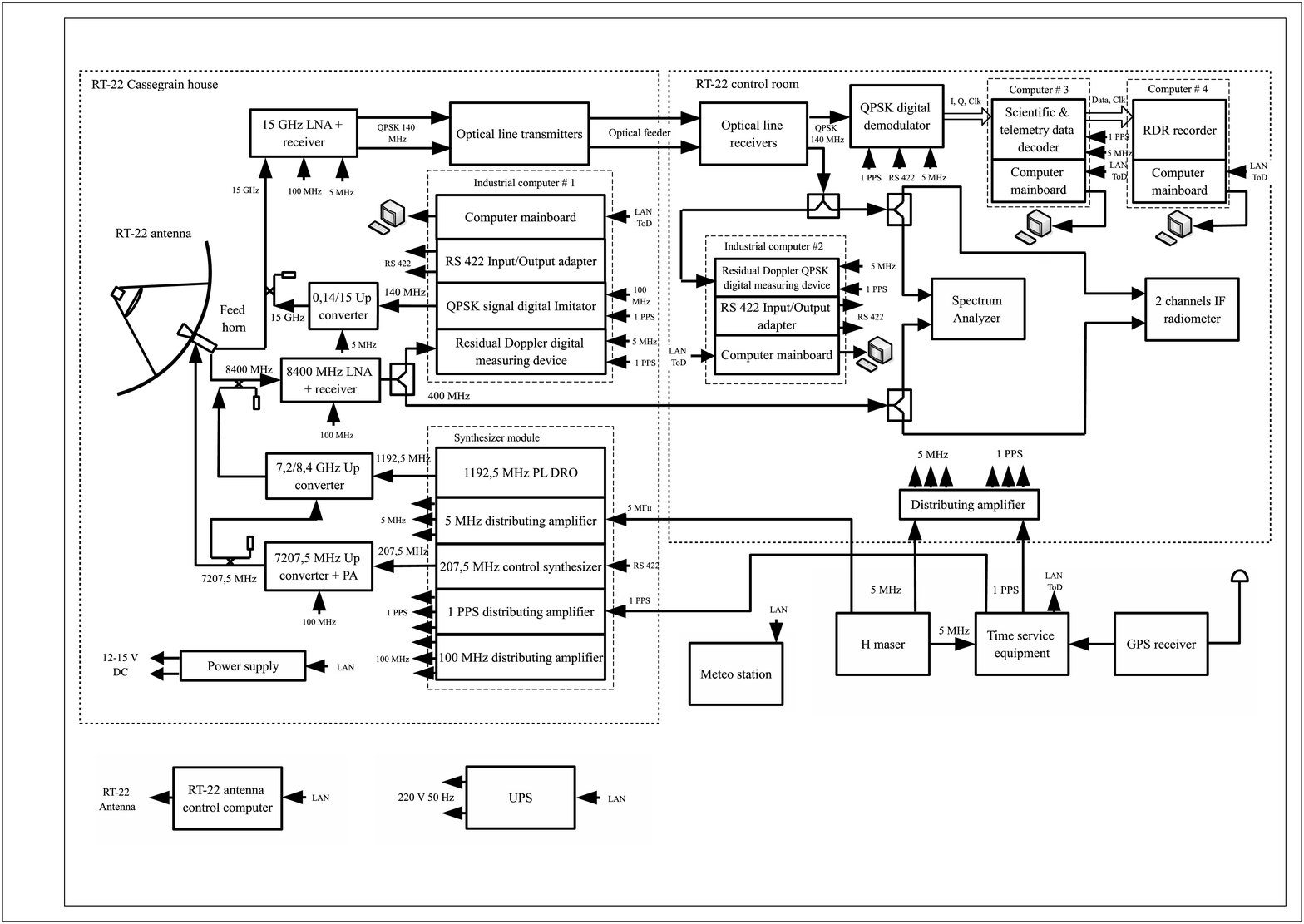}
\caption{Stuctural schematic of the ground tracking station in Pushchino.}
\end{figure*}

\textbf{2.4.2. {Ground Tracking and Scientific Data Acquisition Station.}}
The ground tracking and scientific-data-acquisition station is part of the 
high-data-rate SRT--ground radio link of the {\em RadioAstron}
project. A structural diagram of the tracking station is presented in Fig.~3.
The station is designed to

1) point the PRAO 22-m ground radio telescope toward the SRT and track the
spacecraft during a link session;

2) receive and record the flow of scientific and housekeeping data from the
spacecraft;

3) transmit a phase-stable reference signal synchronized by a ground hydrogen
frequency standard (the tracking station H maser) to the spacecraft;

4) receive the response signal coherently converted onboard the spacecraft,
measure the current frequency of the residual Doppler shift\footnote{The
frequency of the residual Doppler shift refers to the difference between the
measured frequency of the response signal and the frequency predicted taking
into account the Doppler effect.} and the current phase difference between the
response and interrogation signals, and record these measurements with a
current time tag;

5) receive the external data required for the operation of the ground station
and issue information about the status of the ground tracking station and
the data collection to users.

The {\textbf{ground tracking station}} includes:

-- the PRAO 22-m radio antenna, pointing system, feedhorn, and antenna-feeder
tracts at 15, 8.4, and 7.2~GHz;

-- the phase-synchronization transponder system at 7.2/8.4~GHz;

-- the system for the reception of scientific and housekeeping data at 15~GHz;

-- the system for measuring the Doppler residual and variations in the
HDRRC signal phases;

-- the system for recording the scientific and housekeeping data;

-- the system for the reference frequencies, time service, and weather station;

-- the control computer and station software;

-- the apparatus for monitoring the operation of the station;

-- the apparatus for external links and the cable-distribution network.

The effective area of the 22-m antenna with the tracking-station antenna-feeder
system and the receiver system noise temperature were measured. The measured
effective area of the 22-m antenna is 170~m$^2$ at 15~GHz. The system noise
temperature is about 100~K at both 8.4 and 15~GHz.

{\textbf{Apparatus for Measuring the Frequency Doppler Shift.}} The operation
of the ground tracking station's system for measuring the frequency Doppler
shift was tested during the inflight tests. Since these measurements are carried
out at each of the downlink frequencies of 8.5 and 15~GHz, there are two
such measuring systems at the station. The 8.4-GHz measuring device operates
using a tone signal emitted by the onboard HDRRC phase-synchronization loop,
while the 15-GHz measuring device operates using a phase-modulated signal
(produced using the ``quadrature phase manipulation'' method) emitted by the
HDRRC VLBI-data transmitter. During these tests, these signals were fed to the
inputs of the Doppler-shift measurement device and sent to the screen of a
spectrum analyzer in parallel, which was used to carry out independent
measurements of the frequency, signal-to-noise ratio, and other parameters of
the input signals. The 8.4-GHz Doppler-shift measurement device can operate
in one of three regimes:

{\textbf{1) ``Without control''}}: the frequencies of the downlink, Doppler
residual, and integrated phase are measured; the uplink signal from the
ground tracking station is not transmitted; the ballistic-software data are
used only to obtain the Doppler residual; used in the ``H-maser'' regime
of the HDRRC;

{\textbf{2) ``Ballistical''}}: data from a ballistic file (the delay and its
first and second derivatives) are used to control the uplink frequency;
measurements of the downlink frequency, Doppler residual, and integrated
phase are recorded; used with the ``Coherent-B'' regime of the HDRRC;

{\textbf{3) ``Autonomous''}}: used for independent control of the uplink
frequency based on measurements of the downlink frequency, Doppler residual,
and integrated phase; the ballistic data are used only to obtain
the initial delay; used with the``Coherent-A'' regime of the HDRRC.

The 15-GHz measuring device always operates only in the ``without control''
regime, measuring and recording the frequencies of the downlink, Doppler
residual, and integrated phase. The RefFreq program controlling the measuring
device was used to verify the operation of this device at 8.4~GHz. Analogous
verification was carried out for the measuring device at 15~GHz using the
RefFreqM program. The test results confirmed the full operability of the
measuring devices at 8.4~GHz and 15~GHz and their software in all operational
regimes.

{\textbf{Apparatus for the Reception of Scientific Video and Telemetry
Information.}} This apparatus includes instruments in the channels for the
reception of scientific video and telemetry (TM) information. The former
consists of the science data decoder and the RadioAstron data recorder (RDR).
The science data decoder extracts the useful signal from the data flow
arriving at the decoder from the science data demodulator (the scientific
data are subject to a special form of phase modulation onboard). The science
data decoder also decodes and operatively monitors its input data. The
data recorder writes the scientific data in the operational mode. The
start of each recording is synchronized by short pulses with a period of
1~s using a 5-MHz reference signal from a hydrogen maser at the ground tracking
station. The duration of a recording is six to nine hours at the highest
recording speed. The recorder is controlled either directly or via an Ethernet
channel with remote access. The data recorder at the ground station is used
together with control software.

The apparatus in the channel for the reception of telemetry data includes
a decoder for the specialized telemetry data (10~bytes at the beginning of
the frame headers) from a dedicated data-transmission line at the Flight
Control Center (FCC) of the Lavochkin Association sent from the tracking station
in Pushchino. The TM data decoder extracts from each frame of the HDRRC the
10~bytes of telemetry data from the standard telemetry system of the spacecraft,
and saves this information to a hard disk and/or directly transmits the
telemetry data to the FCC or the ASC via an Ethernet port.

Experience has been obtained with the reception, decoding, and transmission of
telemetry data from the spacecraft to the FCC during scientific and test
observations. In all scientific SRT--ground link sessions, the frequency and
Doppler residual were measured at 8.4 and 15~GHz along the HRCRC channel to
the tracking station in Pushchino. These measurements were then fed to an ftp
server at the ASC data-reduction center for further processing and analysis.

Studies of the operation of the high-data-rate radio line, consisting of
the HDRRC complex and the tracking station, were carried out during link
sessions between the spacecraft and the ground tracking session, both with
the SRT operating as a single dish and as an element of a multi-antenna
radio interferometer together with ground radio telescopes. To enhance the
level of the signal arriving at the Earth, the program used to point the onboard
HDRRC antenna toward the 22-m ground radio telescope in Pushchino was refined.
The agreement of the polarizations of the onboard HDRRC antenna and the
antenna-feeder system of the 22-m telescope in the 15-GHz receiver channel
was verified and corrected, leading to an increase in the received power by
nearly a factor of ten. The high potential of the radio line providing
stable operation of the entire complex in Pushchino was confirmed for various
distances of the spacecraft in its orbit. At relatively nearby distances
(less than 150\,000--200\,000~km), the transmitter power is chosen to be
4~W, while this power is increased to 40~W at greater distances. The
connections between the ground station, the Lavochkin FCC, and the Ballistic
Center, Science Scheduling Center, and Science Data Reduction Center (SDRC)
of the ASC have all been debugged.

\section{LAUNCH, INFLIGHT TESTS OF {\em SPEKTR-R} AND THE GROUND CONTROL
COMPLEX}

The {\em RadioAstron} {\em Spektr-R} spacecraft was launched from the
Baikonur cosmodrome on July 18, 2011 at $05:31:17.91$. The spacecraft was
inserted into its orbit using a {\em Zenit~2SB.80} rocket and a {\em
Fregat-SB} booster, and the sequence of activities for the first session was
realized (Figs.~4, 5).

The scheme for introducing the spacecraft [21] into its orbit (with a perigee
height 577~km, apogee height $h_a = 336\,863$~km, and orbital inclination
$i = 51.6^{\circ}$) included successive transfers to a supporting orbit
($h_p = 177$~km, $h_a = 447$~km, $i = 51.4^{\circ}$) and an intermediate orbit
($h_p = 444$~km, $h_a = 3711$~km, $i = 51.5^{\circ}$). The control of the
{\em Spektr-R} spacecraft is carried out by the FCC of the Lavochkin
Association.

The {\em Spektr-R} spacecraft was constructed at the Lavochkin Association,
based on the {\em Navigator} space platform [21, 24], which was successfully
developed for the {\em Elektro-L} spacecraft launched at the beginning of
2011. The {\em Spektr-R} spacecraft is controlled by the Main Operations
Control Group (MOCG) at the Lavochkin Association, with the participation
of specialists from the organizations that developed the onboard systems and,
in particular, the onboard science complex, ground control segment, and
ground science complex. The principles for the organization of the work of
the MOCG are those traditional for the Lavochkin Association. The control and
analysis groups include specialists from the Spacecraft Logic and Control
Division who participated in the planning of the spacecraft and its ground
tests, and also in the preparation and tests of the ground segment for control
of the spacecraft. The staff of the analysis group includes specialists of the
Special Design Bureau supervising the corresponding onboard systems, who
were also involved in all stages of planning and testing of these systems.
Specialists from the Lavochkin Association also make up the Ground-Segment
Control Group, Ballistic Group, FCC Instrument and Software Group, and GS-3.7
Ground-Station Group at the Lavochkin Association. The MOCG also functions as:

-- the Science Operations Group for the Science Scheduling Center of the ASC,

-- the Science Operations Group of IKI for the {\em Plazma-F} project,

-- the Technical Operations Group for the SDRC,

-- the Technical Operations Group for monitoring of the data-conversion block.

The reduction of measurements of the orbital parameters and reconstruction and
prediction of the spacecraft orbit are carried out by the Technical Operations
Group of the Ballistic Center of the IAM, with the participation of specialists
from the Lavochkin Association.

The operation of the MOCG began long before the launch of the spacecraft, and
included preparing the apparatus and software facilities of the FCC,
the spacecraft control software, operational and technical documentation,
training of personnel, conducting autonomous and integrated tests of the
ground control segment, debugging the connections between the FCC facilities
and the control stations and ground science complex, and conducting
practice sessions for the Control Group. This approach to the formation of
a main control group, traditional for the Lavochkin Association, which
also organizes preparation of personnel and the apparatus and software
facilities enabled the preparedness and reliable control of the {\em
Spektr-R} spacecraft from the very first days of flight.

One special characteristic of the organization of the operation of the
{\em RadioAstron} ground--space interferometer is the need to coordinate
the actions of the SRT, ground radio telescopes, the ground tracking stations,
stations for command and control of the spacecraft, the FCC, the Science
Scheduling Center, the Ballistic Center, and science-data reduction centers,
including facilities for communication between these elements. The main task
in the current stage of the project is carrying out a program of science
observations during a number of science sessions. As a rule, an observing
session lasts several hours, but the duration can be days or more in some
cases. A session corresponds to a series of operations providing recording
of the data from an observed source, conversion of the signals
obtained into digital form, transmission of these data to the ground tracking
station, and collection of the scientific data. The scientific
data\footnote{The scientific data also refers to the large volume of telemetry
data produced by instruments in the science complex, including the
low-frequency radiometric outputs of the astronomy receivers, which are
collected onboard by the standard telemetry system into a common data stream
together with data from the housekeeping system and are transmitted to the
Earth along another channel --- the standard radio channel for the
command and control system --- through small, onboard service antennas, and
to the command and control ground stations (see below).} are transmitted to
the ground tracking station via the high-data-rate radio channel in the
Ku band (2~cm) and an onboard narrow-beam, 1.5-m antenna, which is controlled
from the onboard control complex. The {\em Spektr-R} spacecraft is able to
operate with several ground tracking stations. As was noted above, the
station that is currently used for tracking and scientific-data acquisition is
the PRAO 22-m radio telescope of the ASC.

A target source is observed with a network of ground radio telescopes
simultaneously with the SRT. Several dozen radio telescopes have equipment
that is compatible with that of the {\em Spektr-R} SRT, and can in principle
participation in joint VLBI observations. The participation of these
observations is determined in part by the requirements of the specific
science projects to be carried out.

The command and control ground stations used with the spacecraft include
the ``Kobal't-R'' station at Bear Lakes (Moscow region), which has a TNA-1500
antenna complex (Moscow Energy Institute) and an antenna with a 64-m diameter,
and ``Klen-D'' (Ussuriisk), which has a P-2500 antenna complex and a
70-m-diameter antenna. The mean command-session duration is about four hours.
In accordance with a decision by the control group, as a rule, the onboard
command-measurement system transmitter is not turned off at the end of a
session, in order to make it easier to enter into the new link in the following
session. However, this transmitter is turned off during intervals when the
science receivers of the SRT are switched on, i.e., during tests and science
observations. When turned on, the transmitter can also be used to monitor the
telemetry information from the spacecraft at distances to 120\,000~km (near
perigee) using the Lavochkin NS-3.7 ground station with its 3.7-m-diameter
antenna.

The typical program for a control session consists of the following operations:

-- monitoring the current housekeeping information as it is being directly
transmitted;

-- uploading of command sequences for the spacecraft systems for flight and
attitude control, control of the antennas and telemetry system, control of 
the onboard command complex, entering command sequences for ballistic and
navigational use (roughly once in five days), and uploading individual code
commands either directly or with a time lag;

-- monitoring telemetry information recalled from onboard memory and the
electrostatic control system;

-- monitoring of the spacecraft orbit;

-- unloading of the attitude-control reaction wheels;

-- playback of the scientific telemetry information from the science-data
collection system of the {\em Plazma-F} complex;

-- uploading of the command sequences for control of the {\em Plazma-F}
science-apparatus complex.

During a command and control session, an operational program enabling the following
tasks of a typical operational cycle in an autonomous regime is uploaded in the
form of command sequence.

1. Conducting an observing session consisting of the following individual
operations:

-- successive rotations of the spacecraft into a specified attitude, 
enabling pointing of the
SRT toward a target and pointing of the HDRRC narrow-beam 1.5-m antenna
toward the ground tracking station in Pushchino;

-- turning on the required operational regimes of the SRT instrumentation
at the observation time;

-- reverse rotations of the spacecraft to its original attitude.

2. Radio-adjustment of the SRT:

-- operations analogous to an observing session, but with the realization of
a series of successive reorientations of the spacecraft relative to the
direction toward a calibrator radio source without pointing of the HDRRC
antenna at the Pushchino tracking station (the SRT information is written
to an onboard memory unit and recalled at the following control session).

If the required spacecraft attitude is expected to lead to a worsening of
the temperature regime of the structural elements of the HDRRC complex
(girders, antenna drive, transmitter), the HDRRC antenna can be moved to the
position corresponding to the position of the Sun on the spacecraft axes.

3. Laser ranging of the spacecraft:

-- reorientation of the spacecraft over an hour to the position in which
the $-X$ axis of the spacecraft is oriented toward the Earth (i.e., the
SRT is facing away from the Earth);

-- bringing about the spacecraft attitude required for the plasma-energy
monitoring instrument (MEP) from the {\em Plazma-F} complex, with the Sun
located at $100^{\circ}$ to the ${+}X$ axis, for a duration of up to six
hours, with movement of the HDRRC antenna to a specified position.

The sequences of operations during observations of sources, radio adjustments,
and laser ranging are determined by the monthly program of scientific
activities generated by the SRT Science Operations Group (ASC). This
program takes into account scientific tasks, the current ballistic parameters
of the orbit, the current constraints on the activities of the ground radio
telescopes, and constraints on the duration of observational regimes for
specified attitudes of the spacecraft, depending on the position of the
Sun relative to the spacecraft axes and the position of the HDRRC antenna.
The Ballistic Group for Inflight Analysis and the Thermal Regulation System
(TRS) Group applies the operational-analysis software of the FCC to evaluate
the realizability of the monthly science program from the point of view of
all the constraints.

The TRS Group is accumulating a large amount of statistical material, which
can be used to predict variations of the temperature fields in critical
structural elements of the spacecraft as a function of the positions of the
Sun and the HDRRC antenna with the required accuracy. Work is being carried
out on the automation of required calculations for enhancing the efficiency
and reliability of such predictions, and also to facilitate estimated
predictions by specialists at the ASC at the stage of formulation of the
monthy science program. The TRS Group uses a specially developed
three-dimensional model of the spacecraft enabling visual illumination of
the structural elements of the spacecraft for various positions of the Sun
and the HDRRC antenna.

The Command and Control Group of the Lavochkin Association develops the monthly program
for the {\em Spektr-R} spacecraft based on the monthly science program.
In accordance with proposals, and the preferred intervals for the special
attitude of the spacecraft required for optimal operation of the MEP
instrument, the program includes additional operations on the reorientation
of the spacecraft. The {\em Spektr-R} program is confirmed by the operational
technical administration of the MOCG, and becomes the main document facilitating
coordination of the operational work of all systems of the {\em Spektr-R}
spacecraft. The program contains the schedule of sessions for the following
month, the schedule of all main operations with the spacecraft, and the
schedules of operation for the control stations, tracking stations, ground
radio telescopes, and laser-ranging stations. The actions of the command and control
stations are organized by the Ground Segment Control Group at the Lavochkin
Association, of the station in Pushchino and the ground radio telescopes
by the SRT Science Operations Group at the ASC, and of the laser-ranging
stations by the Ballistics and Navigation Group at the IAM.

One day before the following session, the Command and Control Group develops a
plan for the session, in accordance with the monthly schedule for the
{\em Spektr-R} spacecraft and based on template programs. The spacecraft
navigation data and data from the HDRRC antenna are analyzed by the
Ballistic Analysis Instrumental--Software Control Group of the FCC. The
generation of inflight specifications for the control of the SRT
science-equipment complex and the {\em Plazma-F} complex is done automatically
based on command-software information prepared by the SRT and {\em Plazma-F}
Science Operations Groups. The session program is generated in the form of
a control file containing command-software information for controlling the
spacecraft and commands for controlling the ground command--measurement
stations. The correctness of the program is verified using a data--logicical
model for the onboard control complex, which fully corresponds to the
programmatic part of the real onboard complex of the {\em Spektr-R}
spacecraft. This modeling is carried out for successive time intervals, from
the beginning of the session being verified to the beginning of the following
planned session, for one to two days of flight.

The realization of the program for a housekeeping telemetry session occurs in an
automated regime, with the commands directed toward particular instruments
being obtained from the spacecraft and the command--measurement stations.
During such sessions, the telemetry data from the spacecraft arriving at
these stations is processed at the Lavochkin FCC, analyzed by specialists
of the Analysis Group, and transferred to the SDRC of the ASC. The
measurements of the orbital parameters are sent to the FCC
from the command--measurement stations, and further to the IAM.

As the tasks in the program of inflight tests of the onboard systems of the
spacecraft were carried out, the number of Analysis-Group specialists who
were involved in routine operations was reduced. Currently, only specialists
of the Complex Analysis Group, Onboard Control Complex Service, and TRS
Service regularly participation in the routine monitoring of the telemetry
information from the spacecraft. The telemetry analysis uses a program for
the automated monitoring of important parameters of the spacecraft to ensure
compliance with tolerances and with predicted values obtained in simulations
of sessions using the onboard control complex model. The Onboard Systems
Service is able to monitor the telemetry information outside the FCC, at its
work places. Once they have received the housekeeping telemetry information from
the FCC in real time, specialists in the SRT and {\em Plazma-F} Science
Operations Groups at the ASC and IKI monitor the functioning of the
science-equipment complex. When remarks on the operation of the complex
are required during a control session, these groups request an operational
delivery of additional command-software information to the address of the
science-equipment complex.

Before the beginning of the entire cycle of tests with the spacecraft, a
ground tracking and data-collection station was established at the 22-m radio
telescope in Pushchino. The data obtained by the ground tracking station
during interferometric observing sessions is also used to monitor the status
of the spacecraft.  The housekeeping telemetry information extracted from the
headers of the science frames received by the Pushchino ground station,
transmitted through the HDRRC channel, is sent on to the FCC. These data are
processed and used in the same way as the telemetry data transmitted through
the radio channel: through the small onboard antenna, during link sessions
with the command--measurement ground stations, following observations.

The organization of the control of the {\em Spektr-R} spacecraft and of
the Main Control Operations Group described above has provided operational
and reliable control of the {\em Spektr-R}--{\em RadioAstron} complex,
including during the earliest stage of flight, when the first series of
tests were carried out. The inflight tests and organization of the control
of the {\em Spektr-R} spacecraft are described in more detail in~[26].

\begin{table}[t!]
\caption{Parameters of the {\em RadioAstron} orbit on April 14, 2012
(32 orbital revolutions after launch)} 
\begin{tabular}{l|l}
\hline
Major axis                      & $a = 174\,714.234$ km\\
Eccentricity                    & $e = 0.692$\\
Orbital inclination             & $i =  79.69^{\circ}$\\
Ascending node longitude        & $W = 300.55^{\circ}$\\
Argument of perigee             & $w = 303^{\circ}$\\
Time of perigee passage         & 07:12:37.00  UTC\\
                                & 14 April, 2012\\
Orbital period                  & ${\approx }8.5$~d\\
\hline
\end{tabular}
\end{table}

\section{THE ORBIT: PARAMETERS, MEASUREMENTS, AND PRECISION OF RECONSTRUCTION}

After the launch, the major axis of the spacecraft orbit was 173\,400~km,
its perigee height 578~km, its apogee height 333\,500~km, and its orbital
period 8.32~d. The first observations were made from this orbit. Several
months after inserting the spacecraft into its operational orbit, it became
clear that the useful life of the spacecraft could potentially end as early as
the end of 2013, due to the low perigee of the orbit. To avoid further
lowering of the orbit perigee, the system of onboard vernier thrusters was fired
twice in order to correct the orbit. After this correction (March 1, 2012),
the orbit has a calculated ballistic lifetime of more than nine years, with
the interval when the spacecraft is shadowed by the Earth being no more than
two hours. The orbital elements after correction (on April 14, 2012) are
presented in Table~1.

\begin{figure}[t!]
\setcounter{figure}{4}
\includegraphics[width=0.48\textwidth,angle=0,clip=true,trim=0.1cm 0 0 0]{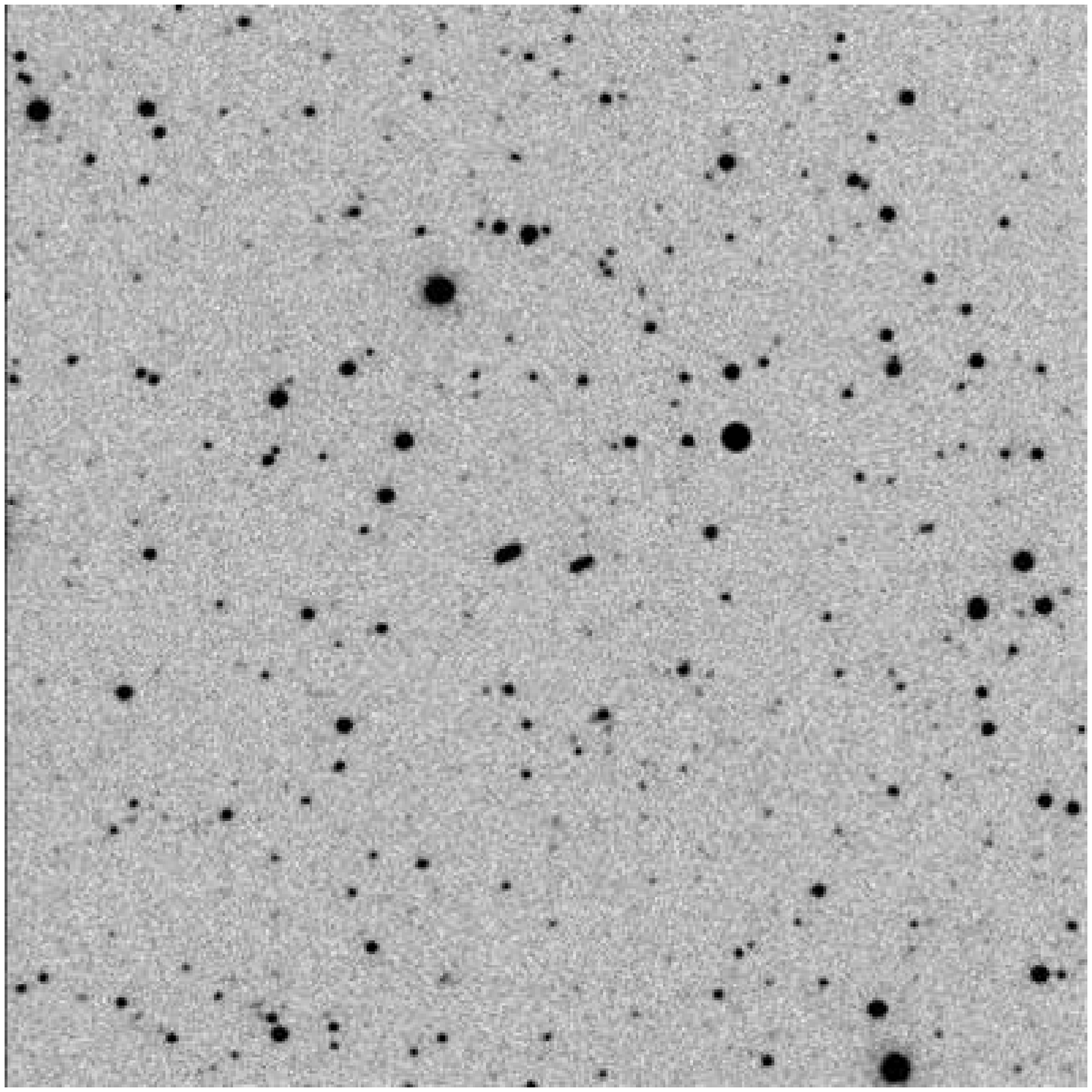}
\caption{
The two elliptical images near the center of the
photograph show the {\em Fregat} booster (on the right) and 
the spacecraft which has separated from it (on the left).
The photograph was made
on July 18, 2011, at 14:25 (Moscow daylight saving time)
using a 45.5-cm optical telescope in New Mexico at the
request of the Keldysh Institute of Applied Mathematics. This telescope is part
of the Scientific Network of Optical Instruments for Astrometric and Photometric
Observations, and is designed to search for asteroids and comets.} 
\end{figure}

The orbit evolves due to the perturbing influence of the Moon and Sun. The
eccentricity will vary from 0.96 to 0.59 during the spacecraft's lifetime, and
the orbital inclination will vary in the range $10^{\circ}$--$85^{\circ}$.
Figure~6a shows the evolution of the radii of perigee and apogee after the
above correction. The radius of perigee varies from 7000~km to 81\,500~km,
and the radius of apogee from 280\,000 to 353\,000~km. Further observations
of the spacecraft motion have shown that the above correction proceeded
normally, and estimates of the actual parameters of the correction are close
to the calculated values. Figures~6b--6e show the calculated evolution of the
projected orbit in 2013--2016. Figures~6f--6k show examples of the corresponding
evolution of the K-band $(u, v)$ coverage obtained for syntheses carried out
over a year using the two edge sub-bands (1.19 and 1.63~cm), for 2013, 2014,
and 2015, for the radio galaxy M87 (Figs.~6f--h) and Cen~A (Figs.~6i--k).
The region encompassed by the observations in the $(u, v)$ plane is appreciably
elliptical for both sources. Therefore, an additional orbital correction may
be applied in the future, in order to realize uniform filling of the $(u, v)$
plane in all directions. More detailed information about these new possibilities
can be found in~[35], and about the evolution of the orbit over the next five
years at the project web site~[25].

\begin{figure*}[p!]
\setcounter{figure}{5}
 \subfigure[]{
  \includegraphics[width=0.25\textwidth,angle=0,clip=true,trim=0.3cm 5cm 1.4cm 4cm]{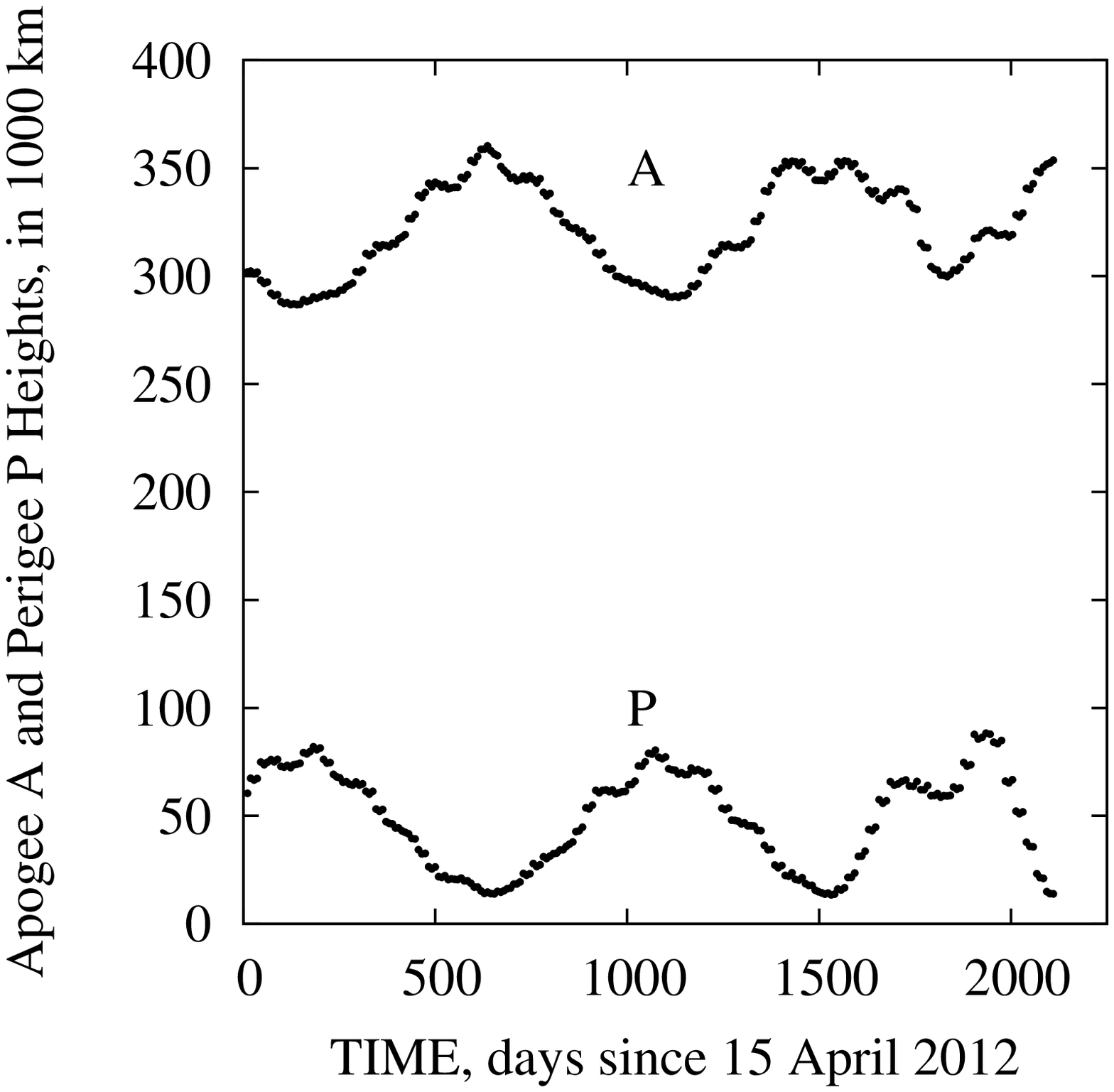}
 }
 \subfigure[]{
  \includegraphics[width=0.25\textwidth,angle=0,clip=true,trim=0.3cm 5cm 1.4cm 4cm]{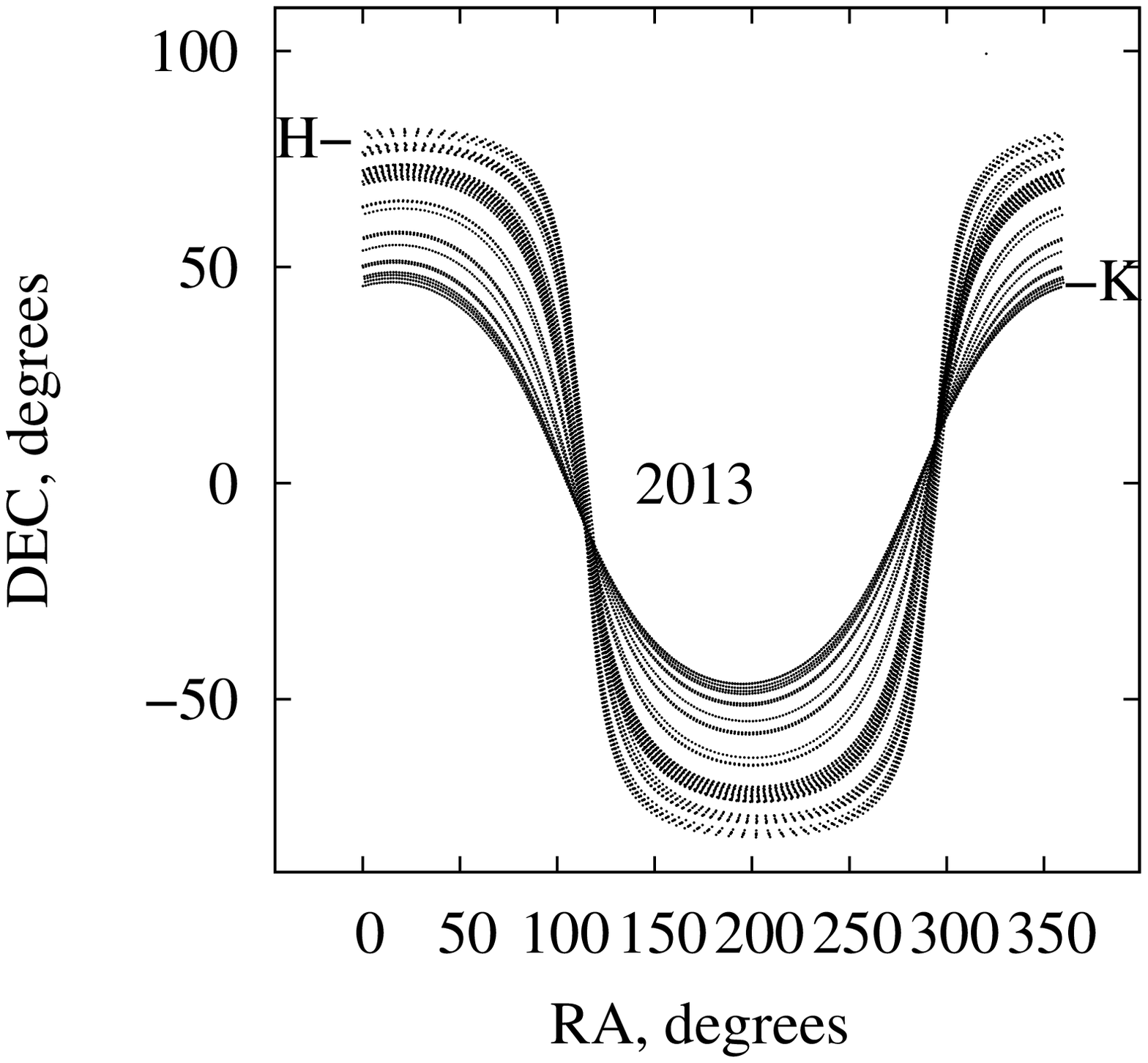}
 }
 \subfigure[]{
  \includegraphics[width=0.25\textwidth,angle=0,clip=true,trim=0.3cm 5cm 1.4cm 4cm]{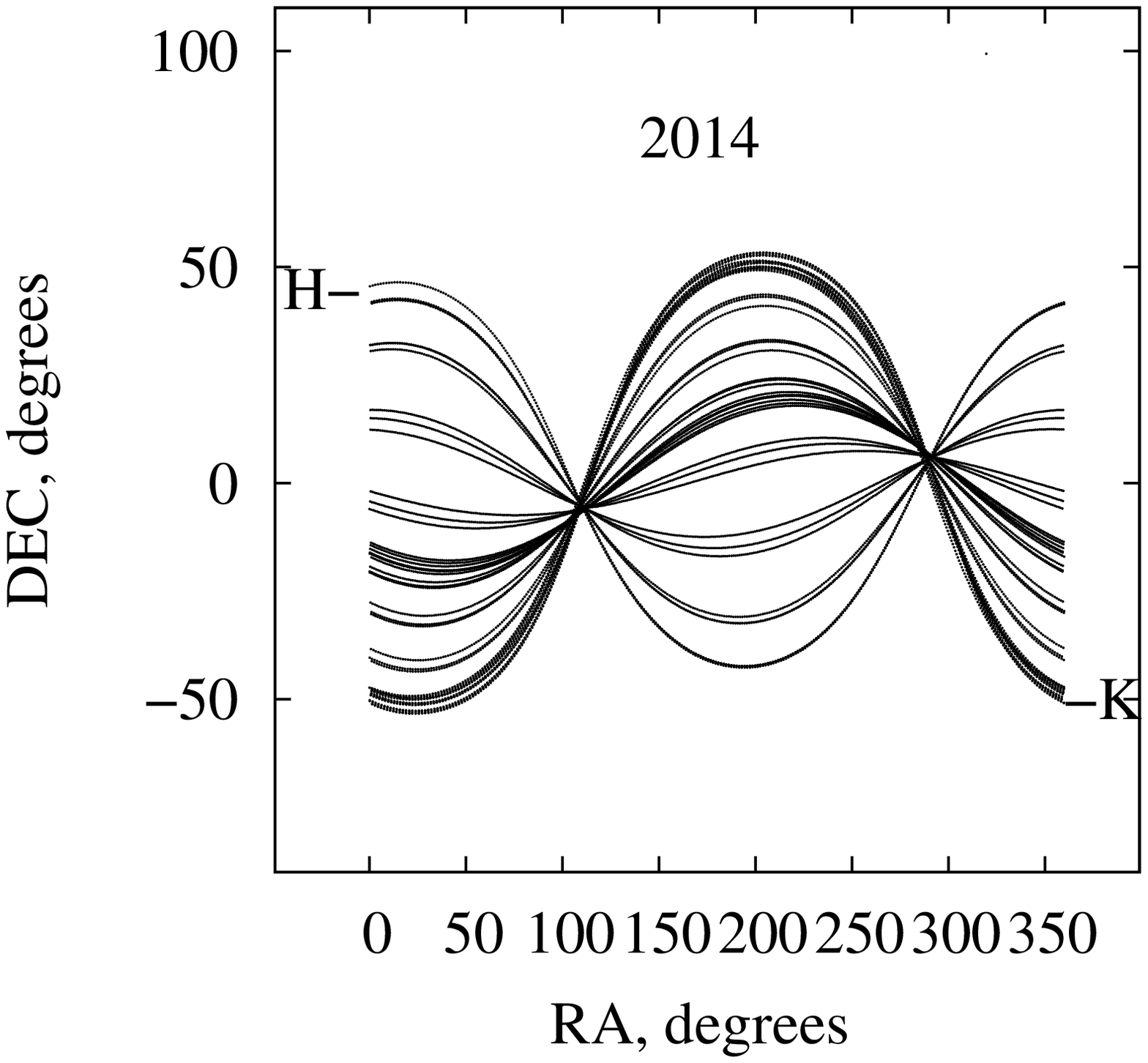}
 }
 \subfigure[]{
  \includegraphics[width=0.25\textwidth,angle=0,clip=true,trim=0.3cm 5cm 1.4cm 4cm]{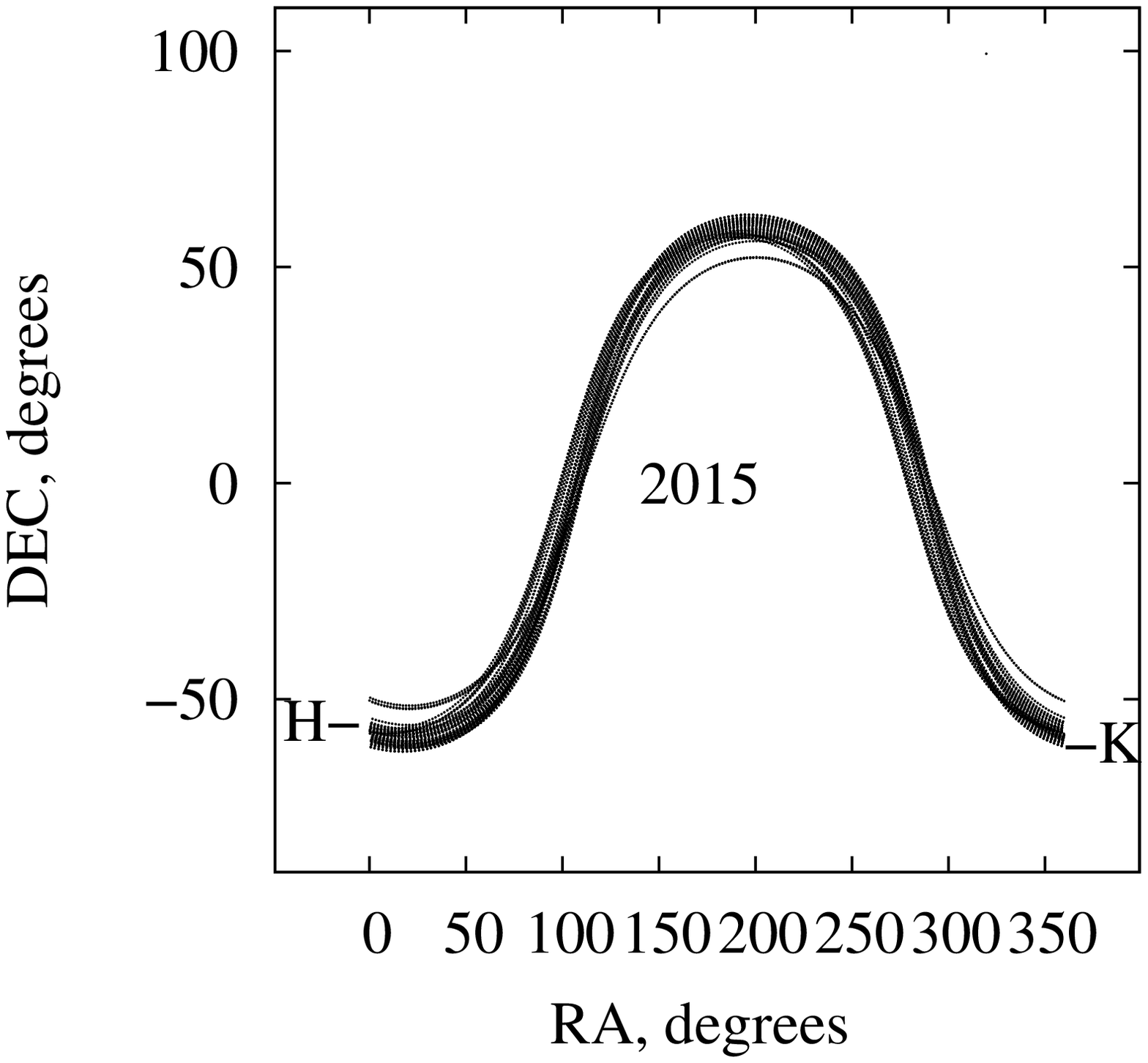}
 }
 \subfigure[]{
  \includegraphics[width=0.25\textwidth,angle=0,clip=true,trim=0.3cm 5cm 1.4cm 4cm]{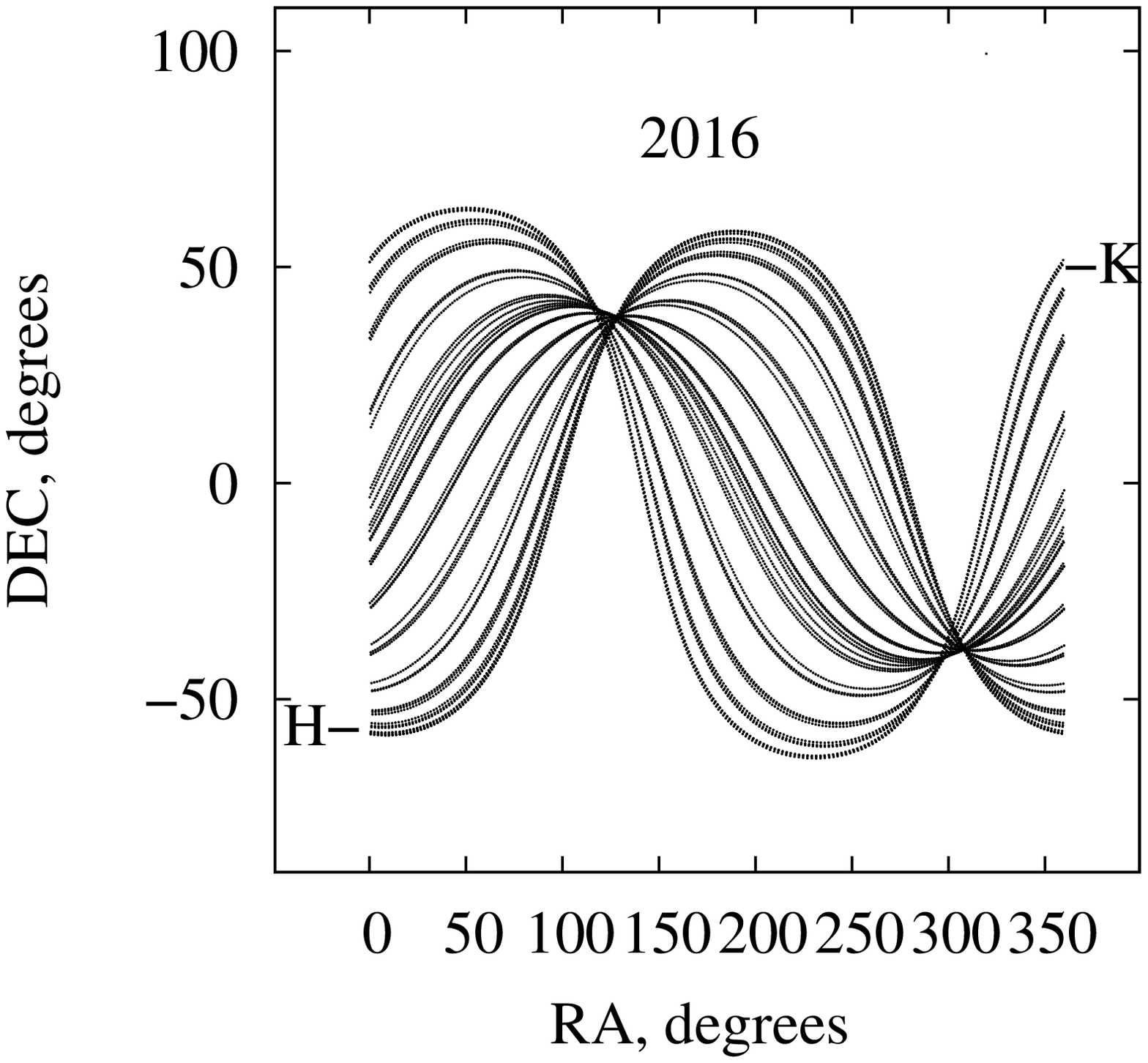}
 }
\\
 \subfigure[]{
  \includegraphics[width=0.25\textwidth,angle=0,clip=true,trim=0 0.3cm 0 0.5cm]{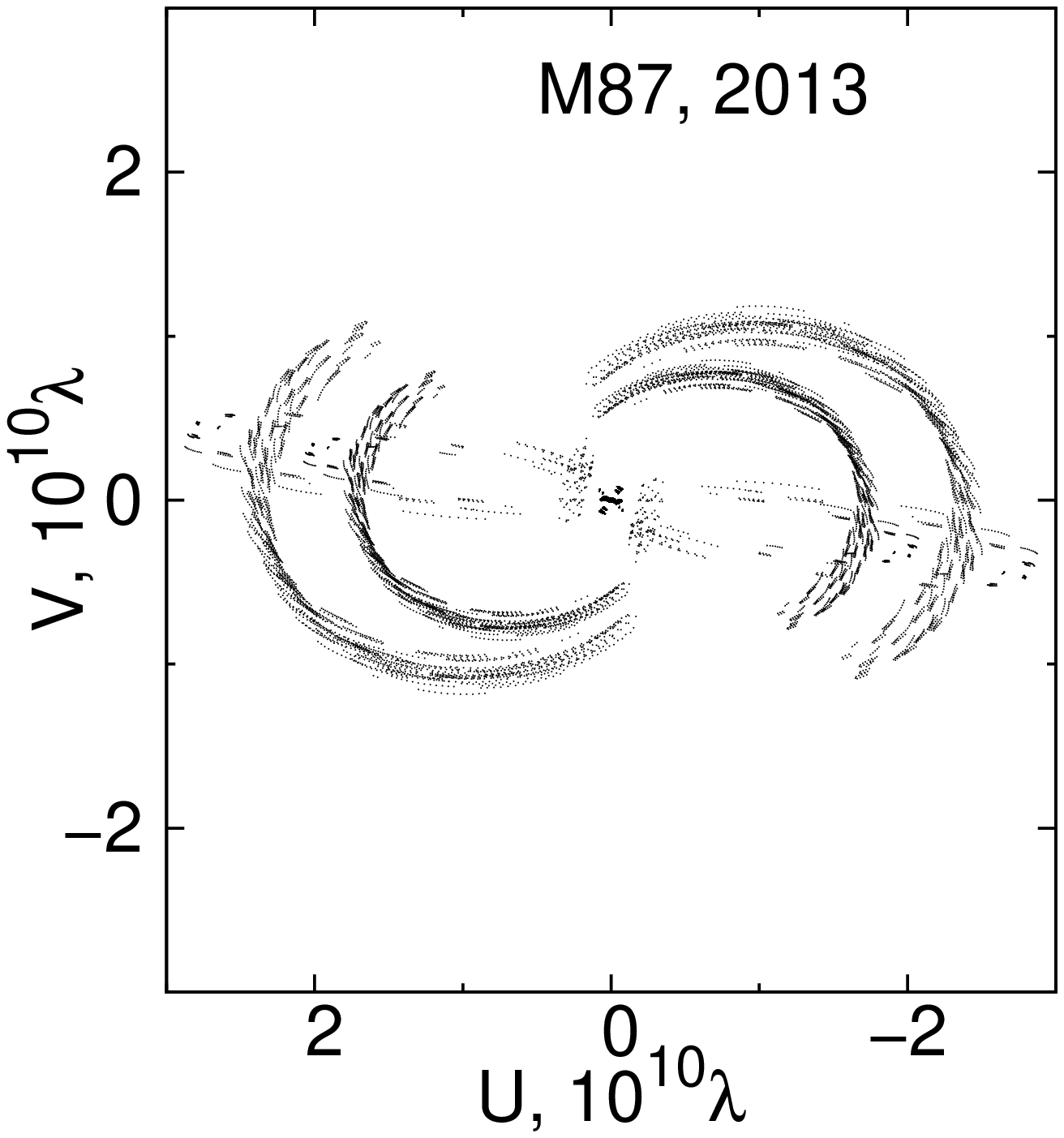}
 }
 \subfigure[]{
  \includegraphics[width=0.25\textwidth,angle=0,clip=true,trim=0 0.3cm 0 0.5cm]{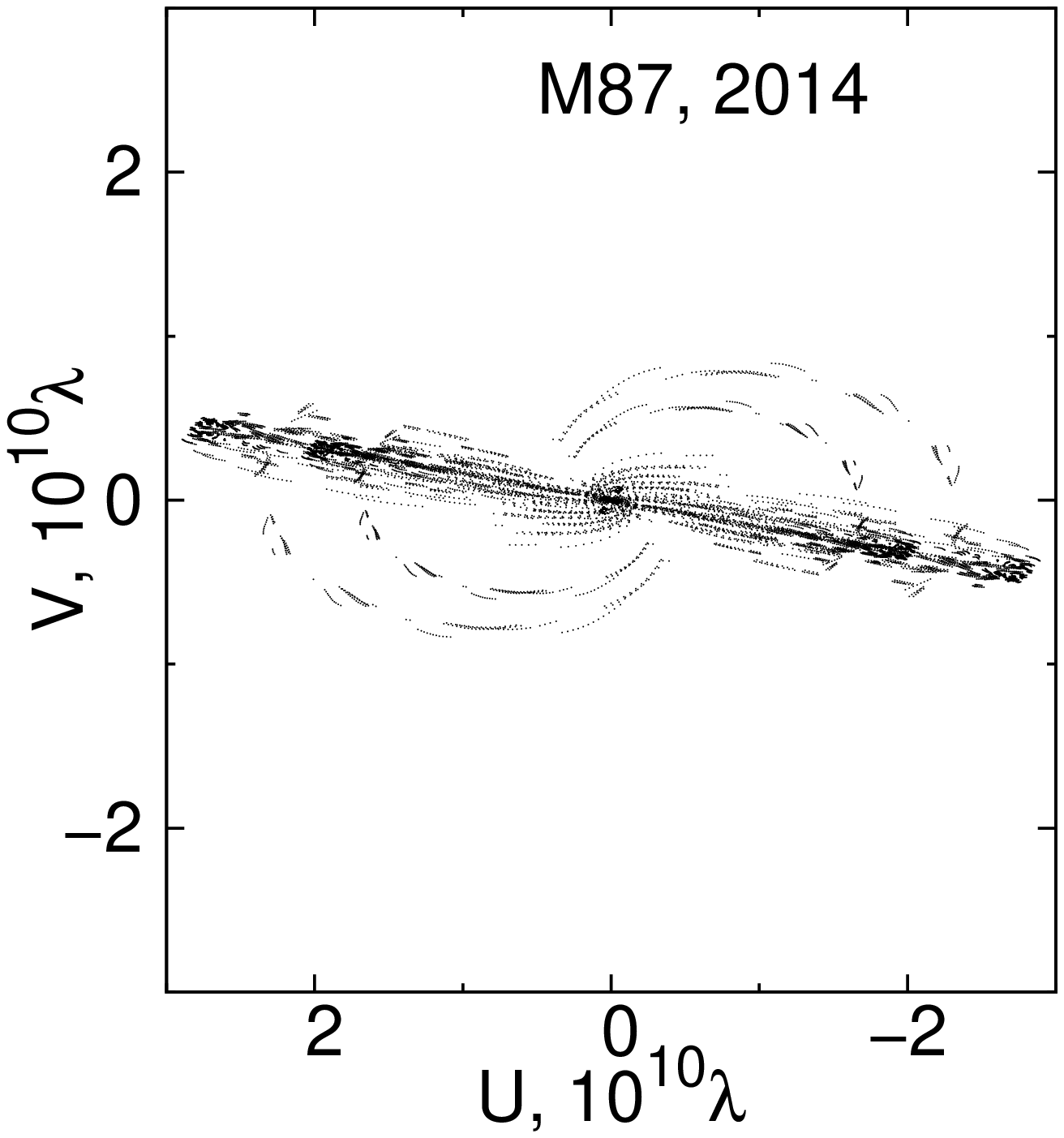}
 }
 \subfigure[]{
  \includegraphics[width=0.25\textwidth,angle=0,clip=true,trim=0 0.3cm 0 0.5cm]{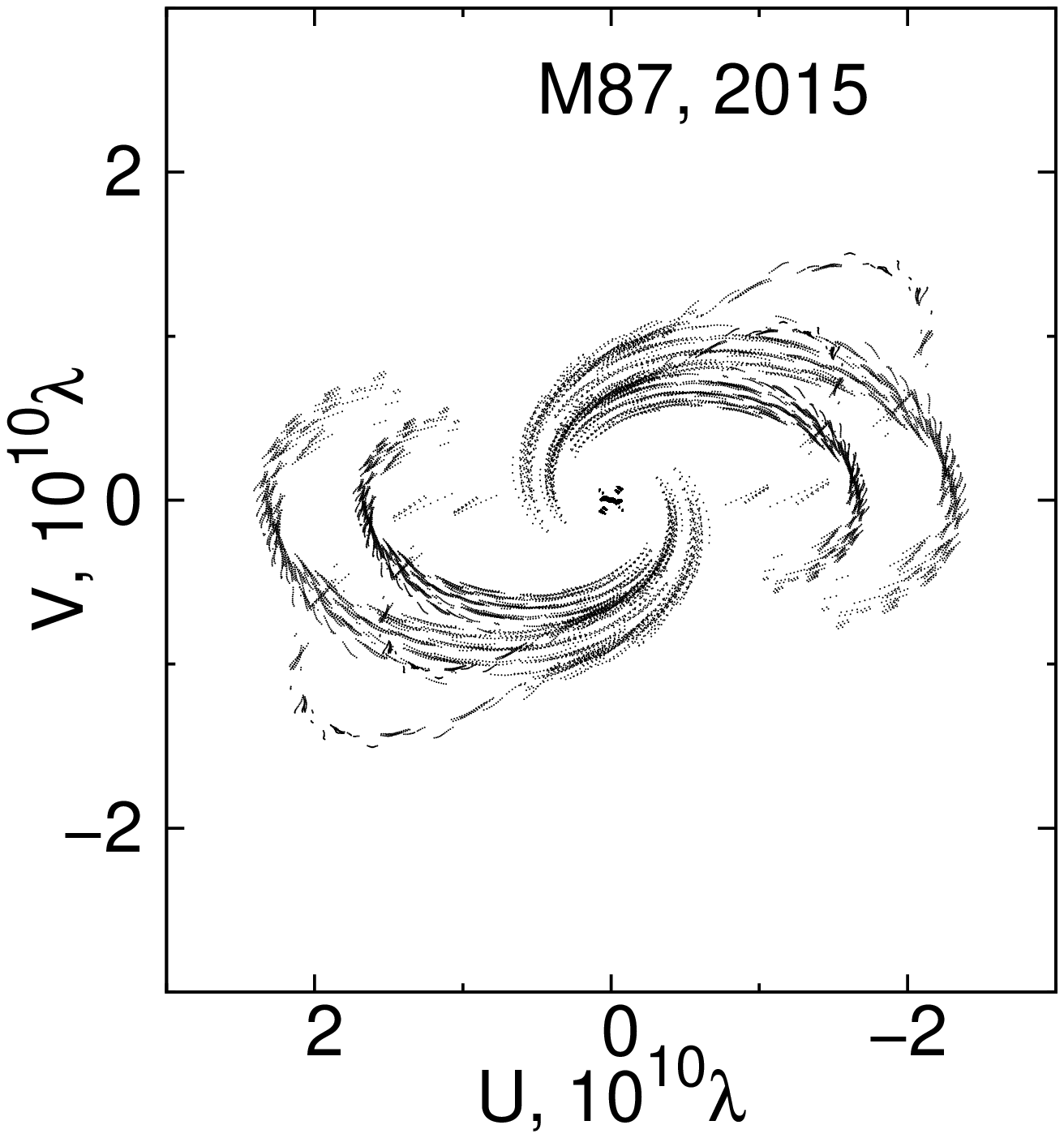}
 }
 \subfigure[]{
  \includegraphics[width=0.25\textwidth,angle=0,clip=true,trim=0 0.3cm 0 0.5cm]{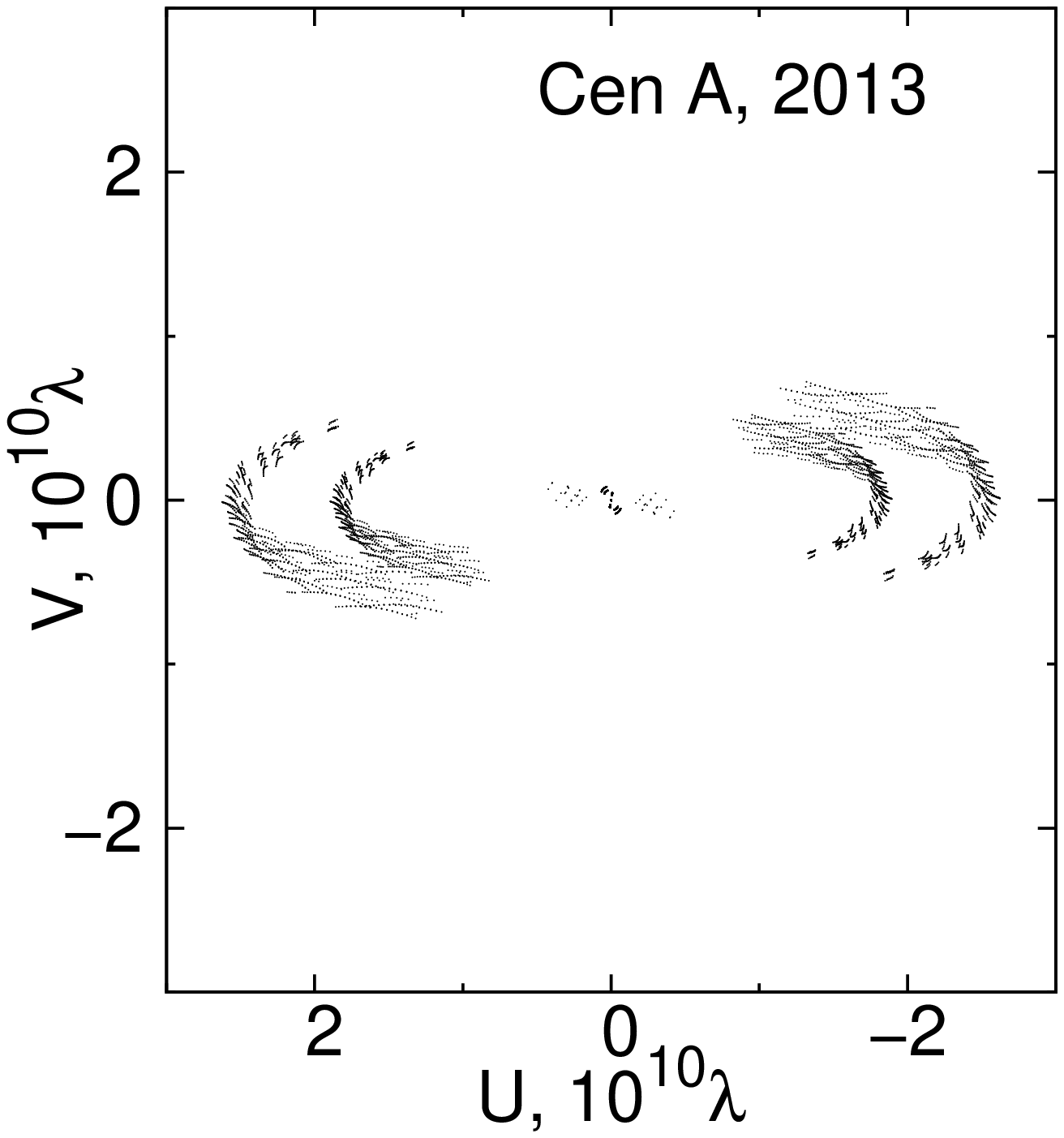}
 }
 \subfigure[]{
  \includegraphics[width=0.25\textwidth,angle=0,clip=true,trim=0 0.3cm 0 0.5cm]{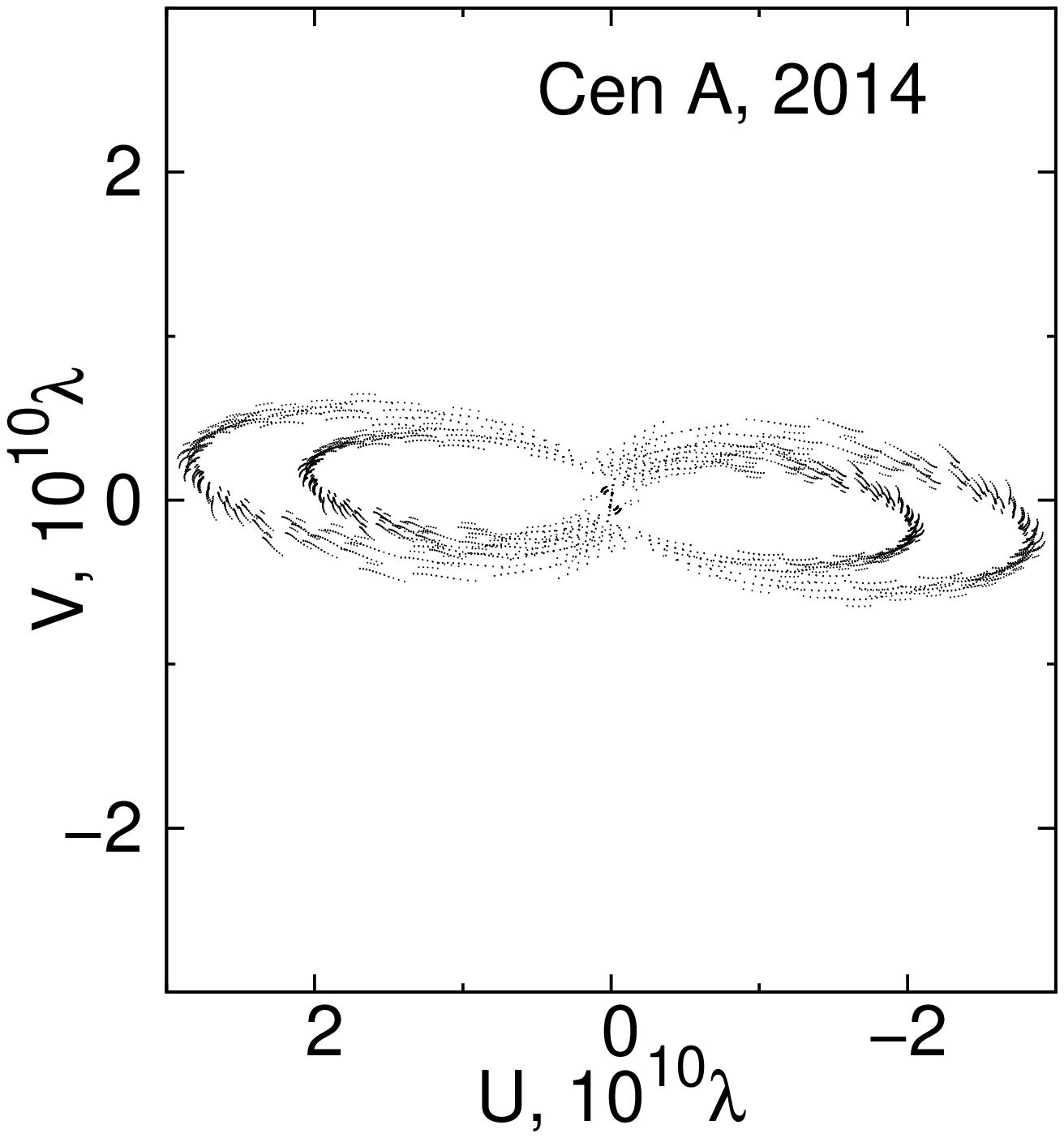}
 }
 \subfigure[]{
  \includegraphics[width=0.25\textwidth,angle=0,clip=true,trim=0 0.3cm 0 0.5cm]{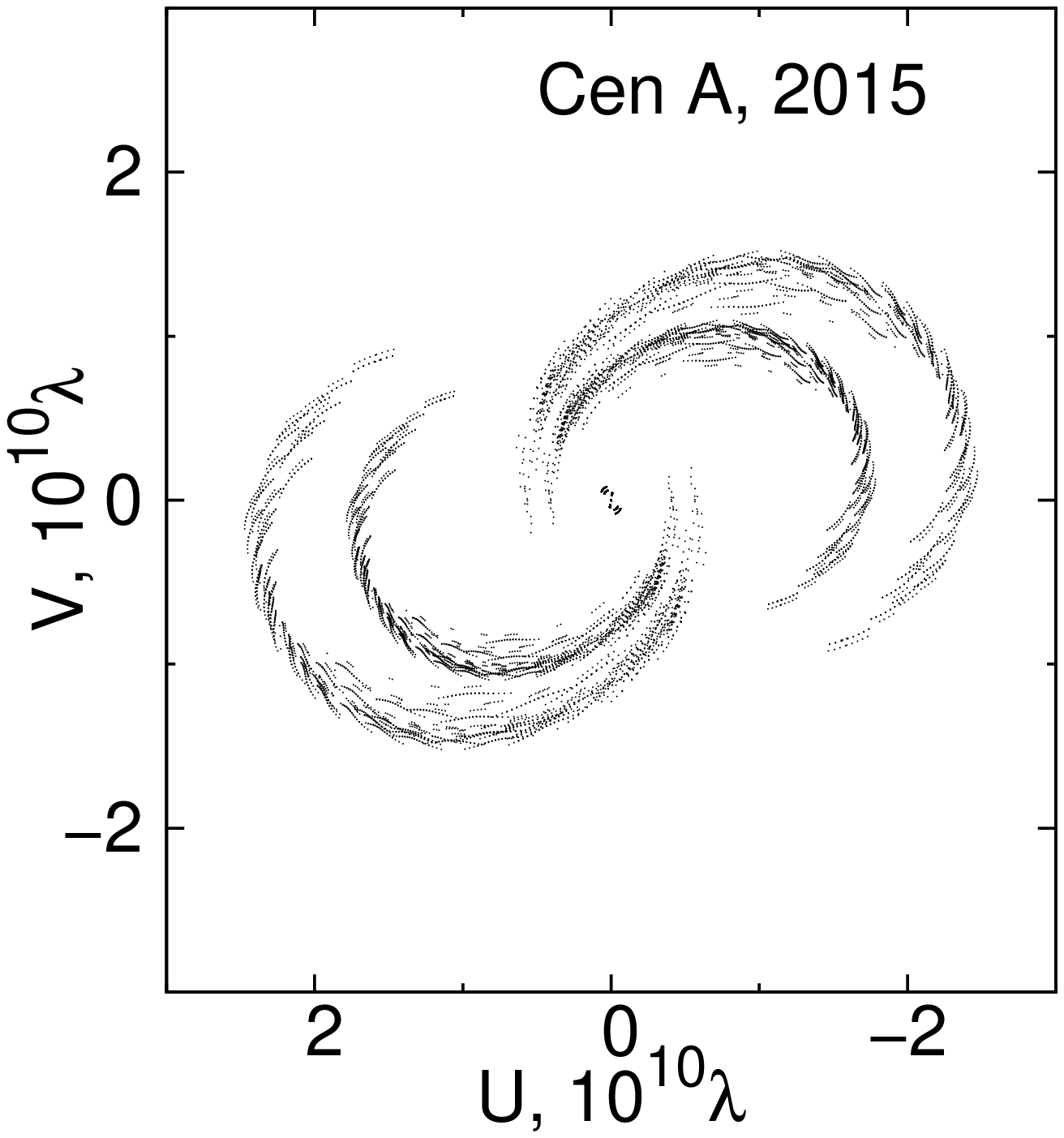}
 }
 \caption{(a)--(e) Calculated time evolution of the SRT orbit and (f)--(k)
examples of the $(u,v)$ coverage obtained for syntheses carried out
using the two edge sub-bands $F_3$ (1.19) and $F_{-4}$ (1.63~cm) in
the 1.35-cm band. Panel (a) shows the perigee height $P$ and apogee height $A$.
Panels (b)--(e) show the coverage of the celestial sphere with possible
observations with the ground--SRT interferometer in (b) 2013, (c) 2014),
(d) 2015, and (e) 2016; the letters ``B'' and ``E'' denote the beginning
and end of the orbital evolution over the year. Panels (f)--(h) show the
$(u,v)$ coverage obtained for observations of the galaxy M87 with the SRT
and the Green Bank (USA), Goldstone (USA), Effelsberg (Germany), Jodrell
Bank (UK), Evpatoria (Ukraine), Parkes (Australia), Tidbinbilla (Australia),
and Robledo (Spain) ground stations carried out in (f) 2013, (g) 2014,
and (h) 2015; the position angle of the jet in M87 is ${-}77^{\circ}$.
Panels (i)--(k) show the same as (f)--(h) for the galaxy Cen~A, whose jet
position angle is $51^{\circ}$). Synthesis of the broad frequency band with
eight sub-bands from 1.63 to 1.19~cm leads to appreciable additional filling
of the $(u,v)$ plane between the tracks shown for these edge wavelengths.\hfill}
\end{figure*}

Carrying out interferometric observations requires determining the ground--SRT
baseline with very high precision. The {\em Spektr-R} spacecraft is very
complex from the point of view of nagivation support. One of the factors
influencing the ballistics of the spacecraft is solar light pressure. The
pressure of the solar radiation acts on elements of the spacecraft surface
differently at different times during flight, leading to appreciable
perturbations of the orbit. In addition to direct perturbation of the motion
of the center of mass, which depends strongly on the current attitude of
the spacecraft, this light pressure exerts a torque about the center of mass.
The specified attitude is maintained by a system of reaction wheels. The long-term
action of perturbing torques in a single direction leads to a constant increase
in the angular velocity of the reaction wheels, which, in turn, leads to the need
to unload them; i.e., to decrease their angular velocity of rotation by
switching on the reactive engines of the stabilization system. This gives rise
to perturbations of the motion of the spacecraft center of mass. The increase
in velocity caused by these perturbations is 5--10~mm/s per unloading. The
accumulated additional shift in the position of the spacecraft due to this
effect acting over the course of a day is 400--800~m in range, which
exceeds the accuracy of radio range measurements. These perturbations
substantially complicate determination of the spacecraft orbit.

A model for the spacecraft motion taking into account a number of perturbing factors
is used in orbit determination. These perturbing factors include:

-- the non-central nature of the Earth's gravitational field, calculated
in accordance with the EGM-96 model~[36];

-- the gravitational attraction of the Moon and Sun, whose coordinates are
calculated based on the DE421 motion theory~[37];

-- solar light pressure;

-- perturbing accelerations arising during unloading of the reaction wheels;

-- ``rigid tides''; i.e., the correction to the Earth's gravitational field
due to its deformation under the action of the lunar and solar gravitational
forces~[38].

The variable pressure of sunlight substantially influences the spacecraft motion.
Due to the presence of the 10-m SRT antenna, the ratio of the midsection to
the mass of the spacecraft is appreciably higher than for other satellites,
and also depends strongly on the spacecraft attitude. Perturbations are
taken into account in the model using an approximation for the shape of the
spacecraft consisting of the three main components forming its surface: the
SRT antenna, central unit, and solar panels.

Measurements of the orbit of the {\em Spektr-R} spacecraft and the velocity
of its motion are carried out using various methods. These include, in
particular, the usual radio measurements of the range and radial velocity,
which are regularly carried out by the Ussuriisk and Bear Lakes control
stations. Measurements of the radial velocity using the signal from the HDRRC
antenna are carried out at the ground tracking station in Pushchino.
Laser-ranging measurements and optical astrometric measurements of the
spacecraft position on the sky are also used. VLBI measurements of the state
vector of the spacecraft are also conducted using the 8.4-GHz HDRRC signal,
applying the PRIDE method~[39]. Such HDRRC measurements accompany science
experiments. The signal is generated using the onboard hydrogen maser. The
radial velocity of the spacecraft can be determined with high precision based
on the measured frequency shift, taking into account relativistic
corrections~[40].

Laser-ranging measurements are one of the most precise and informative of
all the above sources of orbital information. However, a number of conditions
must be satisfied to obtain such measurements. Since the retroreflectors
are only installed on the bottom of the spacecraft in the $-X$ direction,
laser ranging requires a specific attitude of the spacecraft that can not
always be obtained. In addition to weather conditions and time of day, another
important factor limiting possibilities for laser ranging is the range limit
for such measurements. Most existing laser-ranging stations are designed
to work with low-flying spacecraft, and are not able to detect reflected
signals from spacecraft flying above the level of a geostationary orbit.
{\em Spektr-R} is the first high-apogee man-made satellite of the Earth
outfitted with retroreflectors that orbits at distances comparable to the
distance to the Moon. Laser-ranging measurements are currently possible at two
stations: Observatoire de la Cote d'Azur in Grasse (France) and the
Laser--Optical Radar of the Center for Outer Space Monitoring in the Northern
Caucasus (Russia). Complexities associated with laser ranging, the limited number of
stations able to work with such distances, and the strong dependence on weather
conditions hinder the acquisition of such measurements and their use for
refining the orbit on a regular basis. Another important application of
laser measurements is calibration of the regular radio systems.

Astrometric observations (optical measurements of the position of the
spacecraft relative to stars) are carried out by observatories in the
Scientific Network of Optical Instruments for Astrometric and Photometric
Observations~[41], as well as individual observers who submit measurements
to the IAU Minor Planet Center~[42]. More than 400 entries have been received,
containing 13\,300 measurements. Although such measurements cannot yield
the required precision for determining all parameters, they provide data on
the position of the orbital plane that are independent of the radial
characteristics of the spacecraft, and as such are useful supplements to
ranging and radial-velocity measurements.

The use of the model described above enables reconstruction of the orbit
for the reduction of data at a correlator with positional accuracy no worse
than ${\pm} 500$~m and velocity accuracy no worse than ${\pm} 2$~cm/s in
the three coordinates. Note that these numbers do not reflect the rms
values of random errors, but instead are guaranteed estimates of the residual,
slowly varying difference between the real and reconstructed orbit.

\section{MEASUREMENT OF THE MAIN SRT PARAMETERS USING ASTRONOMICAL SOURCES}

\subsection{{Receivers and Sensitivity of the SRT}}

A description of the {\em Spektr-R} spacecraft carrying the {\em
RadioAstron} radio telescope with the onboard science complex, as well as
the ground segment of the ground--space interferometer, is given in [21--23].
We will summarize by distinguishing the elements that determine the sensitivity
of the SRT in independent antenna measurements and the sensitivity of the
interferometer in the ground--space system.

The position of the radio telescope relative to the spacecraft is rigidly
fixed. The pointing of the SRT toward a target object and scanning of an
object is carried out by moving the spacecraft using the attitude-control
system (without turning on the thrusters). The orientation in space is
monitored using star sensors. Radio emission from an astronomical source
that is incident on the 10-m parabolic dish passes through the antenna-feed
assembly and arrives at the inputs of all the radio-astronomy receivers
simultaneously. After the frequency conversion of the input signal in the
receiver and subsequent scientific instruments, the output low-frequency
and high-frequency data forming the two streams are transmitted to Earth along
two radio channels, where they are archived and processed. The sensitivity
is mainly determined by the effective area of the antenna and the equivalent
noise temperature of the SRT, which includes contributions from the receiver,
antenna-feeder tract, and background sky.

The receiver system of the radio telescope consists of eight receivers at
four wavelengths: 1.35, 6.2, 18, and 92~cm, with two receivers for each
wavelength --- one for left- and one for right-circular polarization. The
signals from the circular-polarization generators for a particular wavelength
arrive at the inputs of the corresponding pair of receivers, which are
structurally joined to the antenna feeds in the four-wavelength co-axial
antenna-feed assembly (AFA).

The 1.35-cm feed forms a two-mode circular waveguide, which makes a transition
to a circular-polarization divider waveguide with two rectangular waveguides
at its output. The feeds in the other wavelength ranges are annular-gap feeds,
whose ring radii increase with wavelength. The annular gaps of the feeds are
co-axial with each other and with the circular waveguide. The polarization
dividers at 6.2, 18, and 92~cm are of the strip type with co-axial outputs.
The eight outputs of the AFA polarization dividers are joined with the input
LNA blocks for the corresponding receivers by waveguides at 1.35~cm and
co-axial segments at the other wavelengths. To enhance the sensitivity, the
AFAs and LNAs at all wavelengths except for 92~cm are radiatively cooled to
temperatures of about 150~K (the AFAs) and 130~K (the LNAs). For this purpose,
the 1.35, 6.2, and 18-cm LNAs are separated from the receivers and are located
in the hermetic focal container, where they are mounted on a separate cold
plate in contact with open space in the shadow of the SRT structure. The
uncooled 92-cm LNA is located at a temperature of about $30^{\circ}$~C inside
the thermostatically controlled receiver in the focal container. The
calibration signals from the internal noise generators arrive at the LNA
inputs from the receivers along individual co-axial lines.

Two types of signal are formed at the output of each receiver, after
amplification and heterodyne conversion of the signals from the input
frequencies to intermediate frequencies (IFs) near 512~MHz: the high-frequency
interferometric signal at the IF and the low-frequency radiometric signal.
The latter is formed in the radiometric tract of the receiver, from the
IF signal after detection by a square-law detector, amplification, and
averaging over a time interval of about a second. The radiometric signal
provides the possiblity of rapid and effective monitoring of the functioning
of the SRT and of antenna measurements in a single-dish regime.

The high-frequency IF signal from the receiver output arrives at the IF
selector, where two of the eight IF outputs from all the receivers are
selected for subsequent heterodyne conversion to lower frequencies and the
generation of a continuous digital flow of interferometric videodata for
transmission to the Earth. The phase stability of all the conversions is
provided by the onboard hydrogen maser or, in the alternative standard
regime, by a closed phase-link loop with a ground hydrogen maser. The
low-frequency radiometric signals from the outputs of all receivers arrive
directly at the onboard telemetry system of the spacecraft. The telemetry
system collects the radiometric and other low-frequency data from the entire
set of science and housekeeping instruments and forms a continuous flow of
telemetry data from the SRT.

The telemetry data are transmitted to the Earth through the telemetry channel
(in a real-time regime, or time-sharing regime, if the data can be temporarily
stored on the onboard memory unit), which uses the onboard wide-beam antennas
and the usual measurement points on the Earth\footnote{When the scientific
receivers are turned on, the service transmitters are turned off, so that the
telemetry data cannot be transmitted through the housekeeping channel in real time,
and are therefore written to the onboard memory unit for temporary storage
and subsequent transmission at a convenient time. However, when the SRT is
operating in the interferometric regime (which also must be used in antenna
measurements, in the single-dish regime), both data flows can be unified and
transmitted in real time through the HDRRC channel, which is what is usually
done. In this case, the flow of telemetry data lies in the frame headers of
the high-data-rate flow of interferometric data.}. The flow of data in the
interferometer regime is transmitted to the Earth in real time through the
15-GHz HDRRC for the science data. The 1.5-m parabolic HDRRC transmitter
antenna is located on the back side of the SRT and spacecraft, at the bottom
of the spacecraft, and can be pointed at the 22-m parabolic PRAO ground
tracking station or another tracking station during limited angular
ranges. These antennas are also used as transmitting ground antennas at
8.4/7.2~GHz for operation in the closed phase-link loop regime.

The sensitivity in terms of the antenna temperature $\sigma_T$ and
flux density $\sigma_F$ of the SRT as a single dish with a super-heterodyne
receiver in the radiometric regime (which is mainly used in antenna
measurements) is given by the known relations expressed in terms of the
equivalent system noise temperature $T_\mathrm{sys}$ and effective area of the radio
telescope $A_\mathrm{eff}$~[43]:
$$
\sigma_T =  T_\mathrm{sys}   \sqrt{\frac{2}{\Delta \nu \Delta t}  +
\left(\frac{\sigma_G}{G} \right)^2}  , \eqno{5.1} 
$$
$$
\sigma_F = \frac{2  k_B  \sigma_T}{A_\mathrm{eff}}  . \eqno{5.2} 
$$

\noindent Here, $\sigma_G/G$ is the relative instability in the gain $G$ for
the receiver tract, $\Delta \nu$ is the frequency bandwidth over the IFs (in
our case, equal to the width of the input frequency band, but close to the
intermediate frequency 512~MHz), $\Delta t$ is the integration time for the
signal after square-law detection (all these quantities refer to the
radiometric tract), and $k_B$ is Boltzmann's constant.

The sensitivity $\sigma_\mathrm{SVLBI}$ for the two-element SRT--ground radio telescope
interferometer can conveniently be expressed in terms of $T_\mathrm{sys}$ and
$A_\mathrm{eff}$ for the SRT ($i = SRT$) and the ground radio telescope ($i = RT$)~[44]:
$$
\sigma_\mathrm{SVLBI} = b    \sqrt{ \frac{F_\mathrm{sys,SRT} F_\mathrm{sys,RT}}{2
\Delta \nu_\mathrm{IF}  \Delta t_c}}  ,  \eqno{5.3} 
$$
$$
F_\mathrm{sys,i} = \frac{2  k_B  T_\mathrm{sys,i}}{A_\mathrm{eff,i}} . \eqno{5.4} 
$$

\noindent Here, $b \approx 1{-}2$ (we adopted $b$ = 1/0.637 for observations
with the SRT for numerical estimates~[25, 44]), $\Delta \nu_\mathrm{IF}$ is the
recording bandwidth, $\Delta t_c$ is the correlator averaging time, and the
product $2 \Delta \nu_\mathrm{IF}  \Delta t_c$ gives the number of independent
measurements over this averaging time. The convenience of using $F_\mathrm{sys}$
(the system equivalent flux density, SEFD) is that this quantity can be
measured in a simple way using observations of quasi-point-like sources
with known flux densities $F_s$:
$$
F_\mathrm{sys} = F_s     \frac{U_\mathrm{sys}}{U_s  g}  , \eqno{5.5} 
$$
$$
g = \frac{\int_{4 \pi}^{}{T_b (\vartheta, \varphi)  d\Omega}}
{\int_{4 \pi}^{}{T_b (\vartheta, \varphi)  D(\vartheta, \varphi)
d\Omega}},  \eqno{5.6} 
$$

\noindent where $ U_\mathrm{sys}/U_s = T_\mathrm{sys}/T_s$ can be directly measured at the
radiometer output of the receiver as the ratio of the detected responses
to the system noise $T_\mathrm{sys}$ and to a source with antenna temperature $T_s$
located in the SRT antenna beam, and $g \geq 1$ is the partial-resolution
coefficient for the source, which can be calculated numerically for a known
antenna beam $D(\vartheta, \varphi)$ and known brightness-temperature
distribution for the object $T_b (\vartheta, \varphi)$ ($g = 1$ for a point
source). The integration in~(5.6) is carried out over the solid angle $\Omega$.
For additional monitoring of the system stability, it is convenient to use
the analogous relations (5.4) and (5.5) for the amplitude $F_\mathrm{NS}$ for the
noise generator in units of the equivalent flux density, $F_\mathrm{NS} =
2\ k_B  T_\mathrm{NS} / A_\mathrm{eff}$, and $F_\mathrm{NS} = F_s  U_\mathrm{NS} / (U_s  g)$~[45], since
in contrast to $F_\mathrm{NS}$, $F_\mathrm{sys}$ generally depends on the direction toward
the source, via the sky antenna temperature $T_\mathrm{sky}$ and sky brightness
temperature $T_\mathrm{b,sky} (\vartheta, \varphi)$ [see (5.7) and (5.7a) below].

The expected equivalent system noise temperature of the radio telescope
$T_\mathrm{sys}$ and the corresponding temperature of the noise generator $T_\mathrm{NS}$,
reduced (``recalculated'') to the antenna aperture using the known certified
values of the equivalent noise temperatures of the receiver $T_\mathrm{LNA}$ and
the noise generator $T_\mathrm{NS,LNA}$ in the receiver LNA block, taking into account
the sky antenna temperature, were calculated using the expressions
(compare with~[25]):
\begin{multline*}
T_\mathrm{sys} = T_\mathrm{sky} + T_1 + \frac{T_2}{K_1} \\
+ \frac{T_3}{K_1  K_2} + \frac{T_\mathrm{LNA}}{K_1  K_2  K_3} , \mathrm{~~~~5.7}
\end{multline*}
\vspace*{5pt}
$$
T_\mathrm{sky} = \frac{\int_{4
\pi}^{}{T_\mathrm{b,sky} (\vartheta, \varphi) D(\vartheta, \varphi)
d\Omega}} {\int_{4 \pi}^{}{D(\vartheta, \varphi)  d \Omega}}  ,
\eqno{5.7a}
$$
\vspace*{5pt}
\begin{multline*}
T_i = t_i (L_{i,a} -1)  , \\  K_i = 1/ L_i = 1 / (L_{i,a}
L_{i,r}) , \quad   i = 1, 2, 3,  
\mathrm{~~~~5.7b}
\end{multline*}
$$
T_\mathrm{NS} = \frac{T_\mathrm{NS,LNA}}{K_1  K_2  K_3} . \eqno{5.7c} 
$$

\noindent Here, $T_\mathrm{sky}$ and $T_\mathrm{b,sky} (\vartheta, \varphi)$ are the sky
antenna and brightness temperatures, $T_i$ the equivalent noise temperature
at the input of element $i$ with a physical temperature $t_i$, $K_i$ is the
power transmission coefficient,  $L_{i,a}$ and $L_{i,r}$ are the active and
reactive loss coefficients for element $i$ of the antenna-feeder tract of
the radio telescope, and $i = 1$ for the dish (``dish''), $i = 2$ for the
feed (``feed''), and $i = 3$ for the waveguide or co-axial link line between
the AFA and the LNA (``cbl'') (for convenience, the corresponding subscripts
used for the analogous parameters in [25] are indicated in parantheses).
The notation in (5.7a) for the sky is analogous to that in (5.6) for a
source, with the integration carried out over the solid angle.

\subsection{{Aim and Process of Antenna Measurements}}

The aim of the autonomous antenna measurements is to derive the main parameters
of the SRT in flight, and also during operation as part of a ground--space
radio interferometer. This is achieved by carrying out the following tasks.

1. Measurement of the noise characteristics of the radio telescope at 92, 18,
6.2, and 1.35~cm: the equivalent system noise temperature $T_\mathrm{sys}$ and
system equivalent flux density $F_\mathrm{sys}$. The system here refers to the radio
telescope, consisting of the antenna, antenna-feed tract, and receiver, with
the noise signal from the integrated sky background arriving at its input;
the system noise temperature is reduced to the radio telescope input, i.e.,
to the plane of the dish aperture. Thus, the system noise temperature
``automatically'' includes the sky noise temperature [compare with~(5.7)].

2. Measurement of the effective area $A_\mathrm{eff}$ at the above wavelengths using
observations of astronomical continuum calibrator sources, together with
the noise temperature of a calibration signal from the noise generator,
determined from ground tests.

3. Measurement of the width and shape of the main lobe of the antenna beam at
all wavelengths.

4. Measurement of radio-astronomical corrections to the SRT pointing relative
to the coordinate system determined using star sensors.

\subsection{{Preparation and Conduction of Measurements}}

The radio-astronomical measurements of the main parameters of the orbiting
telescope (``antenna measurements'') presented below were part of a program
of inflight tests of the spacecraft and SRT carried out in the first six
months after launch. It is planned to conduct such measurements regularly
over the entire period of operation of the SRT in flight. Up to the beginning
of the antenna measurements in the middle of September 2011, the following
technical operations were carried out.

1. The efficiencies of the antenna and receivers in the radiometric regime were
verified for each of the two polarization channels (left- and right-circular
polarizations) at 92, 18, 6.2, and 1.35~cm.

2. The realizability of the control and design conditions for motion of the
SRT in an inertial Cartesian coordinate system $XYZ$ rigidly fixed to the
center of mass of the spacecraft, based on commands sent from Earth and the
onboard memory unit, was verified. The possibility of submitting commands and
obtaining results in the astronomical equatorial coordinates of right ascension
and declination at epoch J2000.0 was also verified. The $X$ axis of the
spacecraft coincides with the geometrical axis of the SRT dish, and the $Y$
axis is parallel to the rotational axis of the solar panels. The motion
regimes of the SRT are analogous to those of ground radio telescopes:
``Pointing'', ``Tracking'', ``Scanning''. The SRT tracking regime is equivalent
to maintaining a constant orientation of the telescope in space; as in the
two other regimes, this is achieved using the attitude-control system of
the spacecraft, monitoring the orientation using the star sensors, without
using the thrusters during observations (see Section~4). Pointing at a source
and scanning across a source are carried out in the Cartesian coordinates of the
spacecraft, via rotations about the $Y$ or $Z$ axis. The recalculation to
astronomical equatorial coordinates was carried out using the telemetry data
from the housekeeping system for coordinate provision. In the general case,
scans of a source in the plane of the sky essentially represent linear sections
of trajectories reflecting a cross section of the celestial sphere by the $X$
axis in the equatorial coordinate system (see the example in Section 5.5 in
Fig.~7b in the color insert and in Fig.~8c, with the corresponding responses
of the radio telescope and scanning trajectories).

The method used to carry out the radio-astronomical antenna measurements
was the same for all the wavelengths. In each session (usually over about
two hours), measurements were conducted in one of the selected scanning regimes,
simultaneously for all the operative wavelengths and polarization channels
(as a rule, for two polarization channels at two wavelengths; see typical
examples in Figs.~7a and 7b in the color insert and in Figs.~8a and 8b).
Depending on the frame-generation program and the sensor-interrogation
speed, which is specified by commands to the telemetry system, all the
telemetry parameters of the operative science and housekeeping instruments
are recorded in a series of several successive frames, including the signals
from the radiometric analog and digital receiver outputs and the codes for
the onboard time scale and coordinates.  The corresponding frame period can
vary from fractions of a second to several seconds. The use of a procedure
for ``automatic'' antenna measurements in a ``sparing'' regime for the
operation of the science receivers\footnote{In other words, with the following
conditions: 1)~a previously agreed inflight task cyclogram; 2)~absence of
interference of an operator when the tasks are carried out; and 3)~with the
transmitter turned of and with the telemetry data written to the onboard
memory unit (instead of transmitting the telemetry data to the Earth in real
time via the housekeeping telemetry channel).}, which became normal, made
it possible to avoid both the danger of the failure of the high-sensitivity
transistor amplifiers of the receivers and the influence of interference on
the action of the normal transmitter signal outside the band, through the
wide-beam housekeeping antennas. The virtual absence of interference from
the antenna measurements of the SRT in flight at all wavelengths was a pleasant
surprise, in contrast to the situation with the ground tests.

\subsection{{Telemetry Data and their Reduction}}

The original telemetry information with the antenna-measurement data for each
session arrive at the SDRC of the ASC for subsequent reduction and analysis.
This information includes:

a) the custom text exchange form for each instrument, transferred via the
special TsITRUS database of the Lavochkin Association;

b) an original binary file with extension tmi, containing the entire set of
telemetry-data frames together with the packaged binary data for all the
instruments, which is transmitted from the onboard memory unit via the usual
telemetry channel; this file is generated at the Lavochkin Association and
IKI, and is located on the ftp server of the SDRC of the ASC;

c) an original binary file with extension tmi containing the same set of
telemetry-data frames as in the previous item, but transmitted in real time
together with the videodata, via the HDRRC science channel; this file is
generated at the ASC, and is likewise located on the ftp server of the
SDRC\footnote{The telemetry data for this file are extracted from the
frame headers by the ASC TMSRT software and transmitted through the HDRRC
channel and the ground tracking station. These data are used to generate a
binary telemetry file with extension tmi, whose structure and format are
fully analogous to the tmi files arriving via the telemetry channel. This
simplifies the reduction process using the standard tmi-file reduction software
tested earlier in SRT receiver--transmitter tests.}.

The reduction of the binary tmi files yields six tabular files in text format
with the values of all 600 scientific telemetry parameters of the SRT,
including the radiometric parameters. This reduction is carried out by the
Automated System for the Reduction and  Visualization of Telemetry Sessions
software developed at the ASC~[46]. The individual telemetry parameters in
each text tabular file are distributed across the various columns, and their
time behavior is reflected by the sequence of frames given as rows, which
simplifies graphical analysis of the data.

Usually, for reliability and mutual monitoring, the remote processing of
antenna measurements is conducted using the text files obtained from the
original binary telemetry data using the TsITRUS database of the Lavochkin
Association and the Automated System for the Reduction and  Visualization
of Telemetry Sessions software of the ASC. Data from the standard coordinate
software in the TsITRUS system are always used when obtaining the dependence
of the current coordinates of the trajectory of motion of the SRT axes on
the onboard time. When necessary, the coordinates were interpolated and
the uniformity of the spacecraft motion monitored. Further, the text data
were converted into graphical form and processed in various ways.

The visualization of the text results (for printing or subsequent reduction)
was carried out using the Automated System for the  Reduction and
Visualization of Telemetry Sessions software together with the Gnuplot
package under a Linux operating system. Packages such as Excel were used
under Windows, as well as the specialized program TMI\_VIEWER developed at
IKI. The specialized software package KRTVIZ was developed for the express
monitoring of the observations via the simultaneous visualization of up to
14 radiometric output signals (8 digital and 5 analog) arriving in the binary
flow of data from the HDRRC channel (or alternately from a tmi file). This
software enables monitoring of the recording of onboard observations in
real time on a local network at the ASC.

In spite of a number of specific characteristics, the method used to process
SRT antenna measurements is essentially the same as the standard method used
for antenna measurements at ground radio telescopes~[45]. The results
obtained in various ways were compared, and the origins of any appreciable
differences were identified and eliminated.

\begin{table*}[p]
\caption{Main expected (1.1--1.10) and measured (2.1--2.10) SRT parameters
at K, C, L, and P bands: FWHM of the main lobe of the
antenna beam $\vartheta_{0.5}$ and $\varphi_{0.5}$, effective area $A_\mathrm{eff}$,
aperture efficiency AE, effective system temperature $T_\mathrm{sys}$ and receiver
temperature $T_\mathrm{rec}$, system equivalent flux density $F_\mathrm{sys} \equiv
\textrm{SEFD}$, systematic errors in scanning  $\Delta\vartheta_s$ along the
$\vartheta$ axis and $\Delta \varphi_s$ along the $\varphi$ axis after
introducing the constant correction $\delta \vartheta_p$ to the telescope
pointing, interferometer sensitivity $\sigma_\mathrm{SVLBI}$, ratio $\alpha_{D}$
of the measured beamwidth to its ideal width $\lambda / D$, where $D =
10$~m is the dish diameter}
\begin{tabular}{l|c|c|c|c}
\hline
\multicolumn{1}{c|}{Parameter}              &~K~(1.35 cm)&~C~(6.2 cm)&~L~(18 cm)&~P~(92 cm)\\
\hline
1.~~  SRT in Pushchino, 2003--2004:& & & &\\
1.1~~ $\vartheta_{0.5} \pm 5\%$,
&  $5.6^{\prime}$ & $25.5^{\prime}$  & $74.5^{\prime}$ & $6.2^{\circ}$ \\
1.2~~  $A_{eff} \pm 10\%$,       m$^2$     &   27 &   40 &   40 &  24\\
1.3~~  AE = $A_\mathrm{eff}/A_\mathrm{geom} \pm 10\%$ & 0.34 & 0.51 & 0.51 & 0.31\\
1.4~~  $T_\mathrm{sys}/T_\mathrm{rec}$,     K                                   &80/45 & 70/30 & 50/15 &140/30\\
1.5~~  $T_\mathrm{sys}^\mathrm{(opt)}$,     K                                    &70    & 66    &  33  & 164\\
1.6~~  $SEFD_\mathrm{SRT}^\mathrm{(opt)}$,   Jy                                        &7200    &4600   &2300  &19000\\
1.7~~  $SEFD_\mathrm{GB}$,     Jy                                         & 23     & \phantom{0}8     & 10   & 55\\
1.8~~  $\Delta \nu_\mathrm{IF}$,     MHz                                   & 32     & 32    & 32   &  16\\
1.9~~  $\sigma_\mathrm{SVLBI}^\mathrm{(opt)}$ (for $\Delta t = 5$~min),    mJy & \phantom{0}5   & \phantom{0}2   & \phantom{0}2   & 16\\
1.10~~  $\alpha_D = \vartheta_{0.5} D / \lambda$                        & 1.21   & 1.20  & 1.20 & 1.18\\
\hline
2.~~  SRT in flight, 2011--2012: & & & &\\
2.1~~  $(\vartheta_{0.5} \pm 5 \%) \times (\varphi_{0.5} \pm 5
\%)$,
& $6.0^{\prime} \times 13^{\prime}$ & $25^{\prime}$    & $72^{\prime}$   & $6.1^{\circ}$ \\
2.2~~  $A_\mathrm{eff} \pm 13\%$,     m$^2$;                            & 7.5       & 35    & 41   & 30\\
2.3~~  AE = $A_\mathrm{eff}/A_\mathrm{geom} \pm 13\%$                            & 0.1       & 0.45  & 0.52 & 0.38\\
2.4~~  $T_\mathrm{sys} \pm 13\%$,       K                             & 77        & 130   & 45   & 200\\
2.5~~  $F_\mathrm{sys} \pm 10\%$  (SEFD$_\mathrm{SRT})$,     Jy              & 30000 & 10500 & 3400 & 19000\\
2.6~~  $| \Delta \vartheta_s |$,      &$1.2^{\prime} \pm 0.2^{\prime}$            & & & \\
2.7~~  $| \Delta \varphi_s |$,       &${< }1.5^{\prime}$            & & & \\
2.8~~  $\delta \vartheta_p$,                 &~~~$2.5^{\prime}$              & & & \\
2.9~~  $\sigma_\mathrm{SVLBI}$ (for $\Delta t = 5$~min),     mJy        & \phantom{0}9        & \phantom{0}4     & \phantom{0}2    & 16\\
2.10~~  $\alpha_D = (\vartheta_{0.5}\times \varphi_{0.5})D / \lambda$ &$1.29 \times 2.80$ &1.17 &1.16 &1.16\\
\hline
\end{tabular}
\vskip 5pt

\flushleft{\footnotesize 1. The values of the parameters 1.1--1.3 and 1.10
are based on the results of ground tests of the SRT in Pushchino in
2003--2004~[47]. Prior to the ground tests, the SRT dish was fixed to a
special retaining frame installed on a rotating turntable and adjusted
geodetically. Two noise temperatures are separated by a slash in row 1.4
($T_\mathrm{sys}/T_\mathrm{rec}$), for 1)~the SRT system (the first value, giving the
theoretical estimate for a height above the Earth exceeding 10\,000~km)
and 2)~the receiver (the second value, taken from the receiver documentation).
The SRT noise temperature obtained from an alternative estimate of the
loss in the antenna-feeder tract of the SRT is given in row 1.5 [see (5.7)].
This is used to estimate the expected value of SEFD$_\mathrm{SRT}$ in row 16,
according to (5.4). The values of SEFD$_\mathrm{GB}$ for the 100-m Green Bank radio
telescope in row 1.7 were taken from [25], as an example of a ground
interferometer element. The parameter in row 1.8 gives the recorded bandwidth.
Parameters 1.6--1.8 were used to estimate the expected sensitivity of the
ground--space interferometer given in row 1.9, according to (5.3), with an
integration time of 5~min.
2. Parameters 2.1--2.8 and 2.10 were measured during inflight tests of
the SRT. The value 2.5 for $F_\mathrm{sys}$ (SEFD$_\mathrm{SRT})$ was obtained from
direct measurements using (5.5). These SEFD values agree within the errors
with the calculated estimates obtained using (5.4) together with the parameters
2.2 and 2.4. Row 2.9 presents the calculated sensitivity for the two-element
space--ground interferometer $\sigma_\mathrm{SVLBI}$ for one polarization,
analogous to the parameter 1.9
but estimated using the parameter 2.5.}
\end{table*}

\subsection{{Results}}

The main results are summarized in Table~2 (measurements) and in Table~3
in the Appendix (analysis of measurements). Table~2 also includes the main
results of ground radio-astronomical tests of the engineering model of the
SRT at the PRAO test facility obtained in 2003--2004~[47]. Typical radiometric
responses to some of the first source scans are presented in Figs.~7a (color
insert), 8a, and 8b, which show scans of Cass~A (the first astronomical
source observed using the SRT on September 27, 2011) at 92, 18, 1.35 and
6.2~cm, and Fig.~7b (color insert), which shows scans of the Crab Nebula
at 1.35 and 6.2~cm. An example of a scan of an area of the sky is given
in Fig.~8c. The trajectories presented in this figure correspond to the
radiometric responses in Fig.~7b. Jupiter, the Moon and Virgo~A (extended
objects) were also used for antenna measurements, as well as the quasi-point-like
extragalactic radio sources 3C~84, 3C~273, and 3C~279.

The characteristics of standard calibrators (their fluxes, brightness
distributions, angular sizes, and polarizations) were taken from~[48], and
the fluxes of the strong, quasi-point-like sources 3C~84, 3C~273, and 3C~279,
which are variable, were taken from measurements on the 600-m RATAN-600
annular radio telescope of the Special Astrophysical Observatory of the
Russian Academy of Sciences (Nizhnii Arkhyz, Russia) and the 100-m Effelsberg
paraboloid of the Max-Planck-Institut f\"ur Radioastronomie in Bonn~[49] at
epochs close to the onboard measurement dates. The procedures used for these
ground meaurements are described in~[45, 49].

\begin{figure}[p!]
\setcounter{figure}{7}
 \subfigure[]{
  \includegraphics[height=0.48\textwidth,angle=270,clip=true,trim=0 0 0 0]{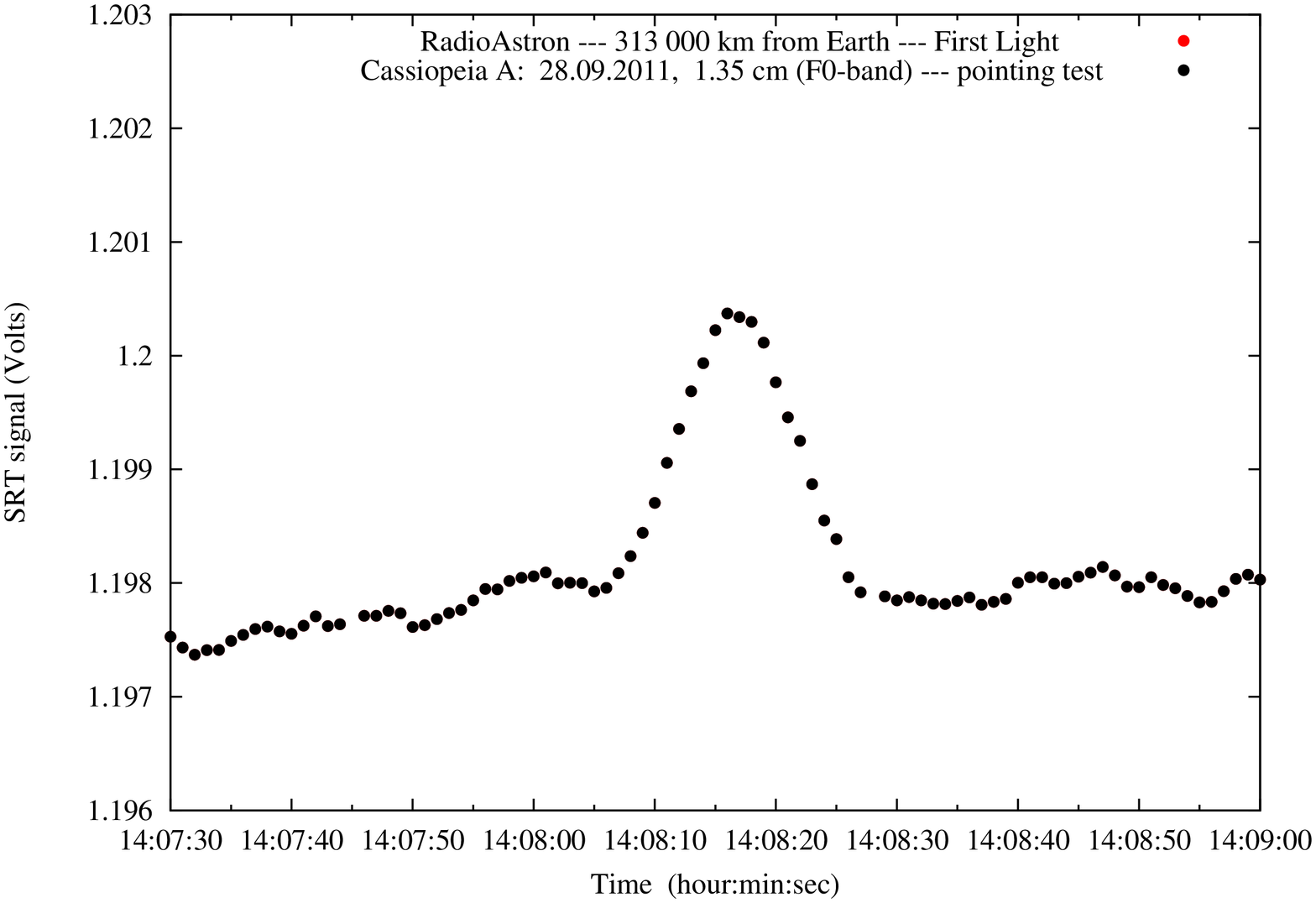}
 }
 \subfigure[]{
  \includegraphics[height=0.48\textwidth,angle=270,clip=true,trim=0 0 0 0]{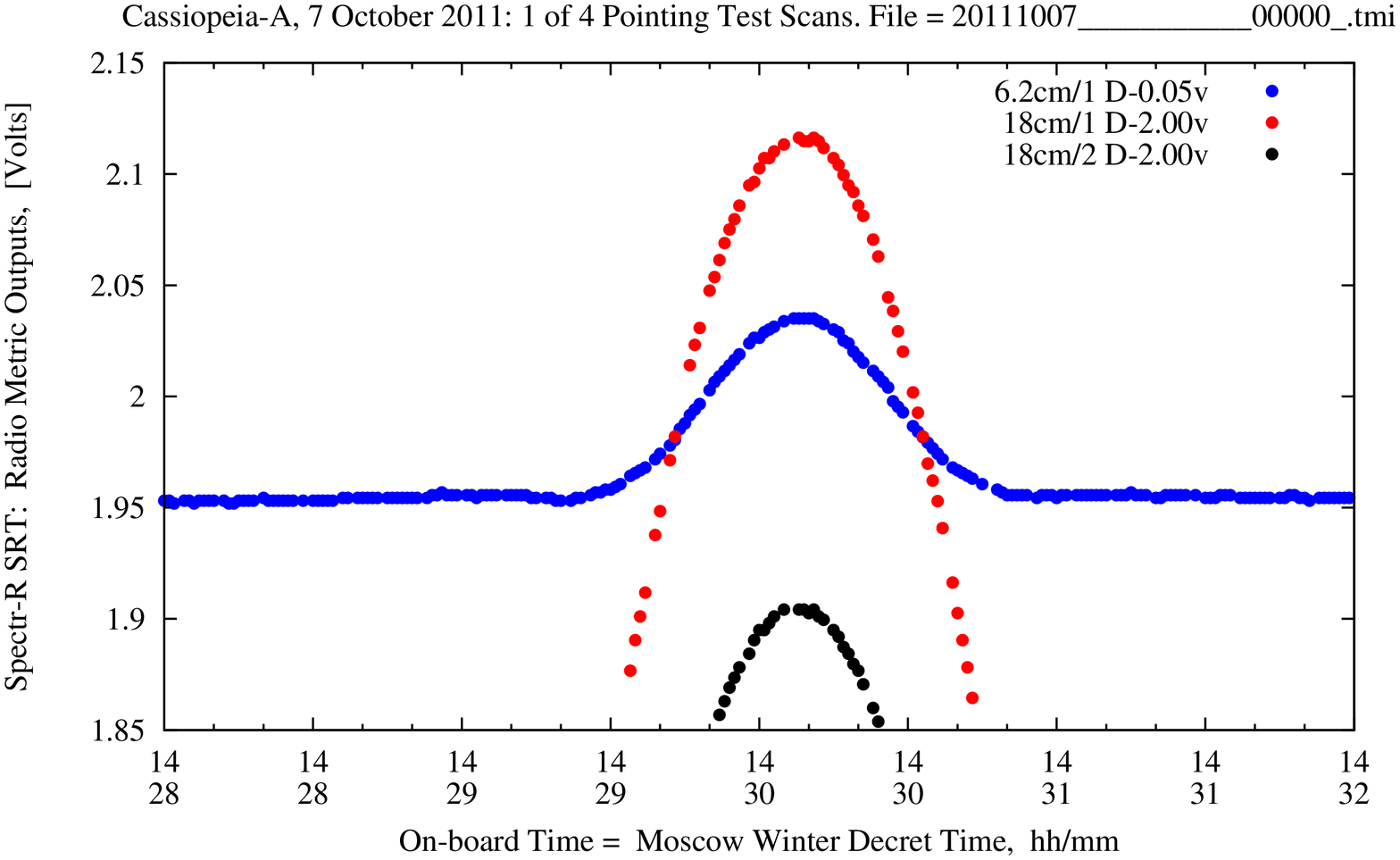}
 }
 \subfigure[]{
  \includegraphics[width=0.48\textwidth,angle=0,clip=true,trim=0 4.5cm 0 5.2cm]{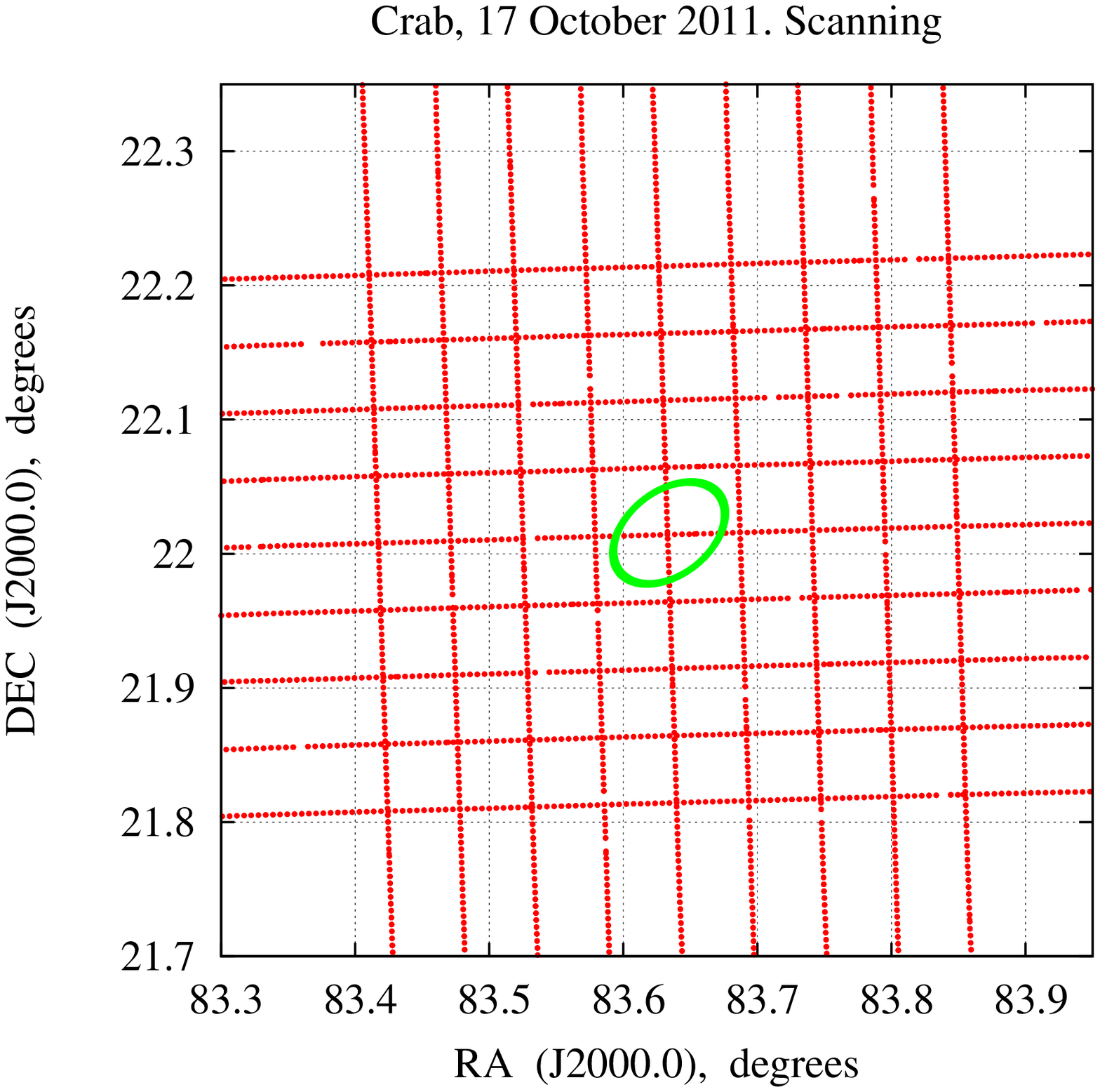}
 }
 \caption{The radiometric response of the SRT for
observations of Cass~A at (a) 1.35~cm on September 28, 2011 and (b) at
6.2~cm on October 7, 2011, the latter against the background of the
simultaneous responses in orthogonal polarizations at 18~cm. (c) Trajectory
for scanning of the part of the sky containing the Crab Nebula on October
17, 2011, corresponding to the responses at 1.35 and 6.2~cm presented in
Fig.~7b of the color insert. The contour shown at the center characterizes
the angular size of the radio source.\hfill}
\end{figure}

\subsection{{Discussion of Results}}

In the absence of phase errors, the full width at half-maximum
(FWHM) of the main lobe of the antenna beam $\vartheta_{0.5}$ and the aperture
efficiency (AE) $\eta$, equal to the ratio of the effective to the geometrical
area, depend on the distribution of the amplitude and phase of the electric
field over the dish aperture and the level of illumination of the dish edge.
For an ideal parabolic reflector with a circular aperture of diameter $D$,
with some {\textit{types}} of theoretical relations for the amplitude
distribution of the co-phased field, the expected beam width $\vartheta_{0.5}$
and aperture efficiency $\eta_0$ at a wavelength $\lambda$ can be estimated
in the co-phased case using the relations~[50--52] $\vartheta_{0.5} =
\alpha_D \cdot \lambda /D$, where $\alpha_D \approx 1.0 {-} 1.5$, and $\eta_0
\approx 1.0 {-} 0.55$, depending on the law for the the field distribution.
The lower the illumination of the dish edge, the lower the value of $\eta_0$
and the higher the value of $\alpha_D$; these values are close to unity only
in the case of uniform field amplitudes all over the aperture. Phase distortions
of the co-phased field in the aperture will additionally increase $\alpha_D$ and
decrease $\eta_0$; i.e., increase the main lobe of the antenna beam and
decrease the effective area.

Comparing these values of $\alpha_D$ with the values $\alpha_D \approx 1.2$
obtained from measurements of the SRT (see parameters 1.10 and 2.10 in Table~2)
shows that the measured beamwidths at 92, 18, and 6.2~cm are close to the
theoretically expected values. The most substantial differences between the
inflight parameter measurements and the predicted or expected values obtained
earlier in tests of the SRT in Pushchino~[47] occurs for the effective area
and the shape of the antenna-beam main lobe at 1.35~cm, and the SRT
system noise temperature at 6.2~cm (Table~2). The use of any of the eight
input bands is possible during operation in the bandwidth-synthesis regime at
1.35~cm. The presented results correspond to the central sub-band, $F_0$ (see
Section~2.2.2).

\textbf{5.6.1. {1.35 cm (central sub-band $\mathbf{F}_{\mathbf{0}}$).}} The
measured FWHM level of the main lobe of the antenna beam corresponds to an
ellipse with axes $\vartheta_{0.5}\times \varphi_{0.5}$ = $6.0^{\prime}\times
13^{\prime}$ with a relative error of $5\%$, compared to the expected circular
beam with a diameter of $5.6^{\prime} \pm 10\%$. This is clearly visible in
the source-scanning responses (Fig.~7b in the color insert and Fig.~8c).
The measured effective area is 7.5~m$^2$~$\pm 13\%$, instead of the predicted
value 27~m$^2$~$\pm 10\%$, or at least the projected value 23~m$^2$~$\pm 15\%$.
These expected values for the beam and area were obtained during ground tests
of the SRT in Pushchino in 2004. The difference between the effective area
obtained in the ground tests and the effective area of an ideal parabolic
surface ($40{-}45$~m$^2$) can be explained in a natural way as an effect of
the random uncertainty in the realization of the dish surface, which has an
rms deviation no worse than the projected value $\sigma = 0.77$~mm, as was
specified via the tolerance $d = \pm 2$~mm ($| d | = 2.6 \sigma$~[43]).
The reduction of the effective area to 7.5~m$^2$ could be explained in this
same way, but with $\sigma \approx 1.4$~mm, or in some other usual way (e.g.
a systematic quadratic phase error in the antenna aperture, with its maximum
value ${\sim} 1.5 \pi$; see the Appendix), if it weren't for the simultaneous
observation of appreciable distortion of the main lobe of the antenna beam.

Such distortions of antenna beams in parabolic antennas are due mainly to
three types of systematic phase distortions in the amplitude--phase distribution
of the field over the dish aperture~[50--53]: quadratic distortions, cubic (coma)
distortions, and astigmatism of the dish and/or feed. Either the dish or feed
could be responsible for quadratic and cubic distortions, or alternately, a
shift of the feed from the dish focus in the directions along (for quadratic
distortions) or transverse (for coma) to the paraboloid axis. Asigmatism
occurs when the points of optimal focus in two main mutually orthogonal planes
perpendicular to the aperture do not coincide~[50]. This means that the optimal
focus point for a feed mounted in the position with minimum aberration (and
therefore with the minimum width for the main lobe of the antenna beam) is
different in these two planes: each has its own focus point, and there is no
single phase center. In this case, there usually exists some common
``equivalent optimal-focus center'' or ``equivalent phase center'' near the
middle of these positions, where it is possible to minimize phase aberrations
and distortion of the antenna beam~[50, 53]. The phase errors in the aperture
are the sum of these three types of errors associated with the dish, feed, and
shift of the feed from the dish focus~[54, 55]. Therefore, it is not possible
to uniquely establish the real origin of phase distortions in the antenna
aperture based purely purely on the results of inflight tests, without
additional data or hypotheses.

A detailed analysis of this problem (see the Appendix) shows that the
observed ellipticity of the main lobe of the antenna beam and the measured
effective area at 1.35~cm can be explained most simply as the effects of
a systematic quadratic error in the distribution of the phase along the
$\varphi$ axis of the order of $1.5 \pi$ on the antenna aperture and
astigmatism due to the feed, in addition to the random error in the realization
of the parabolic SRT dish surface with the projected rms deviation of
$\sigma = 0.77$~mm. Such a systematic phase error along one axis in the
aperture could arise, for example, in the case of astigmatism of the feed
with a quadratic phase error simultaneous with a shift of the optimal-focus
center of the feed relative to the dish focus by about 0.3~cm along the
paraboloid axis, with the distance between the two such centers of focus of
the feed being $b \sim 2$~cm. Evidence supporting this hypothesis is presented
by estimates obtained using one of the phase-distortion models in the Appendix,
obtained using the results of numerical computations of the amplitude--phase
beam of the antenna-feed assembly~[56]. The computations of~[56] suggest possible
astigmatism and quadratic aberrations of the dish illumination with values close
to those required to explain the results of the 1.35-cm measurements.

Attempts to derive a self-consistent explanation for all the antenna measurements
without including astigmatism of the feed were not successful (see the
Appendix). However, this is only one possible explanation. Formally, it is also
possible that the observed systematic phase errors in the aperture are due to
the antenna dish rather than the feed. However, there is currently no firm basis
for this, or sufficient data for a quantitative analysis to justify such a
conclusion. The simple explanation for the asymmetry of the antenna beam as
corresponding to a strong asymmetrical deformation of the dish, such that the
size of the aperture is a factor of two smaller in one plane (to 5~m), should
affect the results at other wavelengths as well. This is in contradiction with
the ``good'' results for the antenna measurements at 6.2, 18, and 92~cm, which
are close to their predicted values.

\textbf{5.6.2. {6.2, 18, and 92 cm.}} In contrast to the other wavelengths,
all measurements at 6.2~cm were conducted separately in the polarization
channels. Turning on both simultaneously led to extremely high output signals
that could not be reduced with attenuators. A distortion of the autospectra
of the output videoband for the channel with right-circular polarization was
also observed in the interferometric regime. These facts suggest a degrading of
the matching between the AFA polarization channels and free space and/or with
the LNA in part of the 6.2-cm AFA input--LNA input antenna-feeder track,
which worsened the isolation between the polarization channels. This led to
an increase in both reactive and active losses in this section when the
channels are turned on separately, causing an increase in the system noise
temperature in accordance with (5.7), and also to self-excitation of the LNA
when both channels are turned on. Analysis of this situation is ongoing.

In all the antenna measurements, the preliminary noise-generator antenna
temperatures $T_\mathrm{NS}$ obtained from pre-flight ground measurements were used,
with ``recalculation'' of $T_\mathrm{NS}$ to the SRT input using formula (5.7c)
together with the measured or calculated~[25] losses in the antenna-feeder
tract. It was planned to correct these values using the results of inflight
tests. This correction was not applied for 92, 18, and 1.35~cm, however, the
$T_\mathrm{NS}$ correction was applied for 6.2~cm: 1)~assuming a corresponding
increase in the losses in the AFA [i.e., a decrease in $K_2$ in (5.7b) and
(5.7c)] and 2)~neglecting distortions from the antenna beam that decrease the
effective area, consistent with the calculations of~[56] and estimates of
parameters 4.2 and 5.2 for 6.2~cm in Table~3\footnote{Since the feeds and
strip or waveguide generators of left- and right-circular polarizations for
each wavelength are structurally joined in the AFA (see Sections 5.1 and 5.2),
the losses in the AFA characterize the losses in both the feeds themselves
(primarily at higher types of waves) and in the polarizers.}. This correction
led to a corresponding increase in the effective area $A_\mathrm{eff}$ and system
noise temperature $T_\mathrm{sys}$ at 6.2~cm (i.e., to an ``improvement'' in
$A_\mathrm{eff}$  and a ``worsening'' of $T_\mathrm{sys}$).

This correction made it possible to obtain a general, self-consistent
explanation for the measured wavelength dependences of the effective area
of the SRT at all wavelengths using a single approach to taking into account
phase errors. It can also simultaneously explain the corresponding increase
in the system noise temperature at 6.2~cm, and possibly partially at 92~cm,
according to (5.7), as being due to an increase in the losses $L_2$ in the
AFA compared to the predicted values\footnote{Due to insufficient resources
and known technical problems with calibrating antenna measurements, including
the problem of manufacturing good-quality, aperture, cooled, matched loads,
the design documentation contains only calculated values of the AFA losses at
92~cm and the results of indirect laboratory measurements or theoretical
estimates of such losses at the other wavelengths.}. The contribution from
the sky background is also appreciable for the 92-cm SRT noise temperature,
and can vary significantly with direction; this can fully or partially
explain the 20\% increase in the measured noise temperature $T_\mathrm{sys} = 200$~K
above the expected value. Based on estimates using published distributions
of the sky brightness temperature and relations (5.7) and (5.7a), the minimum
contribution of the sky background when the antenna is pointed toward the
Galactic pole should be about 60~K, which is included in the expected
temperature $T_\mathrm{sys}^\mathrm{(opt)} = 164 $~K~[25].

We emphasize that the AFA losses $L_2 = 1/K_2$ appear in three terms in (5.7),
which is not always taken into account when obtaining rough estimates. Note
also that, in contrast to calibration using the antenna temperature, which is
necessary for antenna measurements, the astronomical-calibration method
usually used for both the ground radio telescopes and the SRT does not depend
on these characteristics and corrections, since the calibration factors are
proportional to $T_\mathrm{sys}/A_\mathrm{eff}$ (for the SEFD $F_\mathrm{sys}$) or $T_\mathrm{NS}/A_\mathrm{eff}$
(for $F_\mathrm{NS}$). Therefore, this calibration can be carried out for observations
of astronomical sources without knowledge of the absolute values of the
temperatures $T_\mathrm{NS}$ and $T_\mathrm{sys}$ (see Section~5.1 and [45]).

\textbf{5.6.3. {Pointing and Scanning corrections.}} These corrections were
measured by scanning areas containing sources along the $\vartheta$ and
$\varphi$ axes (see the example in Fig.~7b in the color insert and in Fig.~8c).
The results of such scanning were used to find the central cross section of a
source, for which the scanning process was repeated a number of times in
the forward and reverse directions along each axis. The forward and reverse
scans were averaged separately using the telemetry data for the standard
coordinate information, and the difference in the calculated and measured
coordinates for the positions of the signal maxima were calculated, yielding
the desired corrections to the calculated coordinates.

The measurements of the coordinate errors along the $\vartheta$ axis (with the
smaller antenna beam, $\vartheta_{0.5}= 6^{\prime}$, were systematically
different for the ``forward'' and ``reverse'' scans, and the scanning curve
had the ``two-humped'' form characteristic for ground telescopes, with values
of $(3.7^{\prime} \pm 0.2^{\prime})$ for one maximum and $ 1.3^{\prime} \pm
0.2^{\prime}$ for the other. These data were used to introduce a constant
pointing correction $\Delta \vartheta_p = 2.5^{\prime}$, equal to the mean
of these values, which was subsequently applied. The two-humped appearance
of the scans in opposite directions remained with roughly the previous delay,
corresponding to $| \Delta \vartheta_s | = 1.2^{\prime} \pm 0.2^{\prime}$
relative to the calculated value, still with a time delay that was independent
of the forward or reverse direction of the scanning, but now relative to the
zero mean value between the maxima. We interpret this delay in the electrical
axis for motion of the SRT relative to the new calculated position of the axis
as a systematic scanning error (Table~2). No such correction was introduced
for the other axis, since the results in that case were within the uncertainties.

Roughly half the measured time interval between the humps can be explained by
a delay in the response when the signal is integrated at the radiometer output
(see the discussion of this effect in [57]). The main origin of this type of
two-humped curve for ground telescopes (which is absent for the SRT) is believed
to be backlash in the control mechanisms. The origin of the analogous
behavior of the SRT could be related to similar delays during integration of
the signals in the electronic system of the star sensors and the
spacecraft-motion control chains, or some other effect, and requires further
study. The hypothesis of elastic deformations of the boom on which the SRT
focal container is fixed contradicts the telemetry results, which indicate
a fairly uniform speed in sections of the motion.

\textbf{5.6.4. {Telemetry Noise.}} Analysis of the telemetry data showed the
presence of additional ``telemetry noise'' for the digital radiometric outputs
in both channels of the 18 and 92-cm receivers. This has the form of a
background consisting of ``packets'' of short impulsive spikes, and is
characteristic of errors in recording individual bits: the amplitude of their
``variability'' is not random, and repeats systematically, varying in a
step-like fashion ``from packet to packet''.  Comparison with the telemetrized
parameters of ``ADC-ready'' receivers suggests that this noise is due to the
fact that the standard telemetry system does not take the ready signal of
the analog-to-digital converters (ADCs) in the receivers into account when
interrogating all the sensors with fixed velocities. As a result, the
interrogation of the digital sensors sometimes occurs before the ADCs in
these instruments are in the ready state. Such noise was first discovered
during the acceptance tests. A simple and effective means of filtration was
found and applied, which made it possible to eliminate this problem for
the inflight SRT antenna measurements. No such noise is present in the
analog outputs of the 18 and 92-cm receivers or in all outputs of the
1.35 and 6.2-cm receivers.

\begin{figure}[htb!]
\setcounter{figure}{8}
 \subfigure[]{
  \includegraphics[height=0.48\textwidth,angle=270,clip=true,trim=0 0 0 0]{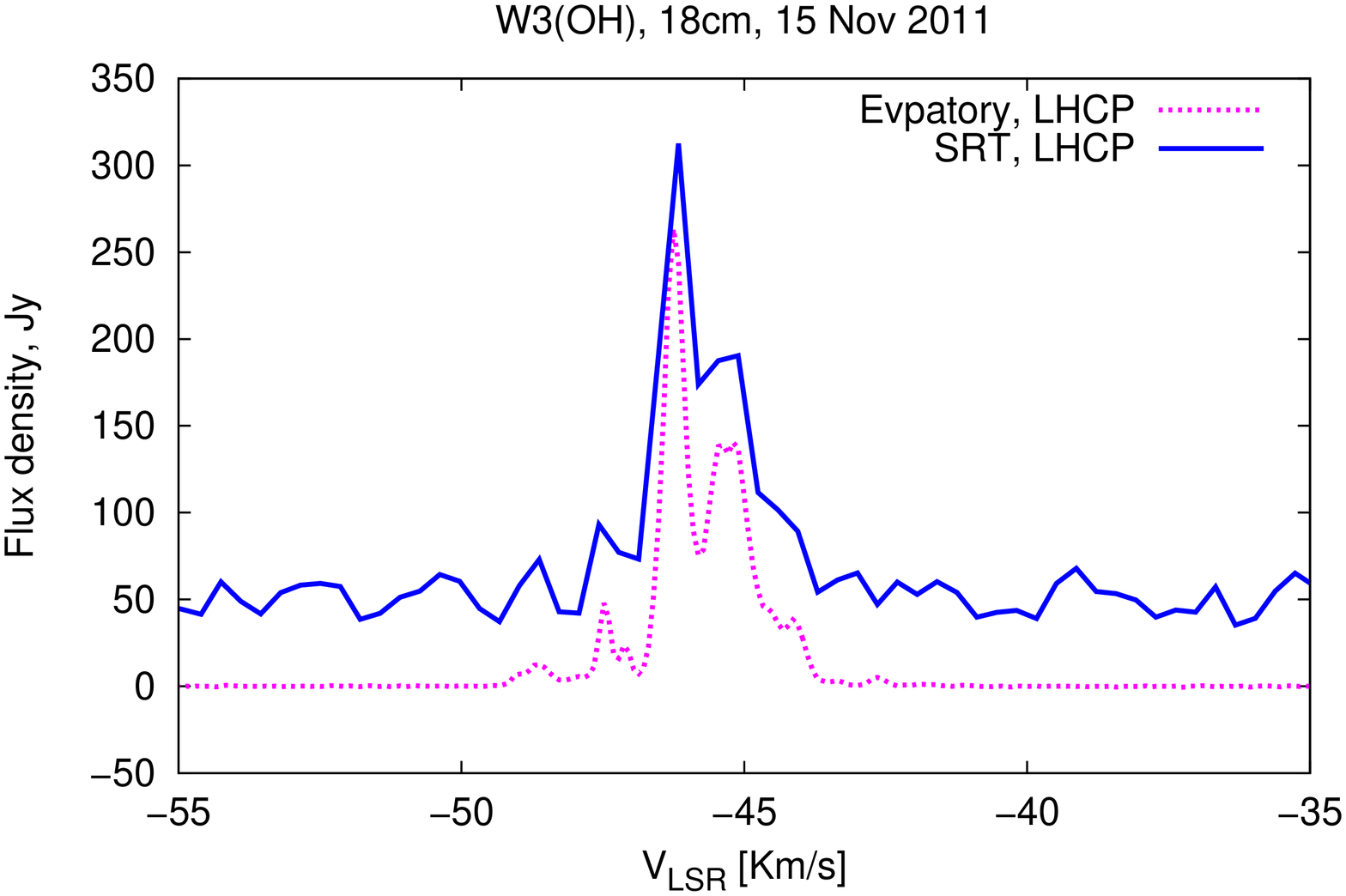}
 }
 \subfigure[]{
  \includegraphics[height=0.48\textwidth,angle=270,clip=true,trim=0 0 0 0]{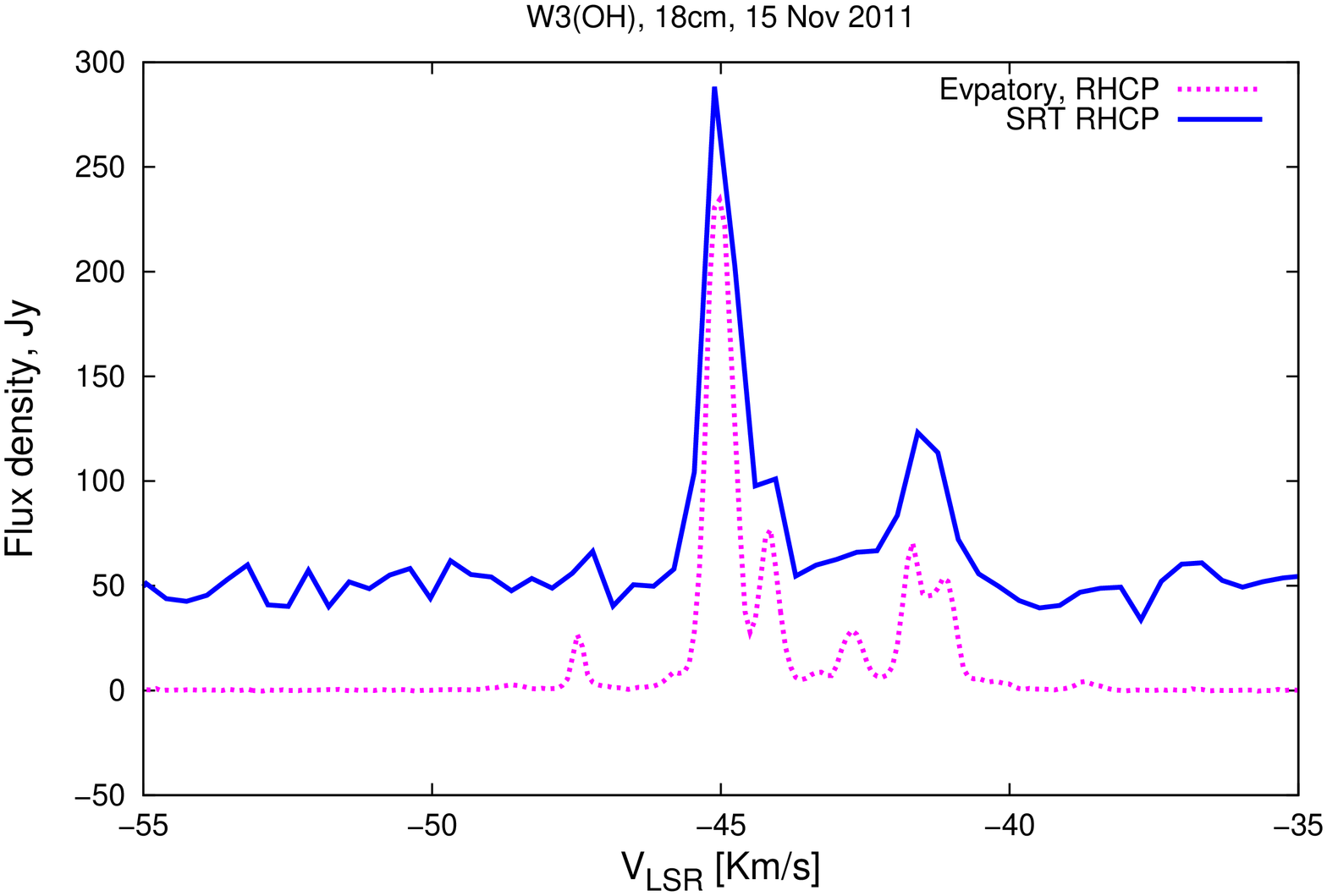}
 }
 \subfigure[]{
  \includegraphics[height=0.48\textwidth,angle=270,clip=true,trim=0 0 0 0]{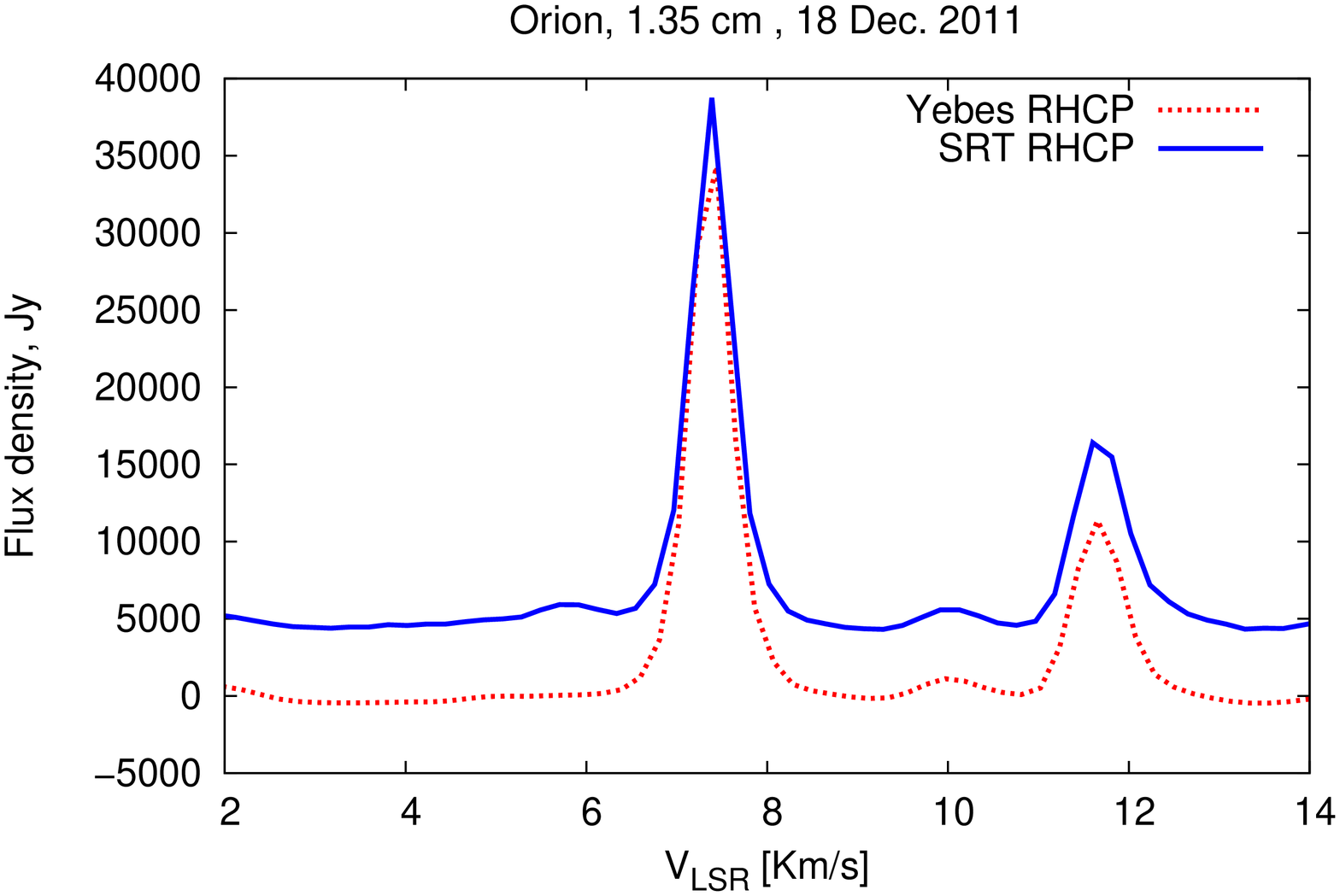}
 }
 \caption{Simultaneous two-antenna observations of cosmic maser sources.
(a, b) 18-cm observations of W3(OH) by the SRT on October 30, 2011 in (a)
left- and (b) right-circular polarization. (c) 1.35-cm observations of the
Orion~KL star-forming region by the SRT and the 32-m Zelenchuk radio telescope
of the Institute of Applied Astronomy on December 18, 2011. The SRT spectra are
shifted in amplitude for clarity.}
\end{figure}

\subsection{{Independent Verification of the Interferometric Regime Using
Radio Lines}}

It is possible to use observations of several strong cosmic OH (18~cm) and
water (1.35~cm) maser sources to test the operation of the receiver equipment,
HDRRC radio line, and correlator in the interferometric regime in K and L
bands. Several measurement sessions were carried out for this purpose,
in which such objects were observed simultaneously in right- and left-circular
polarizations with the SRT and several ground telescopes. The star-forming
region Orion~KL and W3(OH) were chosen for the 1.35-cm observations, and
W3(OH) for the 18-cm observations. The high fluxes of these objects and the
presence of strongly polarized components in their spectra enabled a
comparative analysis of the characteristics of the receiver--recorder
equipment and its suitability for such observations.

\textbf{5.7.1. {Observations.}} {\textbf{1. W3~(OH).}} The star-forming
region W3(OH) is located at a distance of about 2~kpc~[58] in the Perseus
arm of the Galaxy, and is among the most studied objects of this type.
Strong maser emission in both OH lines and water lines (with the latter
approximately $6^{\prime\prime}$ to the East of the OH masers) is observed
toward W3(OH). The spectrum contains polarized features, enabling estimation
of the polarization properties of the SRT by comparing spectral profiles
obtained simultaneously with the SRT and a well-understood ground instrument.

{\textbf{2. Orion KL.}} This well known star-forming region is located at
a distance of $437 \pm 19$~pc~[59] in the constellation Orion. The water
masers undergoes flares, as a result of which individual spectral features
can reach flux densities of several million Jy~[60]. Since the beginning of
2011, Orion KL has been in a phase of enhanced activity~[61], and its flux
at the epoch of our observations was approximately $3.5 \times 10^4$~Jy,
making this maser the strongest object in its class in the sky. The object
is especially convenient for observations, since its high flux density
makes it possible to achieve high signal-to-noise ratios in modest
integration times, even with the low sensitivity of the SRT in K band.

The observations of W3(OH) were carried out in October and December 2011,
and the observations of Orion KL in December 2011, as part of the program
of inflight receiver tests and fringe searches. Below, we analyze the
spectra of these objects obtained with the SRT in an autocorrelation regime
(using data that have passed through the HDRRC channel) and at ground radio
telescopes participating in simultaneous observations.

\textbf{5.7.2. {Analysis of the Spectra.}} Figures 9a, 9b, and 9c show that
the maser-line profiles obtained on the SRT at 1.35~cm for Orion~KL at
the frequency of the 22235.08-MHz water line and at 18~cm for W3(OH) at the
frequency of the 1665.4018~MHz OH line correspond nearly fully to the line
profiles obtained on large ground telescopes. The small observed differences
in the spectrum profiles measured by the ground telescopes and the SRT are
most likely due to the low signal-to-noise ratio of the SRT data. Figures~9a
and 9b for W3(OH) show that, due to the appreciable difference in the
appearance of the spectral features in the different polarizations, it is
easy to identify the received polarization (left- or right-circular
polarization; compare with Fig.~2 of~[62]). Such sources can also serve
as an additional tool for monitoring the pointing and verifying the correctness
of the frequency tuning of the equipment at 1.35 and 18~cm. Another task that
can be addressed with such studies is estimation of the sensitivity of the
SRT based on spectral observations in the L and K bands. The parameters of
the ground telescope that is used can be used to estimate the flux from a
source at the observing epoch, and thereby the SRT sensitivity. Such estimates
were obtained based on observations of W3(OH) in the K and L bands and of
Orion~KL in the K band. The resulting SEFDs are 3400~Jy for the L band and
about 36\,000~Jy for the K band.

\subsection{{Conclusions}}

1. The derived equivalent system noise temperature of the SRT coincides with
theoretical estimates at 92, 18, and 1.35~cm within $20\%$, but exceeds the
calculated value at 6.2~cm by a factor of two, which lowers the interferometer
sensitivity of this last band by a factor of $\sqrt 2$ (for a fixed integration
time). The origin of this enhanced noise temperature is probably an increase
in losses in the antenna-feeder tract (most likely in the section running from
the AFA to the LNA), compared to the losses calculated based on laboratory
measurements carried out on the Earth.

2. The FWHM of the main lobe of the SRT beam agrees with theoretical
expectations and measurements obtained in ground tests within the errors at
92, 18, and 6.2~cm, but differs appreciably from the expected value at 1.35~cm:
the transverse cross section of the main lobe at the half-maximum level is
close to elliptical, with axes $\vartheta_{0.5} \approx 6.0^{\prime} \pm 5\%$
and $\varphi_{0.5} \approx 13 ^{\prime}  \pm 5\%$, rather than the expected
circular cross section with diameter $5.5 ^{\prime}  \pm 10\%$. The profile
of the longitudinal cross section of the main lobe in the plane in which it
is wider is appreciably asymmetrical.

3. The effective areas of the SRT in flight are close to the calculated values
and the values measured in ground tests at 92, 18, and 6.2~cm. The effective
area at 1.35~cm is a factor of 3.6 smaller than the value obtained in ground
tests, 27~m$^2$~$\pm 10\%$, and a factor of three smaller than the projected
area, 23~m$^2$~$\pm 15\%$, which reduces the sensitivity in the interferometric
regime by nearly a factor of two (for a fixed integration time).

4. Estimates based on calculated values of the amplitude--phase beam of the
AFA in a simple model for the phase errors in the antenna-feeder system lead
to the following conclusions about the SRT in flight.

-- The mean dish surface profile may be close to parabolic, with its rms
deviation equal to the predicted value, 0.77~mm.

-- The antenna feed may have: a) quadratic phase errors with the maximum
error in the feed phase at the edge of the disk equal to roughly
${-}100^{\circ}$ at 1.35~cm and ${-}35^{\circ}$ at 6.2 and 18~cm (according
to the calculated phase beams for these feeds); b) astigmatic aberrations
at 1.35~cm, approximated by the presence of two equivalent centers of focus
of the feed --- centers 1 and 2 in orthogonal planes 1 and 2, respectively ---
shifted from the dish focus by approximately 7~mm toward the dish in plane~1
and 13~mm away from the dish in plane~2; c) a common center of optimal focus
of the feed (between centers 1 and 2), which is shifted from the focus of the
unfurled antenna by approximately 3~mm from the dish along the longitudinal
axis of the antenna.

-- These phase errors in the illumination of the dish surface may provide
the main reason for the measured decrease in effective area and the ellipticity
of the main lobe of the SRT antenna beam at 1.35~cm compared to expected values.

5. The results of radio adjustments show that the mean pointing error along
the axis with the smaller beam width is $3.7^{\prime} \pm 0.2^{\prime}$  for
scanning of a source in one direction and $1.3^{\prime} \pm 0.2^{\prime}$  for
scanning in the reverse direction. This can partially be explained by a delay
of the response due to integration of the signal in the radiometer output,
and requires further study. Based on these measurements, a constant pointing
correction was introduced, $\Delta \vartheta_p = 2.5^{\prime}$. The remaining
scanning error, $| \Delta \vartheta_s | = 1.2^{\prime} \pm 0.2^{\prime}$,
characterizes a delay in the signal maximum relative to its calculated position
during scanning in either direction, and is due to a systematic error in the
antenna motion. The pointing error along the axis with the larger antenna beam
lies within the measurement uncertainties, and does not exceed $1.5^{\prime}$;
this error was not corrected for in subsequent observations.

6. Comparison of the autocorrelation spectra for two strong maser sources
in L and K bands indicates that the spectral observational regime of the SRT
is functioning normally. Such observations can be used to monitor the frequency
tuning and polarization regime during observations. Estimates of the SRT
sensitivity based on observations of radio lines are in agreement with the
results of continuum observations. The full determination of the polarization
parameters of the radio telescope, which is possible using specially planned
observations of several bright maser sources such as Orion~KL, remains
incomplete.

\section{VERIFICATION OF THE FUNCTIONING OF THE GROUND--SPACE INTERFEROMETER
(FIRST FRINGES) AND FIRST OBSERVATIONAL RESULTS}

In this section, we present a brief survey of the first results obtained
in the interferometric regime. These results will be discussed in more
detail in future articles by various international groups involved with the
fringe searches and the Early Science Program, after a more thorough analysis
of the data.

The detection of the interferometer signal between the {\em RadioAstron}
SRT and ground radio telescopes has demonstrated the overall successful
operation of the space--ground VLBI system at all four wavelengths: 92, 18,
6.2, and 1.35~cm (Figs.~7c--7g in the color insert). The first signal
from the space interferometer was obtained for observations made on November
15, 2011, of the quasar 0212+735 at 18~cm, when the spacecraft was about
100\,000~km from the Earth and the projected baseline between {\em RadioAstron}
and the 100-m Effelsberg telescope was $B = 8100$~km (Fig.~7c in the color
insert). In all, 20 test sessions of the interferometer have been conducted
thus far. Interferometric observations of the pulsar PSR 0950+08 at 92~cm
were carried out with the spacecraft at the record distance of 300\,000~km
(a projected baseline of about 220\,000~km) on January 25, 2012, with the
participation of the largest ground radio telescope --- Arecibo, with a diameter
of 300~m (Figs.~7f, 7g in the color insert). Most interferometric observations
have used the onboard hydrogen maser, but successful test sessions have also
been carried out in a regime where a coherent signal from the hydrogen maser
at the tracking station in Pushchino is sent to the spacecraft and then sent
back to the tracking station (a so-called closed phase loop regime).

In the interferometric regime, the sensitivity of the system of two telescopes
is proportional to the square root of the product of the effective areas
of these telescope; thus, the combination of the 10-m SRT and a 100-m ground
radio telescope has a sensitivity equivalent to a pair of two 30-m telescopes.
It is not possible to obtain the results of a ground--space interferometer
measurement immediately after the measurement itself. The scientific data
recorded at the various radio telescopes are first transmitted to a reduction 
center for correlation (detection of the interferometer response). This
correlation can be carried out only after high-precision reconstruction of
the spacecraft orbit at the ballistic center.

\begin{figure}[ht!]
\setcounter{figure}{9}
\includegraphics[width=0.5\textwidth,angle=0,clip=true,trim=0.5cm 1cm 0 0]{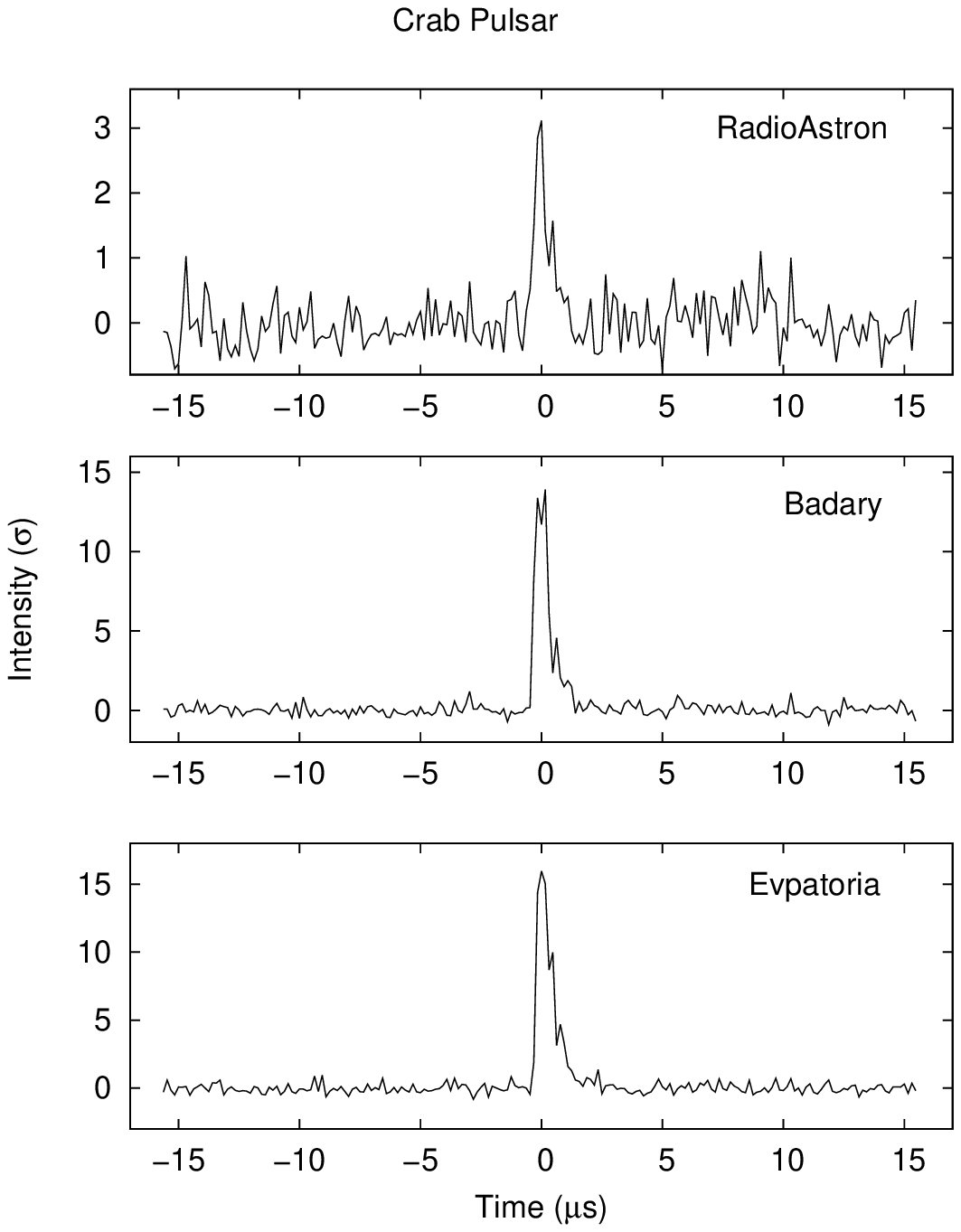}
\caption{Simultaneous three-antenna observations of the giant pulses of the
Crab Nebula pulsar on November 14, 2011, carried out by the SRT, the
Badary 32-m telescope (near Irkutsk, Russia), and the 70-m Evpatoria telescope
(Ukraine) at 18~cm.}
\end{figure}

\begin{figure}[t!]
\setcounter{figure}{10}
\includegraphics[width=0.48\textwidth,angle=0,clip=true,trim=0 0 0 0]{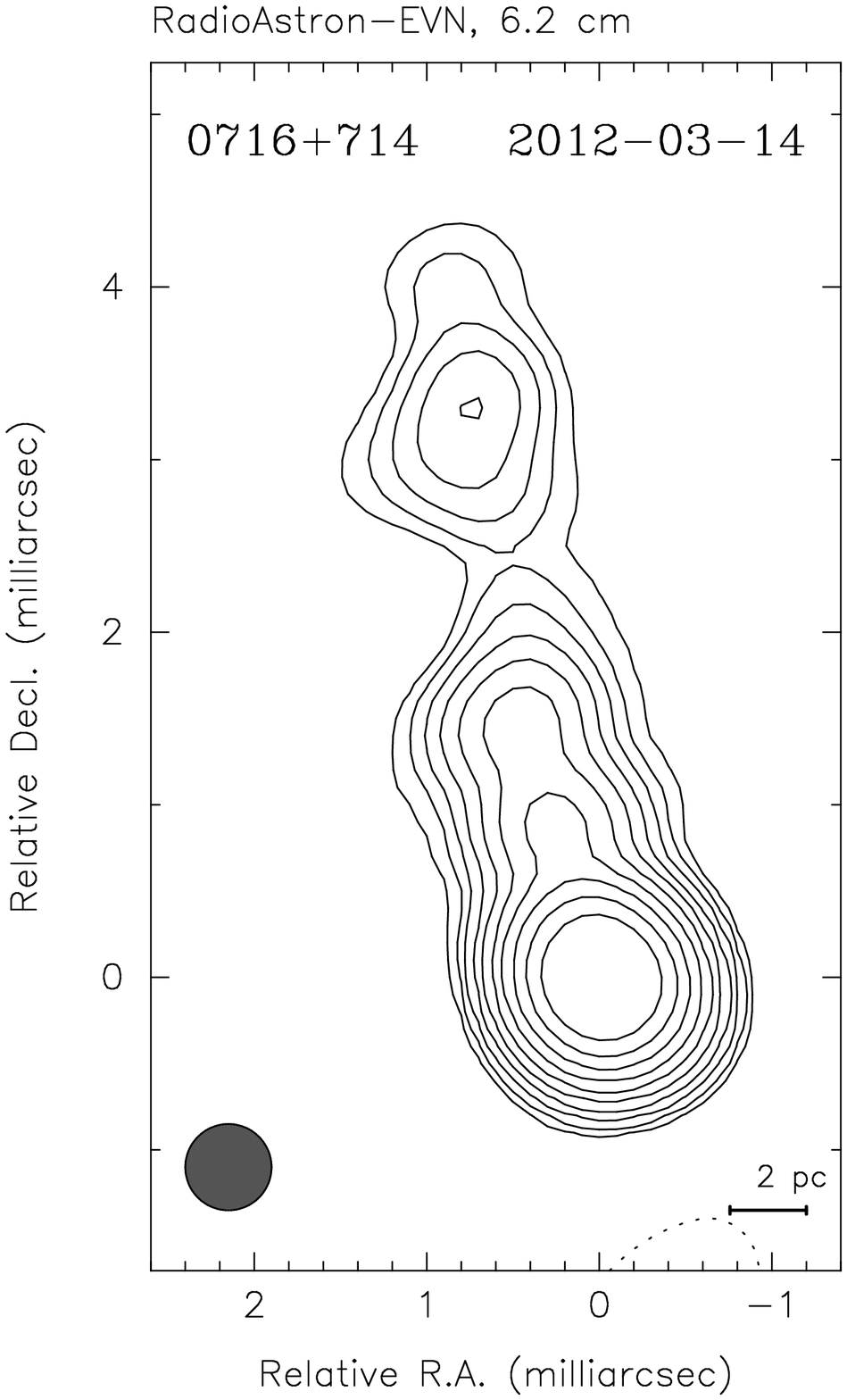}
\caption{6.2-cm image of the rapidly variable BL Lac object 0716+714.
The observations were carried out using the SRT and the EVN on March 14-15, 2012
as part of the {\em RadioAstron} Early Science Program on AGNs.
This image was constructed using an antenna beam with a FWHM of 0.5~mas,
whose cross section is shown in the lower left-hand corner of the figure.
The lowest contour is 0.25~mJy/beam, and the contours increase in steps of
a factor of two. The peak intensity is 0.43~Jy/beam. The ``beam'' is the
solid angle of the shown cross section.\hfill}
\end{figure}

The reduction and analysis of data obtained using the {\em RadioAstron}
ground--space interferometer are carried out at the ASC in collaboration with
other participants of the project. First and foremost, this includes the
correlation of the data flows recorded at the individual
radio telescopes at rates of 128 or 256~Mbits/s, including the space segment
of the SRT (128~Mbits/s), using the RDR-1 recording system created at the
ASC~[63] and the Mark5 recording system developed in the USA~[64]. The
FX correlator of the ASC is based on a computing cluster with a performance
of 1~Tflop/s and a RAID data-storage system with a volume of up to 200~Tbyte.
The technical characteristics of the processor cluster of the ASC SDRC
is able to handle data flows from 10 stations including the
SRT with integrated data rates of up to 2.56~Gbits/s; accordingly, it can
process up to 45 interferometer baselines. This can be done essentially
at the rate at which the data were recorded in real time.

The detection of an interferometer response is not in itself a final scientific
result. However, with certain assumptions about the structure of a compact
feature, it can be used to estimate its angular size and brightness temperature.
Multiple observations with various telescope configurations, most importantly
with various positions of the spacecraft in its orbit, are necessary to derive
trustworthy conclusions about the structure of a studied object. The required
set of such configurations can be realized in observations over no less than
one calendar year. The planned observing strategy is to study a sample of radio
sources over the course of a year (and sometimes several years), after which
a multi-faceted analysis is applied to draw basic conclusions about the
structure and physical conditions in the studied objects.

The {\em RadioAstron} Early Science Program has been underway since February
2012, and is overseen at the ASC and carried out by international groups of
researchers formed in the framework of the project. Thus far, the interferometer
responses from the pulsars PSR~0950+08, PSR~0531+21 (Crab), PSR~0833-45 (Vela),
and PSR 1919+21, the AGNs 0212+735, 0716+714, 0748+126, 0754+100, 2013+370,
0851+202 (OJ287), 1954+513, and 2200+420 (BL~Lac), and the Galactic maser
W51 have been measured (see the {\em RadioAstron} newsletters for
2011--2012~[65]). These responses were obtained for projected baselines for
the ground--space interferometer of less than 10\,000~km to about 250\,000~km,
roughly 20 Earth diameters.

After confirmation of the possibility of observing giant pulses from the
Crab Nebula pulsar (at a distance of 1~kpc) with the SRT, correlations
between the 18-cm pulses recorded at the SRT and the Evpatoria, Svetloe,
Zelenchuk, and Badary ground radio telescopes were detected for observations
made on November
14, 2011, for projected baselines of up to $B =  40\,000$~km (Fig.~10).
This testifies that scattering of the image in the interstellar medium
along the path from the pulsar to the observer at 18~cm was no greater than
the angular resolution of the interferometer, 400~$\mu$as.

Interferometer observations of ordinary pulses of the nearby pulsar PSR~0950+08
(260~pc; Fig.~7f in the color insert) did not detect interstellar scattering
even at 92~cm with projected baselines of up to $B=220\,000$~km. In this way,
an angular resolution of 370~$\mu$as was achieved at meter wavelengths.
Observations of this pulsar over one hour (Fig.~7g in the color insert)
enabled the detection of variability of the visibility function with such long
baselines, opening possibilities for studying the parameters of turbulence
of the interstellar plasma and the achievement of even higher angular
resolution using the ``interstellar interferometer'' principle~[66].
Scattering in the interstellar medium was detected for another nearby pulsar
Vela, at a distance of ${\sim} 300$~pc. This pulsar was observed at 18~cm
jointly by the SRT and the largest radio telescopes of the Southern hemisphere
in May 2012. The Parkes, Mopra, Hobart (all Australia), Hartebeesthoek (South
Africa), and Tidbinbilla (Australia) radio telescopes participated in these
observations. The reduction of these data showed that the structure of the
interferometer response changes completely at a projected baseline of
100\,000~km (numerous narrow brightenings are observed), indicating that we are
observing the results of a multi-path propagation of the radio waves through
inhomogeneous interstellar plasma.

An example of the interferometer response to an H$_2$O maser source (1.35~cm)
in the star-forming region W51 obtained during observations with the SRT and
the 100-m Effelsberg radio telescope is presented in Fig.~7e (color insert).
These observations were carried out on May 12, 2012, with a projected
ground--space baseline of 14\,500~km (1.14 Earth diameters), yielding an
angular resolution of 80~$\mu$as.

The largest number of interferometric observations have been associated with
studies of the structure of AGNs. Interferometry responses at 6.2~cm with
projected values for the SRT--Effelsberg baseline $B = 92\,000$~km, or 7.2
Earth diameters, have been obtained for the two quasars OJ~287 (which some
authors have suggested may harbor a binary supermassive black hole; see,
for example, [67]) and BL~Lacertae (the prototype of the class of BL~Lac
objects). This yields brightness temperature estimates for the dominant
feature in the compact jet --- the core --- of about $10^{13}$~K or somewhat
higher. This exceeds the well known inverse-Compton limit for the brightness
temperature [68], but the emission can still be explained in the standard
model with incoherent synchrotron emission from a Doppler-boosted relativistic
jet~[69].

For the AGN 0716+714, which is among the most rapidly variable extragalactic
objects, interference fringes were detected at 6.2~cm at multiple projected
ground--space baselines from roughly 1.5 to more than 5 Earth diameters.
The international group working on the Early Science Program constructed an
image of this object (Fig.~11) and estimated the parameters of the core.
The width at the base of the jet in the core region is approximately
70~$\mu$as, or 0.3~pc, and the brightness temperature is $2 \times 10^{12}$~K.
Note that these parameters were measured at an epoch of minimum activity of
this object.

\section{CONCLUSION}

The {\em RadioAstron} space radio telecope has been inserted into the
nominal orbit, and successfully tested in both autonomous and ground--space
interferometric regimes at all four operational wavelengths (1.35, 6.2, 18,
and 92~cm). The main parameters of the SRT and interferometer have been
determined. Record angular resolutions, more than a factor of 10 better than
is attainable on Earth, have been achieved. The sensitivity derived from
inflight tests is sufficient to enable a full-scale program of scientific
studies. Further measurements will be used to refine the limits of possible
observations in several directions: 1)~determination of the flux limit for
the detection of linearly polarized emission, 2)~use of evolutionary changes
in the orbit to obtain fuller information about the structure of objects
(in paraticular, to attain ultra-high angular resolution both along and
across jet structures in AGNs), 3)~application of multi-frequency synthesis
in image construction, 4)~analysis of Faraday rotation of the plane of
polarization based on multi-frequency observations, 5)~study of variability
of the source structures, 6)~high-precision astrometric measurements,
7)~analysis of possibilities for high-precision determination of the spacecraft
motion, etc.

Brief reports about current results are regularly issued in 
{\em RadioAstron} newsletters~[65].

\section*{ACKNOWLEDGEMENTS}

The development and modernization of the {\em RadioAstron} project was
carried out over many years. We consider it our duty to express thanks for the
contributions to this work by all our colleagues, who are too numerous to
list here individually. Among them, we are especially grateful for the
support and participation of Roal'd Sagdeev, Victor Troshin, Anatolii Trubnikov,
and Oleg Andreev (IKI) and Gerard Mersch and Kees van't Klooster (ESA/ESTEC)
in the early stages of the project, and for the participation of Jan Buiter
(Netherlands),  Sam Wongsowijoto (Germany), Yurii Onopko and Tanya Downs
(Australia), Petri Jukkala, Juha Mallat, and Petri Piironen (Finland) in
planning and the creation of the first onboard science receivers as part of
the international receiver development group in 1985--1994.

We are grateful to the scientific and technical staff of the Special Astrophysical
Observatory of the Russian Academy of Sciences (Nizhnii Arkhyz, Russia) and
Effelsberg (Germany) for ground support of the antenna measurements of the
SRT parameters based on multi-frequency observations of quasi-point-like
variable raido sources, and the staff of the observatories whose radio
telescopes have participated in the inflight tests to carry out fringe
searches, namely the Svetloe, Zelenchuk, and Badary antennas of the Kvazar
Network (Russia), Evpatoria (Ukraine), Effelsberg (Germany), the NRAO Green
Bank Telescope and Arecibo Observatory (USA), Westerbork (The Netherlands),
Yebes (Spain), Medicina (Italy), and Usuda (Japan). Subsequent observations
carried out as part of the Early Science Program included joint observations
of the SRT together with some individual ground telescopes, as well as the
Kvazar, EVN, and LBA VLBI networks. We thank the staff of the Svetloe,
Zelenchuk, and Badary (Russia), Evpatoria (Ukraine), Effelsberg (Germany),
NRAO Green Bank Telescope and Arecibo (USA), Westerbork (Netherlands),
Yebes and Robledo (Spain), Medicina and Noto (Italy), Usuda (Japan),
Jodrell Bank (UK), Onsala (Sweden), Shangai and Urumqi (China), ATCA,
Parkes, Mopra, Hobart, and Tidbinbilla (Australia), Hartebeesthoek (South
Africa), and Ooty (India) observatories.

We thank the referee for comments that have made it possible to improve this
paper, L.S.~Chesalin for help in setting up the tracking station at the
Pushchino Radio Astronomy Observatory, E.P.~Kolesnikov for useful discussions
of the electromagnetic compatibility of scientific receivers and the
housekeeping transmitters of the SRT in the case of out-of-bandwidth reception
and the results of measurements of the main parameters of the SRT, P.G.~Tsybulev
for discussions of the origin of ``telemetry noise'' and the development of
the KRTVIZ program for visualizing observations, G.V.~Lipunova for help in
preparing the figures with the test results.

Work on the Early Science Program of the {\em RadioAstron} project has
been partially supported by the Basic Research Programs of the Presidium of
the Russian Academy of Sciences P-20 (``The Origin, Structure, and Evolution
of Objects in the Universe'') and P-21 (``Non-stationary Phenomena in Objects
of the Universe''), the Basic Research Programs of the Division of Physical
Sciences of the Russian Academy of Sciences OFN-16 (``Active Processes and
Stochastic Structures in the Universe'') and OFN-17 (``Active Processes in
Galactic and Extragalactic Objects''), the Ministry for Education and Science
of the Russian Federation, in the framework of the Federal Targeted Program
``Science and Scientific Staff of Innovative Russia'' for 2009--2013
(State contract  16.740.11.0155; Agreement 8405), the Russian Foundation for
Basic Research (projects 10-02-0076, 10-02-00147, 11-02-00368, 12-02-33101), 
and the ``Dinastiya'' Foundation for Non-Commercial Programs. The RATAN-600
observations used in the analysis of the antenna measurements were supported
by the Ministry for Education and Science of the Russian Federation
(State contracts 16.518.11.7062 and 16.552.11.7028). The European VLBI
Network is a joint facility of European, Chinese, South African and other
radio astronomy institues funded by their national research councils.
The National Radio Astronomy Observatory is a facility of the National
Science Foundation operated under cooperative agrement by Associated
Universities, Inc.

Authors thank Denice Gabuzda for translating the original manuscript
(Astronomicheskii Zhurnal, 2013, Vol.~90, No.~3, pp.~179--222)
from Russian to English.

\begin{flushright}
 {\textit{APPENDIX}}
\end{flushright}

\section*{POSSIBLE INTERPRETATION OF THE 1.35-CM ANTENNA MEASUREMENTS}

The measured ellipticity of the main lobe of the SRT antenna beam at the
half-maximum power level could be due to comparatively large phase errors along
the $\varphi$ axis combined with relatively modest distortions in the orthogonal
$\vartheta$ axis. We will refer to the orthogonal azimuthal planes corresponding
to longitudinal cross sections of the main lobe of the SRT beam with widths
$\vartheta_{0.5}$ and $\varphi_{0.5}$ along these axis as ``plane 1'' and
``plane 2'', and to the corresponding phase errors over the dish aperture
as $\delta \varphi_1$ and $\delta \varphi_2$. The main lobe of the antenna
beam in plane~2 (along the $\varphi$ axis) turned out to be a factor of 2.2
wider than in plane~1, appreciably asymmetric, and somewhat shifted relative to
the calculated geometrical axis of the SRT; these properties are characteristic
for distortions that arise due to combinations of all three types of
aberration --- quadratic, cubic, and astigmatic.

Physically, the maximum quadratic and cubic phase errors in ground radio
telescopes usually indicate defocusing of radiation reflected from the edge
of the dish that arises due to transverse or longitudinal shifts of the
feed from the focus; such errors are characteristic even for ideal
paraboloids with modest ratios of the focal distance $F$ to the diameter of
the dish aperture $D$ ($F/D \approx 0.43$ for the SRT). In this case, assuming
that the feed was mounted at the calculated geometrical focus of the
paraboloid with the projected accuracy (${\pm} 1$~mm), such errors could
appear, for example, if the real focus of the unfurled dish did not precisely
coincide with the calculated position. However, other origins are also
possible, including peculiarities of the illumination of the dish (see below).

Rough quantitative estimates yield for the phase error in plane~1 at the
dish edge (relative to the aperture center) $\delta \varphi_1 \approx 0$
(since $\vartheta_{0.5}$ is close to the predicted value for the undistorted
beam within the errors, see parameters 1.10 and 2.10 in Table~2), while the
error in plane~2 is $\delta \varphi_2 \sim (2 \pi / \lambda) R_0
(\varphi_{0.5} - \vartheta_{0.5}) = 1.5 \pi$\footnote{Estimates based on
plots presented in~[51, 53] give similar values: $\pi < \varphi_q < 2\pi$
for a broadening of the main lobe of the beam by a factor of $\approx
1.5{-}3$.}, where $\lambda = 1.35$~cm is the wavelength, $R_0 = 500$~cm the
radius of the SRT dish, $\varphi_{0.5} = 13^{\prime}$, and
$\vartheta_{0.5} = 6^{\prime}$.

The difference in the effective area can be explained quantitatively by the
combination of a systematic error in the dish aperture $\varphi_q = \delta
\varphi_1 + \delta \varphi_2 \sim 1.5 \pi$ and random errors with the projected
rms $\sigma = 0.77$~mm, using the following formulas for the coefficients
$\eta_\sigma$ and $\eta_\varphi$ for the decrease in the effective area of
the antenna aperture~[43, 55]:
$$
\eta_\sigma = \exp [{-}(4 \pi \sigma / \lambda)^2],
\eqno{\mbox{A}.1}
$$
$$
\eta_\varphi  = 6.55 (1.01 - 0.2 \cos \varphi_q) / (5.3 +
\varphi_q^2). \eqno{\mbox{A}.2} 
$$
Whence, $\eta_\sigma = 0.60$, $\eta_\varphi = 0.24$, $\eta =\eta_\sigma
\eta_\varphi = 0.15$, and we obtain for the effective area $A_0$ ``corrected''
for these losses $A_0 = A_\mathrm{eff} /\eta_\varphi = 31$~m$^2$ and $A_0 = A_\mathrm{eff} /
\eta = 50$~m$^2$ for $A_\mathrm{eff} = 7.5$~m$^2$. These values of $A_0$ confirm that
the two types of errors considered could, in principle, be the main causes of
the reduction in $A_\mathrm{eff}$ and the aperture efficiency (reductions of AE from
$1$ to $0.5 {-} 0.6$ are typical for single-dish antennas, and are usually
determined by other well known factors, most importantly over-illumination of the
dish and incidence of radiation toward the dish edge (depending on the feed
beam) and errors in the dish surface; see, for example, Appendix~5 in~[43]).
Note that, since formula~(A.2) was obtained assuming a Gaussian amplitude
distribution over the aperture, with illumination of the aperture edge to the 0.1
level~[55], we expect that this formula is more applicable for our estimates,
in spite of the fact that this takes into account only quadratic, not cubic,
phase errors.  We will neglect area losses due to cubic phase errors
(or coma-type aberration) here.

The reduction of the effective area can also formally be explained using only
$\eta_\sigma$ with an rms deviation $\sigma \approx 1.4$~mm. This value of
$\sigma$ would then be considered some kind of ``effective'' random error. In
our case, it is possible to distinguish the similar contributions of $\sigma$
and $\varphi_q$ only because of the observed ellipticity of the antenna beam.

The observed scatter in the measured effective areas about the mean area
(more than $10\%$) could indicate variability of the random and systematic
errors ($\sigma$ and $\varphi_q$) due to variation in the temperature
distribution over the dish surface with variation in the SRT orientation
relative to the Sun. For example, it follows from (A.1) that, with $\sigma
\approx 0.77$~mm, varying $\sigma$ by even 0.1~mm could lead to appreciable
variations in the effective area of the SRT at 1.35~cm. This strong dependence
on the random error is due to the closeness of the central wavelength at
1.35~cm to the so-called ``limiting'' wavelength of the radio telescope
$\lambda_\mathrm{min}$, which is equal to $\lambda_\mathrm{min} \approx (20 {-} 16) \sigma$
(according to the practical criterion for estimating this limiting wavelength
given in~[43]), and depends on the accuracy of the realization of the dish
surface, usually relative to an ideal parabolic surface. We have for the
SRT $\lambda / \sigma = 18$ for $\lambda = 13.5$~mm and the projected rms
deviation $\sigma = 0.77$~mm.

\begin{table*}[p!]
\caption{Modeling of phase errors in the antenna system: three models for
the contribution of feed phase errors~[56] to the decrease in $A_\mathrm{eff}$ and
the aperture efficiency AE, as well as the ellipticity of the cross section
of the main lobe of the SRT beam} 
\begin{tabular}{l|c|c|c|c}
\hline
\multicolumn{1}{c|}{Parameter}            &\multicolumn{1}{c|}{~K (1.35 cm)}&\multicolumn{1}{c|}{~C (6.2 cm)}&\multicolumn{1}{c|}{~L (18 cm)}&\multicolumn{1}{c}{~P (92 cm)}\\
\hline
SRT in flight, 2011--2012: & & & &\\
1. ~~  $A_\mathrm{eff}$,     m$^2$                                & 7.5     & 35    & 41   & 30\\
2. ~~  AE = $A_\mathrm{eff} / A_\mathrm{geom}$                            & 0.1     & 0.45  & 0.52 & 0.38\\
3. ~~  $\eta_\sigma = \eta_\sigma (\sigma = 0.77$ mm)                & 0.60    & 0.98  & 1.00 & 1.00\\
\hline
4. ~~  {\textbf{Model 1:}}    $\varphi_q = 2  \delta \Psi_\mathrm{max}$.              & & & & \\
4.0~~   $\delta \Psi_\mathrm{max}$ (in degrees of phase)                  & -100     & -35    & -35   & \multicolumn{1}{c}{--}\\
4.1~~   $\varphi_q$ (in degrees of phase)     & -200     & -70    & -70   & \multicolumn{1}{c}{--}\\
4.2~~   $\eta_\varphi = \eta_\varphi (\varphi_q)$                 & 0.45    & 0.91  & 0.91 & \multicolumn{1}{c}{--}\\
4.3~~   $\eta = \eta_\sigma  \eta_\varphi$                       & 0.27    & 0.89  & 0.91 & \multicolumn{1}{c}{--}\\
4.4~~   $A_0 = A_\mathrm{eff} / \eta$,     m$^2$                   & 28      & 39    & 45   & \multicolumn{1}{c}{--}\\
4.5~~  AE$_0 = A_0 / A_\mathrm{geom}$                                     & 0.35    & 0.50  & 0.57 & \multicolumn{1}{c}{--}\\
\hline
5. ~~   {\textbf{Model 2:}}    $\varphi_q = 2  \delta \Psi_\mathrm{max} + \delta \varphi_x$.   & & & & \\
   ~~~~~~~  $\delta_1 = 5.3$~mm                                    & & & & \\
5.0~~   $\delta \varphi_x = k \delta_1/2 $ (in degrees of phase)           & -70      & -15     & -5    & \multicolumn{1}{c}{--}\\
5.1~~  $\varphi_q$ (in degrees of phase)                   & -270     & -85    & -75   & \multicolumn{1}{c}{--}\\
5.2~~   $\eta_\varphi = \eta_\varphi (\varphi_q)$                 & 0.24    & 0.87  & 0.89 & \multicolumn{1}{c}{--}\\
5.3~~   $\eta = \eta_\sigma  \eta_\varphi$                       & 0.15    & 0.85  & 0.89 & \multicolumn{1}{c}{--}\\
5.4~~   $A_0 = A_\mathrm{eff} / \eta$,     m$^2$                     & 50      & 41    & 46   & \multicolumn{1}{c}{--}\\
5.5~~   AE$_0 = A_0 / A_\mathrm{geom}$                                & 0.64    & 0.52  & 0.59 & \multicolumn{1}{c}{--}\\
\hline
6. ~~   {\textbf{Model 3:}}    $\varphi_q = \delta \varphi_1 + \delta \varphi_2$; & & & & \\
   ~~~~~~~   $\delta \varphi_1 = \delta \Psi_\mathrm{max} + k \delta_{11}/2 = 0$;             & & & & \\
   ~~~~~~~   $\delta \varphi_2 = \delta \Psi_\mathrm{max} -k \delta_{12}/2 = -1.5 \pi$;        & & & & \\
6.0~~   $\delta \varphi_x = k (\delta_{11} -\delta_{12}) /2$ (in degrees of phase)  &-70 & --& --& \multicolumn{1}{c}{--}\\
6.1~~   $\varphi_q$ (in degrees of phase)                   & -270     & \multicolumn{1}{c|}{--}& \multicolumn{1}{c|}{--}& \multicolumn{1}{c}{--}\\
6.2~~   $\eta_\varphi = \eta_\varphi (\varphi_q)$                 & 0.24      & \multicolumn{1}{c|}{--}& \multicolumn{1}{c|}{--}& \multicolumn{1}{c}{--}\\
6.3~~   $\eta = \eta_\sigma  \eta_\varphi$                       & 0.15      & \multicolumn{1}{c|}{--}& \multicolumn{1}{c|}{--}& \multicolumn{1}{c}{--}\\
6.4~~   $A_0 = A_\mathrm{eff} / \eta$,     m$^2$                     & 50        & \multicolumn{1}{c|}{--}& \multicolumn{1}{c|}{--}& \multicolumn{1}{c}{--}\\
6.5~~    AE$_0 = A_0 / A_\mathrm{geom}$                                & 0.64      & \multicolumn{1}{c|}{--}& \multicolumn{1}{c|}{--}& \multicolumn{1}{c}{--}\\
6.6~~   $\delta_{11}$, mm                                        & 7.5       & \multicolumn{1}{c|}{--}& \multicolumn{1}{c|}{--}& \multicolumn{1}{c}{--}\\
6.7~~   $\delta_{12}$, mm                                        & 13        & \multicolumn{1}{c|}{--}& \multicolumn{1}{c|}{--}& \multicolumn{1}{c}{--}\\
\hline
\end{tabular}
\flushleft{\footnotesize{}1. No modeling of phase errors was carried out at
92~cm. 2. The area-loss coefficients are denoted $\eta =\eta_\sigma
\eta_\varphi$, where  $\eta_\sigma$ characterizes losses due to random errors
with the rms deviation $\sigma$ and $\eta_\varphi$ characterizes losses due
to systematic quadratic errors $\varphi_q$, due primarily to the feed.
3. Model 1: AFA at the focus. The main phase errors in the dish aperture
are determined by a) random errors with the projected rms deviation for the
actual dish surface from an ideal parabolic surface $\sigma = 0.77$~mm and
b) systematic quadratic phase errors over the dish with the maximum value
$\varphi_q$, due to the calculated phase beam of the feed having an error
$\delta\Psi_\mathrm{max}$ for illumination of the antenna edge. 
4. Model 2: AFA not at the focus. A modest quadratic phase error
$\delta \varphi_x$ is added to Model~1, due to a longitudinal shift of the
feed from the dish focus by 5~mm.
5. Model 3: astigmatism of the feed at 1.35~cm. The single feed phase
center at 1.35~cm in Model~2  ``splits'' into two equivalent phase centers
separated by a distance $\delta_{11} + \delta_{12}$ in the orthogonal planes.
More precisely, the phase center splits into two ``optimal centers of focus
in the main planes'', since the concept of a phase center refers to the
idealized co-phased case of illumination by a spherical wave (by
definition)~[50, 51, 53], whereas the phase front is usually far from co-phased
in practice. The additional phase error with the same magnitude $\delta
\varphi_x$ is due to the longitudinal shift of the average between these
centers from the dish focus.}
\end{table*}

We emphasize that it is not possible to unambiguously establish the physical
origins of phase distortions in the antenna aperture and distinguish the
contributions to the errors of the dish, feed, and geometry of the
antenna-feeder system based on the results of inflight tests of the SRT alone,
without additional data or assumptions. We can only be sure that these are
due to some total phase error over the aperture due to 1)~differences between
the real antenna reflecting surface and an ideal parabolic surface,
2)~longitudinal and transverse shifts of the feed from the dish focus, and
3)~peculiarities of the amplitude--phase beam of the feed (including, for
example, different widths of the real amplitude beam of the feed in
orthogonal planes, with the effective width of the illumination of the
ideal dish being equal to the projected value in one plane, but a factor of
two worse in the other plane, or astigmatism of the feed, when the dish focus
and feed center of focus are close to each other in one plane but in
appreciably different positions in the other plane).

The phase distribution of the field $\varphi (x)$ at a point $0 \leq x \leq
R_0$ in a dish aperture with radius $R_0$ can be written~[54] $\varphi (x) =
\Phi (x) - \Phi_0$, where $\Phi(x)$ and $\Phi_0$ are the initial field phases
at the point $x$ and at the center of the aperture. $\Phi(x) = \Psi (\psi)
+ k (\rho +t)$ and is determined by the phase of the feed beam $\Psi (\psi)$
($0 \leq\psi \leq \psi_0$; here, $\psi$ is the angle from the dish focus
between the points $x = 0$ and $x \leq R_0$ ) and the lengths of the path
$\rho$  from the feed to the dish and the path $t$ from the dish to the
aperture ($k$ is the wavenumber). Therefore, $\Phi(x)$ is a function of the
phase beam of the feed, the dish profile (with the deviation $\delta_3$ of
the surface from an ideal surface), and the shift of the feed phase center
relative to the paraboloid focus (by $\delta_1$ in the longitudinal and
$\delta_2$ in the transverse direction relative to the antenna axis). The
deviations of these factors from their projected values give rise to a
total deviation of the phase distribution at the dish aperture from the
projected (in the ideal case, close to co-phased) value $\delta \varphi (x)$,
and can be estimated with sufficient accuracy for our purposes as~[54]
\begin{multline*}
\delta \varphi (x) = 2\delta \Psi (\psi) + k [  \pm  \delta_1 (1 -
\cos \psi) \\
- \delta_2 \sin \psi - 2 \delta_3 (1 + \cos (\psi/2)].
\mathrm{~~~~A.3}
\end{multline*}
Here, in the general case, $\delta \Psi (\psi)$ contains random and systematic
deviations of the feed phase beam. The term with $\delta_3$ is associated
with inaccuracy of the dish surface, the terms with $\delta_1$ and
$\delta_2$  reflect a systematic ``incursion'' of the phase from the center
of the aperture toward the edge due to the lack of coincidence between the
dish focus and the feed centerof focus (with even and odd functions relative
to the center of the aperture, respectively), and $k = 2 \pi /\lambda$ is
the wavenumber. A plus sign in front of $\delta_1$ in (A.3) corresponds to
a shift $\delta_1$ of the feed phase center from the focus toward the dish,
while a minus sign corresponds to a shift from the dish.

The maximum phase incursion is reached at the edge of the aperture, with
$\psi = \psi_0 = 60^{\circ}$ for the SRT. Thus, we can neglect terms with
$\delta \Psi (\psi)$ and $\delta_3$ in~(A.3). Assuming $k | \delta_2 |
\sin \psi_0 = \varphi_q \sim 1.5 \pi$, we obtain the rough estimate for the
transverse shift $|\delta_2 | \sim 12$~mm. Analogously, assuming
$k | \delta_1 | (1 -\cos \psi_0) = \varphi_q$ yields the estimate of the
longitudinal shift $|\delta_1 | \sim 20$~mm, and the estimate for the
maximum distance between the feed phase center and dish focus
$\delta_\mathrm{max} =(\delta_1^2 + \delta_2^2)^{1/2} \sim 23$~mm. On the other
hand, if the main contribution in~(A.3) is made by phase distortions of the
feed beam, setting $2\delta \Psi (\psi_0) =\varphi_q \sim 1.5 \pi$ yields
the estimate for the maximum feed phase error at the edge of the dish
required for this $\delta \Psi (\psi = 60^{\circ}) \sim 0.75 \pi = 135^{\circ}$.

Let us estimate the phase errors in the antenna-feeder system in a simple
model (Table~3), based on the computational data~[56] for the amplitude--phase
beams of the SRT feeds and the projected rms deviation for the dish surface
compared to an ideal parabolic surface $\sigma = 0.77$~mm. Let us consider
three types of model:

1) the AFA is at the focus,

2) the AFA is not at the focus,

3) the AFA at 1.35~cm has astigmatism.

Astigmatism is disregarded in models 1 and 2. It follows from the data of
[56] that, at each of the wavelengths 1.35, 6.2, and 18 cm

1)  the phase $\Phi_0$ on the beam axis varies with the azimuthal angle
$\alpha$ as $\Phi_0 \propto \alpha$, and is shifted by $\pi/2$ in the
orthogonal planes 1 and 2;

2) in each azimuthal plane, the phase can be approximated using identical
quadratic equations relative to the beam axis (dish center), with the maximum
illumination errors at the dish edge equal to $\delta \Psi_\mathrm{max}\approx
{-}100^{\circ}$ of phase at 1.35~cm and ${-}35^{\circ}$ of phase at 6.2 and
18~cm.

We thus find that

1) there may be astigmatism of the feed that transforms the phase center
into a ``phase line'' along the feed axis;

2) a quadratic phase error in the illumination of the dish edge at 1.35~cm
(${-}100^{\circ}$) may be close to the value required to explain the measured
effective area [$\delta \Psi_\mathrm{max} = \delta \Psi (\psi = 60^{\circ})
\approx {-}135^{\circ}$];

3) if the feed had a single phase center that coincided with the dish focus
($\delta_1 = 0$, $\delta \varphi_x = 0$) or was shifted from the focus away
from the dish along the axis a distance $\delta_1= \delta \varphi_x /
[k \cdot (1 - \cos 60^{\circ})]$ (so that $\delta \varphi_x \approx
{-}70^{\circ}$), then, according to~(A.3), the phase error in the aperture
$\delta \varphi_\mathrm{max}  = 2 \delta \Psi_\mathrm{max} + \delta \varphi_x$ would be
$\delta \varphi_\mathrm{max} \sim{-}200^{\circ}$ or $\delta \varphi_\mathrm{max} \sim
{-}270^{\circ}$, and the corresponding loss coefficient for the effective
area would be $\eta \sim 0.27$ or $\eta \sim 0.15$ (see Models 1 and 2 in
Table~3 for more detail).

Thus, the observed {\textit{reduction in area}} at 1.35~cm (in addition to
the expected reduction from the ideal case due to the fact that $\sigma \ne 0$)
can be explained nearly fully by these calculated quadratic phase distortions
in the SRT feed beam, according to the data of~[56].

However, in both models, the quadratic phase errors are the same and large
in planes 1 an 2. This means that such errors can explain the increase in
the width of the main lobe of the beam, but not its ellipticity.

Model 3 solves the problem of this ellipticity as well. For simplicity, we
will assume that the real astigmatism of the feed can be described using
an approximation in which the feed has two centers of focus (centers 1 and
2 in planes 1 and 2) that are shifted along the focal axis in opposite
directions from the dish focus by distances $\delta_{11}$ (center~1, toward the
dish) and $\delta_{12}$ (center~2, away from the dish). Then, using~(A.3),
the phase errors in planes 1 and 2 will be $\delta \varphi_1 = \delta
\Psi_\mathrm{max} + k  \delta_{11}  c$ and $\delta \varphi_2 = \delta \Psi_\mathrm{max} -
k  \delta_{12}  c$, where $c = (1 - \cos 60^{\circ}) = 1/2$ and the total
phase error in the aperature is $\varphi_q = \delta \varphi_1 + \delta
\varphi_2 =2 \delta \Psi_\mathrm{max} + (\delta_{11} - \delta_{12})  \pi / \lambda$.

Hence, setting $\Psi_\mathrm{max} = -100^{\circ}$, $\delta \varphi_1 = 0$, and
$\delta \varphi_2 = {-} 1.5 \pi$, we finally obtain self-consistent estimates
for the phase distortions required to explain both the measured effective
area and the ellipticity of the SRT beam; the corresponding shifts of the
centers of focus of the feed 1 and 2 are in opposite directions from the
dish focus: $\varphi_q = {-}1.5 \pi$,   $\delta_{11} = 7.5$~mm and
$\delta_{12} = 13$~mm (Table~3). The distance between these centers is
$b = \delta_{11} + \delta_{12} \sim 20$~mm, and the shift of the overall
equivalent center of focus from the dish focus is $\Delta x = (\delta_{11} -
\delta_{12}) / 2 \sim {-}3$~mm (away from the dish), due to the asymmetry
in the positions of center 1 and center 2 relative to the focus.

Since the presence of appreciable asymmetry in the shape of the main lobe
of the beam in plane~2 (where its width is larger) requires a contribution
from coma-type aberration, this suggests the presence of a transverse shift
of one of the feed phase centers relative to the focal axis of the antenna,
with a maximum cubic error ${\sim} \pi/2$. More precise conclusions about
the origins of the observed beam characteristics will require rigorous
calculations and a comparative analysis of the main possible models.

\begin{figure*}[p]
\setcounter{figure}{3}
\includegraphics[width=1.0\textwidth,angle=0,clip=true,trim=2cm 2cm 2cm 2.5cm]{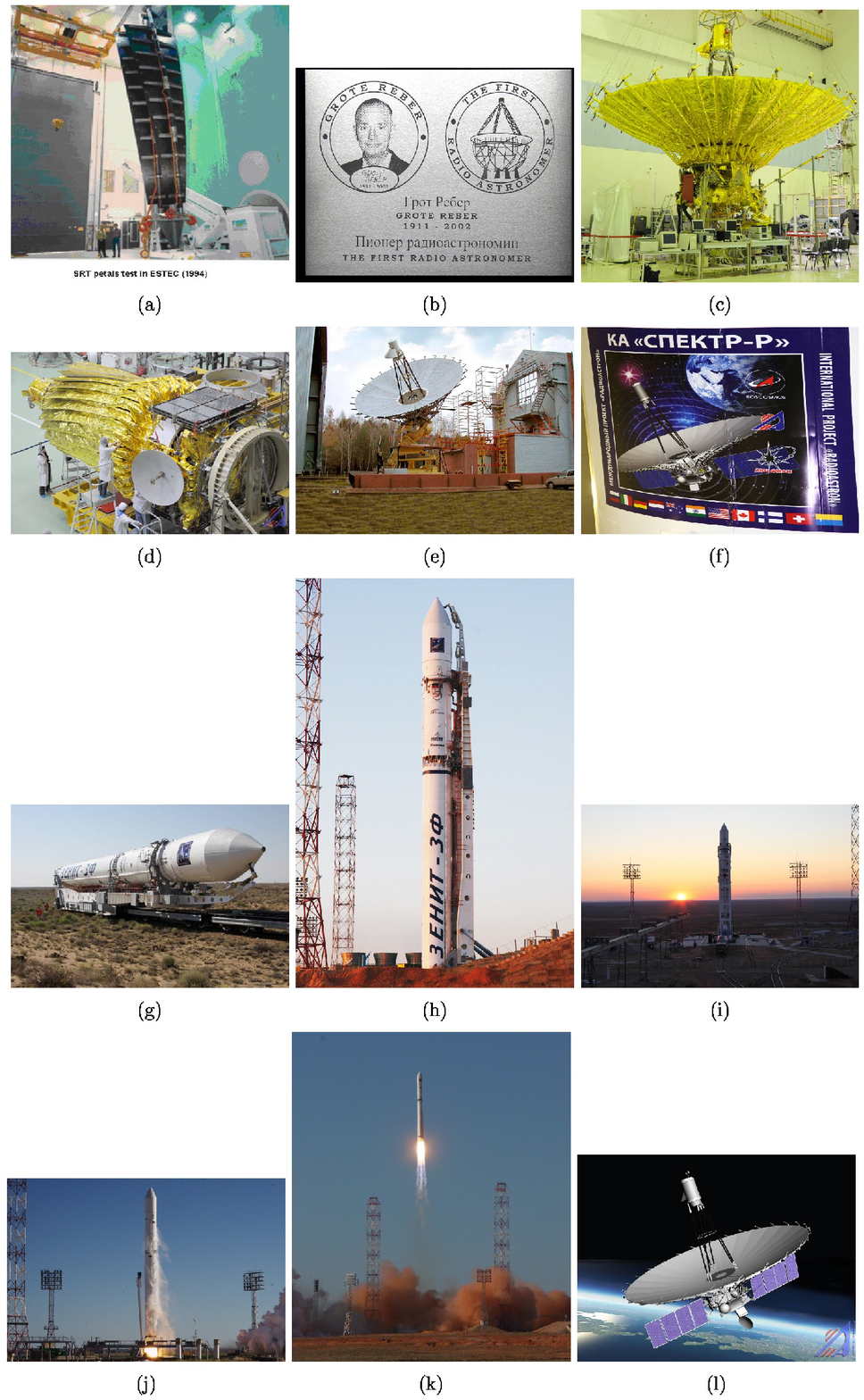}
\caption{\footnotesize (Color inset 1). 
See full caption on a separate page below.
\label{fig_inset1}
}
\end{figure*}
%

\begin{figure*}[p]
\setcounter{figure}{6}
\includegraphics[width=1.0\textwidth,angle=0,clip=true,trim=2cm 3cm 2cm 2cm]{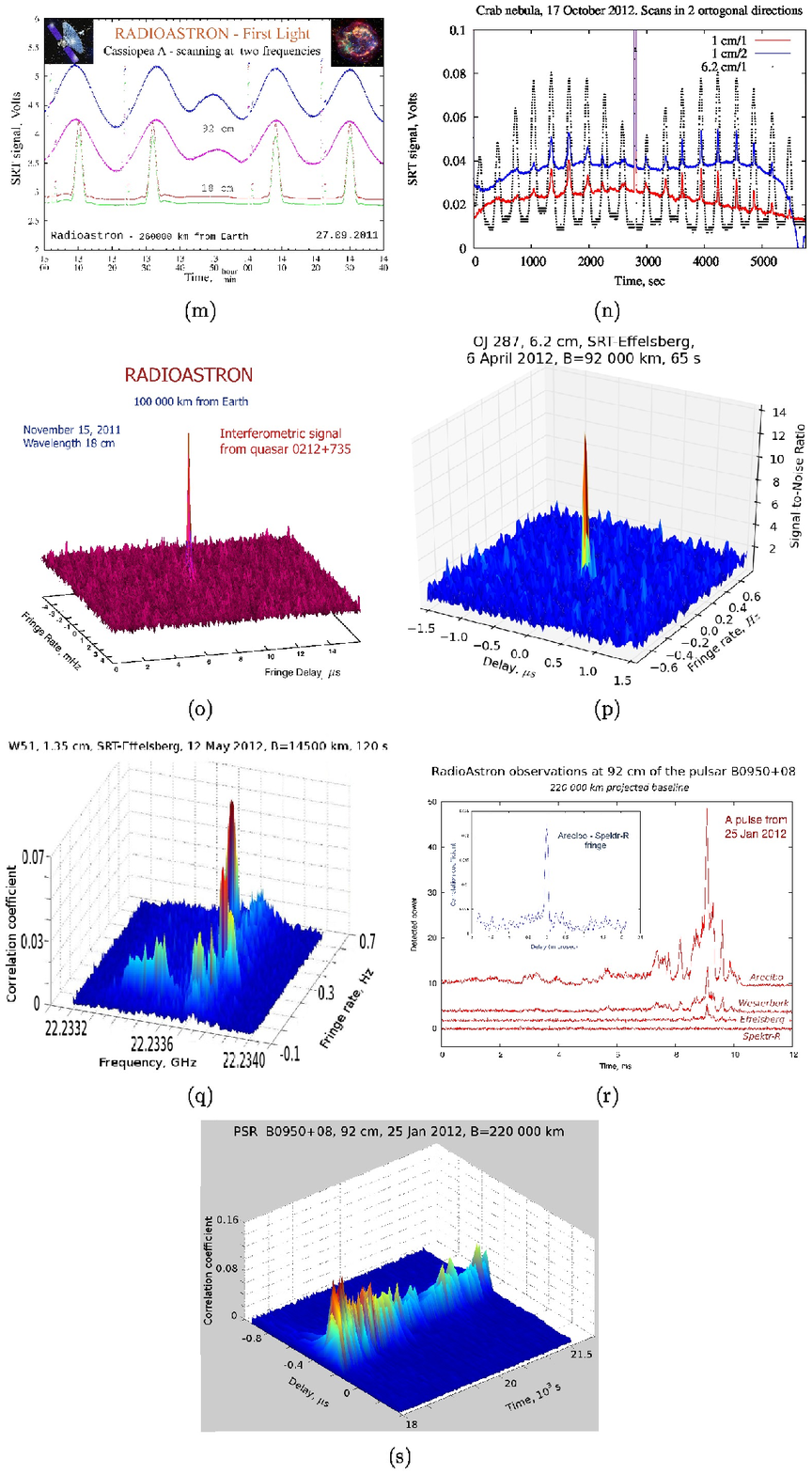}
\caption{\footnotesize (Color inset 2).
See full caption on a separate page below.
\label{fig_inset2}
}
\end{figure*}
%

\clearpage
\onecolumngrid
\noindent
\textbf{Figure~\ref{fig_inset1} (color inster 1) caption:}\\
(a) Tests of the precision carbon-fiber panels of
the radio-telescope reflector at ESTEC in 1994 (Nordwijk, the Netherlands).
(b) Memorial plate with a portrait of the first radio
astronomer Grote Reber (1911--2002) mounted on the SRT.
(c,d) Final ground tests of the SRT with the {\em Navigator}
module were carried out at the Lavochkin Association right up to the transport
of the spacecraft to the cosmodrome. Panel (d) shows the antenna of the SRT in
the folded state; the 1.5-m antenna of the HDRRC and some of the solar panels
are also visible.
(e) Tests of a model of the SRT and interferometer at the PRAO (ASC) in
2003--2004. 
(f) Image of the SRT with logos of the organizations and flags of the countries participating
in the {\em RadioAstron} project on the rocket fairing of the {\em
Zenit-3F} complex. (g)--(h) Transport of the rocket with the {\em Spektr-R}
spacecraft and the {\em Fregat} booster at the launch pad of site No.~45
of the Baikonur cosmodrome. 
(i) Launch on July 18, 2011 at 5:31:19.91
Moscow daylight savings time.
(j) Artist's impression of the SRT in orbit after it had
successfully deployed on July 23, 2011.

\bigskip
\noindent
\textbf{Figure~\ref{fig_inset2} (color inset 2) caption:}\\
(a) Radiometric responses of the telescope during
the first observation of the radio source Casseoipeia A with the SRT in
flight, on September 27, 2011, at 92 and 18~cm in left- and right-circular
polarizations. The two scans to the left were made for one cut
through the source (in the forward and reverse directions), the two scans to
the right were made perpendicular to this but, and the middle scan shows
a cut through the edge of the object made during repointing of the
antenna. The short pulses are responses to calibration signals from the
receivers. (b) Example of radiometric responses of the SRT to a source during
scanning of a portion of sky containing the Crab Nebula simultaneously at
two wavelengths~--- 1.35~cm (in both polarizations) and 6.2~cm (in one
polarization)~--- for observations with the SRT in a single-dish regime
obtained on October 17, 2011. (c) First signal obtained at the ASC from
the space--ground radio interferometer at 18~cm, for observations of the
quasar 0212+735 made by the SRT and the 100-m Effelsberg radio telescope
(Germany) on November 15, 2011. (d) Response of the space--ground interferometer
(in units of the signal-to-noise ratio) for 6.2-cm observations of the quasar
OJ287 on April 6, 2012. The projected baseline between the SRT and Effelsberg
was $B = 7.2$ Earth diameters. (e) Interferometer response for observations
of narrow water lines at 1.35~cm in maser sources in the star-forming region
W51 obtained for observations made by the SRT and the Effelsberg telescope
on May 12, 2012. The projected baseline is $B = 1.14$ Earth diameters
(14\,500~km), corresponding to an angular resolution of about 0.0002~arcsec.
The integration time was 120~s. The vertical axis plots the correlation
coefficient for the interferometer response as a function of the observing
frequency and the fringe rate. (f) Four-antenna interferometric observations
of a single pulse from the pulsar PSR~B0950+08 made on January 25, 2012,
involving the SRT and the Arecibo (Puerto Rico, USA), Westerbork (Netherlands),
and Effelsberg (Germany) ground radio telescopes. The projected baseline
is 220\,000~km. (g) Same as (f) for the accumulation of a long series of
pulses. The variation of the signal with time over an hour is clearly
visible.

\end{document}